\documentclass[a4paper,10pt]{article}

\pdfoutput=1 
\usepackage{jheppub} 
\usepackage[T1]{fontenc} 
\usepackage[utf8]{inputenc}

\usepackage{slashed}

\newcommand{\secn}[1]{Section~\ref{#1}}
\newcommand{\appn}[1]{Appendix~\ref{#1}}
\newcommand{\beq}{\begin{eqnarray}}
\newcommand{\eeq}{\end{eqnarray}}

\newcommand{\as}{\alpha_{\mbox{\tiny{S}}}}
\newcommand{\eps}{\epsilon}

\newcommand{\sig}{\sigma}
\newcommand{\Gam}{\Gamma}
\newcommand{\gam}{\gamma}

\newcommand{\lam}{\lambda}

\newcommand{\npo}{{n+1}}
\newcommand{\npt}{{n+2}}

\newcommand{\pnpoh}{\widehat{\Phi}_\npo}

\newcommand{\ppon}{\Phi_1}

\newcommand{\mc}{\mathcal}

\newcommand{\ep}{1/\eps}

\newcommand{\dt}{\!\cdot\!}

\newcommand{\lra}{\leftrightarrow}

\newcommand{\one}{\, (\mathbf{1})}
\newcommand{\two}{\, (\mathbf{2})}
\newcommand{\otwo}{\, (\mathbf{12})}
\newcommand{\twosc}{(\mathbf{2, \, SC})}
\newcommand{\RV}{(\mathbf{RV})}
\newcommand{\LO}{{\mbox{\tiny{LO}}}}
\newcommand{\NLO}{{\mbox{\tiny{NLO}}}}
\newcommand{\NNLO}{{\mbox{\tiny{NNLO}}}}

\newcommand{\nnb}{\nonumber}
\newcommand{\sub}{\mbox{\tiny{sub}}}

\newcommand{\kt}{\tilde{k}}

\newcommand{\Si}{{\bf S}_i}
\newcommand{\Sj}{{\bf S}_j}

\newcommand{\Cj}{{\bf C}_{ij}}
\newcommand{\Cl}{{\bf C}_{kl}}

\newcommand{\bSi}{\overline{\bf S}_i}
\newcommand{\bSj}{\overline{\bf S}_j}
\newcommand{\bCj}{\overline{\bf C}_{ij}}
\newcommand{\SCj}{{\bf SC}_{ijk}}
\newcommand{\SCl}{{\bf SC}_{ikl}}
\newcommand{\CCj}{{\bf C}_{ijk}}

\newcommand{\CCl}{{\bf C}_{ijkl}}
\newcommand{\SSk}{{\bf S}_{ik}}
\newcommand{\SSj}{{\bf S}_{ij}}

\newcommand{\q}{{q}}

\def\eq#1{Eq.~(\ref{#1})}

\newcommand{\bL}[2]{{\bf L}^{#1}_{#2}}
\newcommand{\bS}[1]{{\bf S}_{#1}}
\newcommand{\bC}[1]{{\bf C}_{#1}}
\newcommand{\bSC}[1]{{\bf SC}_{#1}}
\newcommand{\bCS}[1]{{\bf CS}_{#1}}

\newcommand{\bbS}[1]{\overline{\bf S}_{#1}}
\newcommand{\bbC}[1]{\overline{\bf C}_{#1}}
\newcommand{\bbSC}[1]{\overline{\bf SC}_{#1}}
\newcommand{\bbCS}[1]{\overline{\bf CS}_{#1}}

\newcommand{\W}[1]{\mc{W}_{#1}}
\newcommand{\bW}[1]{\overline{\mc{W}}_{#1}}
\newcommand{\Wab}[1]{\mc{W}^{(\alpha\beta)}_{#1}}

\newcommand{\kkl}[1]{\{\bar k\}^{(#1)}}
\newcommand{\kk}[2]{\bar k_{#1}^{(#2)}}
\newcommand{\sk}[2]{\bar s_{#1}^{(#2)}}

\newcommand{\Norm}{\mc{N}_1}
\newcommand{\hc}{{\rm \, hc}}
\newcommand{\so}{{\rm \, s}}
\newcommand{\onehc}{\, (\mathbf{1}, \hc)}
\newcommand{\ones}{\, (\mathbf{1}, \so)}
\newcommand{\otwohc}{\, (\mathbf{12}, \hc)}
\newcommand{\otwos}{\, (\mathbf{12}, \so)}
\newcommand{\Bn}{B}
\newcommand{\Vl}{V}
\newcommand{\Rl}{R}
\newcommand{\RR}{RR}
\newcommand{\RVl}{RV}
\newcommand{\VV}{VV}

\newcommand{\varsi}{\varsigma}

\title{\boldmath Local Analytic Sector Subtraction at NNLO}

\author{L.~Magnea, E.~Maina, G.~Pelliccioli, C. Signorile-Signorile, 
P.~Torrielli, and S.~Uccirati}

\affiliation{Dipartimento di Fisica and Arnold-Regge Center, Universit\`a di Torino,\\
                 and INFN, Sezione di Torino, Via P. Giuria 1, I-10125 Torino, Italy}

\emailAdd{magnea@to.infn.it}
\emailAdd{maina@to.infn.it}
\emailAdd{gpellicc@to.infn.it}
\emailAdd{signoril@to.infn.it}
\emailAdd{torriell@to.infn.it}
\emailAdd{uccirati@to.infn.it}

\abstract{We present a new method for the local subtraction of infrared divergences 
at next-to-next-to-leading order (NNLO) in QCD, for generic infrared-safe 
observables. Our method attempts to conjugate the minimal local counterterm 
structure arising from a sector partition of the radiation phase space with the 
simplifications following from analytic integration of the counterterms. In this 
first implementation, the method applies to final-state massless particles. We 
show how our method compactly organises infrared subtraction at NLO, we deduce
in detail the general structure of the subtraction terms at NNLO, and we provide
a proof of principle with a complete application to a simple process at NNLO.}

\begin{document} 
\maketitle
\flushbottom

%%%%%%%%%%%%%%%%%%%%%%%%%%%%%%%%%

\section{Introduction}
\label{intro}

The increasing precision of experimental measurements at the Large Hadron 
Collider (LHC), together with the complexity of the final states currently probed in 
hadronic collisions, constitute a severe challenge for theoretical calculations.
This challenge has driven the development of a number of novel techniques,
for precision calculations of scattering amplitudes to high orders, for the study 
of final-state hadronic jets, and for the accurate determination of parton distribution 
functions (see, for example, Ref.~\cite{Bendavid:2018nar} for a review of recent 
developments). In particular, a consequence of the current and expected 
precision of experimental data is the fact that the next-to-next-to-leading 
perturbative order (NNLO) in QCD is rapidly becoming the required accuracy 
standard for fixed-order predictions at LHC. A crucial ingredient for the 
calculation of differential distributions to this accuracy is the treatment of 
infrared singularities, which arise both in virtual corrections to the relevant 
scattering amplitudes, and from the phase-space integration of unresolved 
real radiation.

In principle, the problem is well understood. Infrared singularities (soft and 
collinear) arise in virtual corrections as poles in dimensional regularisation, 
and all such poles are known to factorise from scattering amplitudes in terms 
of universal functions, which admit general definitions in terms of gauge-invariant 
matrix elements \cite{Collins:1989bt,Sterman:1995fz,Catani:1998bh,Sterman:2002qn,
Dixon:2008gr,Gardi:2009qi,Gardi:2009zv,Becher:2009cu,Becher:2009qa,Feige:2014wja}. 
These functions are in turn determined by a small set of anomalous dimensions which,
in the massless case, are fully known up to three loops~\cite{Almelid:2015jia,
Almelid:2017qju}. General theorems then ensure that, when considering infrared-safe 
cross sections, virtual infrared poles must either cancel, when combined with 
singularities arising from the phase-space integration of final-state unresolved 
radiation~\cite{Bloch:1937pw,Kinoshita:1962ur,Lee:1964is,Grammer:1973db}, or 
be factored into the definition of parton distribution functions, in the case of collinear 
initial-state radiation~\cite{Collins:1989gx}. Real-radiation matrix elements have 
also been shown to factorise in soft and collinear limits, and the corresponding 
splitting kernels are fully known at order $\as^2$~\cite{Campbell:1997hg,Catani:1998nv,
Bern:1999ry,Catani:1999ss,Kosower:1999xi,Catani:2000pi}, with partial information 
available at $\as^3$ as well~\cite{DelDuca:1999iql,Duhr:2013msa,Li:2013lsa,
Banerjee:2018ozf,Bruser:2018rad}.

Even with this detailed knowledge of the relevant theoretical ingredients, the
practical problem of constructing efficient and general algorithms for handling
infrared singularities for generic infrared-safe observables beyond next-to-leading 
order (NLO) proves to be highly non-trivial. The origin of the difficulty lies in the 
fact that typical hadron-collider observables have a complicated phase-space 
structure, nearly always involving jet-reconstruction algorithms as well as 
complex kinematic cuts; furthermore, real-radiation matrix elements become 
increasingly intricate, and they cannot be analytically integrated in $d$ dimensions.
Integration over unresolved radiation must therefore be performed numerically in $d=4$, 
and all infrared singularities must be cancelled before this stage of the calculation 
is reached. This cancellation involves a careful use of approximations to the 
real-radiation matrix elements in the singular regions, and requires a remapping 
of the real-radiation phase space to match the Born-level configurations.

At NLO, the first fully differential results for jet cross sections were 
obtained~\cite{Giele:1993dj,Giele:1994xd} by isolating singular phase-space
regions and treating them separately, performing the pole cancellation 
by integrating approximate matrix elements within those regions (a procedure 
usually described as `slicing'). Subsequently, two general algorithms were 
developed, the {\it FKS}~\cite{Frixione:1995ms} and {\it CS}~\cite{Catani:1996vz} 
subtraction methods, based on the idea of introducing local counterterms 
for all singular regions of phase space, and then integrating them exactly 
in order to achieve the cancellation of poles without need of slicing 
parameters (which is usually described as `subtraction' in a strict sense). 
These algorithms are currently implemented in full generality in fast and 
efficient NLO generators~\cite{Campbell:1999ah,Gleisberg:2007md,
Frederix:2008hu,Czakon:2009ss,Hasegawa:2009tx,Frederix:2009yq,
Alioli:2010xd,Platzer:2011bc,Reuter:2016qbi}, so that the `subtraction 
problem' can be considered solved to this accuracy.

At NNLO, numerical and conceptual challenges related to the proliferation
of overlapping singular regions become much more significant. This has led
to the development of several different methods, which have been successfully 
applied to a number of simple collider processes. NNLO differential distributions
for hadronic final states in electron-positron annihilation were first computed 
in~\cite{GehrmannDeRidder:2008ug,Weinzierl:2008iv}, while among the first 
hadronic processes involving coloured final-state particles to be studied 
differentially at NNLO there were the production of top-antitop  quark pairs, 
achieved in~\cite{Czakon:2015owf,Czakon:2016ckf} within the {\it Stripper} 
framework~\cite{Czakon:2010td}, and the associated production 
of a Higgs boson and a jet, achieved with the {\it N-Jettiness} slicing 
technique~\cite{Boughezal:2015dra,Gaunt:2015pea,Boughezal:2015ded,
Boughezal:2016wmq}. A number of hadronic processes with up to two 
final-state coloured particles  at Born level have since been studied at 
the differential level with various approaches, including $q_T$ 
slicing~\cite{Catani:2007vq,Grazzini:2017mhc,Grazzini:2017ckn,
deFlorian:2016uhr}, and {\it Antenna} subtraction~\cite{Ridder:2015dxa,
Currie:2016ytq,Currie:2017eqf}. 

There are several reasons to surmise that existing methods for NNLO 
subtraction can be generalised and improved: on the one hand, current 
applications have been computationally very demanding, either in terms 
of the analytic calculations involved, or because of the large-scale numerical
effort required; on the other hand, it is clear that precise NNLO predictions
will soon be needed for more complicated processes, such as the production
of more than two jets, and it will similarly be useful to compute simple processes
at the next order in perturbation theory, N$^3$LO. The need for improved 
and efficient subtraction algorithms is in fact leading to the development 
of other methods, or the refinement of existing ones: examples include 
the {\it CoLoRFulNNLO} framework~\cite{Tulipant:2017ybb,DelDuca:2016csb,
DelDuca:2016ily}, currently applied to processes with electroweak initial states, 
the {\it Projection to Born} method~\cite{Cacciari:2015jma}, and the technique 
of {\it Nested Soft-Collinear subtractions}~\cite{Caola:2017dug,Caola:2017xuq}. 
New ideas are also being introduced~\cite{Sborlini:2016hat,Herzog:2018ily}, 
and the first limited applications to differential N$^3$LO processes have 
appeared~\cite{Dreyer:2016oyx,Dulat:2017prg,Currie:2018fgr}.

In this paper, and in a companion paper devoted to the underlying factorisation
framework \cite{Magnea:2018ebr}, we present a new approach to the subtraction
problem beyond NLO, which attempts to re-examine the fundamental building 
blocks of the subtraction procedure, take advantage of all available information, 
and build a minimal structure which will hopefully help to streamline and simplify 
future applications. The ideal subtraction algorithm, in our view, should aim
to achieve the following goals: complete generality across infrared-safe 
observables; exact locality of infrared counterterms in the radiative phase 
space; independence from `slicing' parameters identifying singular regions 
of phase space; maximal usage of analytic information in the construction
and integration of the counterterms; and, of course, computational efficiency
of the numerical implementation. These are, clearly, overarching goals,
and in this paper we present the first basic tools that we hope to use in 
future more general implementations. In particular, we focus for the moment 
on the case of massless final-state coloured particles.

In order to achieve the desired simplicity, we attempt to take maximal advantage
of the available freedom in the definition of the local infrared counterterms,
exploiting and extending ideas that have been successfully implemented at NLO.
In particular, a key element of our approach is the partition of phase space in
sectors, each of which is constrained to contain a minimal subset of soft and
collinear singularities, in the spirit of {\it FKS} subtraction~\cite{Frixione:1995ms}.
A crucial ingredient is then the choice of `sector functions' used to build the
desired partition: these functions must obey a set of sum rules in order to 
simplify the analytic integration of counterterms when sectors are appropriately 
recombined. A second crucial ingredient is the availability of a flexible family 
of parametrisations of momenta within each sector, allowing for simple 
mappings to Born configurations in different unresolved regions. Finally, 
it is necessary to take maximal advantage of the simple structure of factorised 
kernels in multiple singular limits, which follows in general from the factorised 
structure of scattering amplitudes: a detailed analysis of this structure will 
be presented in~\cite{Magnea:2018ebr}.

With this general strategy in mind, we begin in \secn{NLO} by revisiting the NLO
subtraction problem. We define sector functions satisfying our requirements, 
we introduce local counterterms and appropriate parametrisations, and 
we integrate the counterterms on the unresolved phase space. Effectively,
\secn{NLO} constructs a complete NLO subtraction algorithm for massless final 
states, which stands out for the simplicity of the required integrations. In 
\secn{NNLO} we attack the NNLO problem, displaying the general structure 
of subtractions in our approach, defining sector functions, and constructing
all local counterterms relevant for massless final states. We then perform
the relevant integrations for a specific subset of singularities, and, in 
\secn{Proof},  we use the results to complete a proof-of-concept calculation
of NNLO subtraction for the leptonic production of two quark pairs. We conclude in
\secn{sec:conclusions}, outlining the status of our method and the forthcoming
steps needed to construct a competitive algorithm. Four appendices contain 
a number of technical details.

%%%%%%%%%%%%%%%%%%%%%%%%%%%%%%%%%

\section{Local analytic sector subtraction at NLO}
\label{NLO}

%%%%%%%%%%%%%%%%

\subsection{Generalities}
\label{NLOsetup}

We restrict our analysis to reactions featuring only massless particles, with $n$ 
partons appearing in the final state at Born level. We assume a
colour-singlet initial state, and we allow for coloured and colourless particles
in the final state, the latter not affecting our arguments. Scattering
amplitudes involving  $n$ final-state partons with momenta $k_i$, $i = 1, \ldots, n$,
with $k_i^2 = 0$, are expanded in perturbation theory as
\beq
  {\cal A}_n (k_i) \, = \, {\cal A}_n^{(0)} (k_i) \, + \, {\cal A}_n^{(1)} (k_i) \, + \,
  {\cal A}_n^{(2)} (k_i) \, + \, \ldots \, ,
\label{pertexpA}
\eeq
with ${\cal A}_n^{(0)}$ describing the Born process. Correspondingly, differential 
cross sections with respect to any infrared-safe\footnote{We use the term {\it 
infrared} (IR) to indicate collectively soft {\it and} collinear singularities.} observable 
$X$ are schematically written as
\beq
  \frac{d \sig}{d X} \, = \, \frac{d \sig_\LO}{d X} \, + \, \frac{d \sig_\NLO}{d X} \, + \, 
  \frac{d \sig_\NNLO}{d X} \, + \, \ldots \, ,
\label{pertexpsig}
\eeq
where, up to NLO,
\beq
  \frac{d \sig_\LO}{dX} & = & \int d \Phi_n \, \Bn \, \delta_n (X) \, , 
\label{eq:LO_struct} \\
  \frac{d \sig_\NLO}{d X} & = & \int d \Phi_n \, \Vl \, \delta_n (X) + 
  \int d \Phi_\npo \, \Rl \, \delta_\npo (X) \, .
\label{eq:NLO_struct}
\eeq
In Eqs.~(\ref{eq:LO_struct}) and (\ref{eq:NLO_struct}), $\Bn$, $\Rl$, and $\Vl$ 
denote the Born, real, and virtual contributions, respectively, with
\beq
  && \hspace{2mm} \Bn \, = \, \left| {\cal A}_n^{(0)} \right|^2 \, , \qquad 
  \Rl \, = \, \left| {\cal A}_\npo^{(0)} \right|^2 \, ,  
  \qquad
  \Vl \, = \, 2 \, {\bf Re} \left[ {\cal A}_n^{(0) *} \, 
  {\cal A}_n^{(1)} \right] \, ,
\label{pertsigNLOparts}
\eeq
where the virtual correction has been renormalised in the $\overline{\rm MS}$ 
scheme. Furthermore, $\delta_i (X) \equiv \delta(X - X_i)$, with $X_i$ representing 
the observable under consideration, computed with $i$-body kinematics.

In dimensional regularisation, in $d = 4 - 2 \eps$ space-time dimensions, the 
virtual contribution features up to double IR poles in $\eps$, while the real 
contribution, finite in $d = 4$, is characterised by up to two overlapping singular 
limits of soft and collinear nature in the radiation phase space. The phase-space
integration of such singularities in $d$ dimensions results in explicit 
poles in $\eps$, which cancel those of virtual origin if $X$ is infrared
safe, ensuring the finiteness  of the cross section~\cite{Kinoshita:1962ur,Lee:1964is}.

The NLO-subtraction procedure avoids analytic integration of the full real-radiation
amplitudes by adding and subtracting to \eq{eq:NLO_struct} a counterterm
\beq
  \left. \frac{d \sig_\NLO}{d X} \right|_{\rm ct} \, = \, \int d \pnpoh \, \overline K \,\, 
  \delta_n (X) \, .
\label{Ct}
\eeq
The combination $d \pnpoh \, \overline K$ must reproduce all singular limits of 
the real-radiation contribution $d \Phi_\npo \, \Rl$, and must be sufficiently simple 
to be analytically integrated in $d$ dimensions. Note that we allow for the possibility
of simplifying the phase-space measure $d \Phi_\npo$ to $d \pnpoh$ in the counterterm,
under the assumption that the two coincide in all singular limits. Defining now the
(single) radiation phase space as $d \widehat{\Phi}_{\rm rad} = d \pnpoh/d \Phi_n$,
we may introduce the integrated counterterm
\beq
  I \, = \, \int d \widehat{\Phi}_{\rm rad}  \, \overline K \, ,
\label{eq:int_count}
\eeq
and rewrite identically the NLO cross section in \eq{eq:NLO_struct} in subtracted 
form as
\beq
  \frac{d \sig_\NLO}{dX} & = & \int d \Phi_n \, \Big( \Vl +  I \Big) \, 
  \delta_n (X) \nonumber \\ 
  && \hspace{5mm} + \, \int \left( d \Phi_\npo \, \Rl \, \delta_\npo (X) - 
  d \pnpoh \, \overline K \, \delta_n (X) \right) \, ,
\label{sigNLOsub}
\eeq
where the first and the second lines are separately finite in $d = 4$ and do not 
present any phase-space singularities, allowing an efficient numerical integration.

%%%%%%%%%%%%%%%%

\subsection{Sector functions}
\label{NLOsecfun}

Our first step in setting up the subtraction formalism at NLO is to introduce a 
partition of the real-radiation phase space by means of {\it sector functions} 
$\mc W_{ij}$, inspired by the {\it FKS} method~\cite{Frixione:1995ms}, and
satisfying the following properties
\beq
  \sum_{i, \, j \neq i} \mc W_{i j} & = & 1 \, , 
\label{eq:sec_prop1}
\eeq
\beq
  \Si \, \mc W_{ab} & = & 0 \, , \qquad \forall \, i \ne a \, ,
\label{eq:sec_prop2} \\ [4pt]
  \Cj \, \mc W_{ab} & = & 0  \, , \qquad \forall \, a b \notin \pi(i j) \, , 
\label{eq:sec_prop3}
\eeq
\beq
  \Si \, \sum_{k \neq i} \mc W_{ik} \, = \, 1 \, , 
\qquad
  \Cj \hspace{-3mm} \sum_{ab \, \in \, \pi(i j)} 
  \! \! \! \! \! \mc W_{ab} \, = \, 1 \, ,
\label{eq:sec_prop4}
\eeq
where $\pi(ij) = \{ij,\,ji\}$. $\Si$ and $\Cj$ are projection operators on 
the limits in which parton $i$ becomes soft ({\it i.e.}~all components of its 
four-momentum approach zero), and partons $i$ and $j$ become collinear 
({\it i.e.}~their relative transverse momentum approaches zero), respectively: 
the action of these operators on matrix elements and sector functions will 
be described in detail below. \eq{eq:sec_prop1} is a normalisation condition 
that recognises the $\mc W_{ij}$ functions as a unitary partition of phase 
space. \eq{eq:sec_prop2} and \eq{eq:sec_prop3} express the fact that a 
given sector function $\mc W_{ij}$ selects \emph{only} one soft and one 
collinear singular configurations, $\Si$ and $\Cj$, respectively, among all 
those present in the real-radiation matrix element. The sum rules in 
\eq{eq:sec_prop4} imply that, upon summing over \emph{all} combinations 
of indices associated to sectors that survive in a given soft or collinear limit, 
the corresponding sector functions reduce to unity. This fact proves crucial 
for the analytic integration of the subtraction counterterms, as is well known 
in the \emph{FKS} method, and as we will further discuss in the following; 
analytic counterterm integration in turn makes it possible to show in closed 
form the correctness of the singularity structure of the subtraction terms.

There is ample freedom in the choice of sector functions, the only requirement 
being that they satisfy the relations (\ref{eq:sec_prop1}) to (\ref{eq:sec_prop4}). 
In order to provide an explicit definition of $\mc W_{ij}$, let us introduce some  
notation: let $s$ be the squared centre-of-mass energy, $\q^\mu = (\sqrt s,\vec 0\,)$ 
the centre-of-mass four-momentum, and $k_i^\mu$ $(i = 1, \ldots, n+1)$ the 
$n + 1$ final-state momenta of the radiative amplitude. We set
\beq
  s_{\q i} & = & 2 \, \q \cdot k_i \, ,\qquad s_{ij} \, = \, 2 \, k_i \cdot k_j \, , \nnb\\ 
  [3pt] 
  e_i & = & \frac{s_{\q i}}{s} \, , \qquad \quad \hspace{1pt}  
  w_{ij} \, = \, \frac{s \, s_{ij}}{s_{\q i} \, s_{\q j}} \, .
\label{eq:Ewdef}
\eeq
We now define NLO sector functions as (see also~\cite{Frederix:2009yq})
\beq
  \mc W_{ij} \, = \, \frac{\sigma_{ij}}{\sum\limits_{k, \, l \neq k} \sigma_{kl}} \, , 
  \qquad \quad 
  {\rm with} \qquad \sigma_{ij} \, = \, \frac{1}{e_i \, w_{ij}} \, .
\label{eq:sfunNLO}
\eeq
The double sum in \eq{eq:sfunNLO} runs over all massless final-state partons, 
including those that are not associated with singular limits. This choice is 
made in order to ease NNLO extensions, as detailed below. With the definition 
in \eq{eq:sfunNLO}, it is easy to verify that all properties in Eqs.~(\ref{eq:sec_prop1}) 
to (\ref{eq:sec_prop4}) are satisfied, and in particular one finds that
\beq
  \Si \, \mc W_{ab} \, = \, \delta_{ia} \, \frac{1/w_{ab}}{\sum\limits_{l \neq a} 
  1/w_{a l}} \, , \quad \quad
  \Cj \, \mc W_{ab} \, = \, \left( \delta_{ia} \delta_{jb} + \delta_{ib} \delta_{ja} \right)
  \frac{e_b}{e_a + e_b} \, ,
\label{limsecnlo}
\eeq
from which the desired properties follow.

%%%%%%%%%%%%%%%%

\subsection{Definition of local counterterms}
\label{sec:NLOcntdef}

As discussed above, properties~(\ref{eq:sec_prop2}) and~(\ref{eq:sec_prop3}) 
ensure that, in a given sector $ij$, only the $\Si$ and the $\Cj$ limits (as well as 
their product) act non-trivially. A \emph{candidate} local counterterm $K_{ij}$ 
for the real matrix element $\Rl$ in this sector can thus be built collecting all 
terms in the product $\Rl \, \mc W_{ij}$ that are singular in such soft and collinear 
limits, and taking care of correcting for the double counting of the soft-collinear 
region. We define therefore
\beq
  K_{ij} & = & \left( \Si + \Cj - \Si \, \Cj \right) \, \Rl \, \mc W_{ij} 
  \,  \equiv \,  \bL{\one}{ij}\, \Rl \, \mc W_{ij} \, ,
  \label{eq:NLOK}
  \\[5pt]
  K & = & \sum_{i, \, j \neq i} K_{ij}  \, = \,  
  \sum_{i,\,j\neq i} 
  \left( \Si + \Cj - \Si \, \Cj \right) \, \Rl \, \mc W_{ij}
  \nnb\\
  & = & \sum_{i} \Big[\sum_{j\neq i} \Si \,\W{ij}\Big] \, \Si \, \Rl+ \sum_{i, \, j > i} 
  \Big[\Cj \big(\W{ij}+\W{ji}\big)\Big] \Cj\, \Rl
  -\sum_{i,j\neq i} \Big[\Si\,\Cj\,\W{ij}\Big]\Si\,\Cj\,\Rl \, .\qquad
  \label{eq:NLOKbis}
\eeq
Here and in the following, projection operators are understood to act on \emph{all} 
quantities to their right, unless explicitly separated by parentheses: for instance 
in the expression $( \Si \, A ) \,B$ the soft limit is meant to act only on $A$, and
not on $B$. In \eq{eq:NLOK}, the term featuring the composite operator $\Si \, \Cj$ 
removes the soft-collinear singularity, which is double-counted in the sum 
$\Si + \Cj$; the order in which the projectors act is arbitrary, as they commute 
(see \appn{app:commutationNLO}). As will be detailed in \secn{sec:NLOcntint}, 
and can be deduced from the sum rules in Eqs.~(\ref{eq:sec_prop4}), the content 
of each square bracket in \eq{eq:NLOKbis} is equal to 1 upon summation over 
sectors, a crucial property for counterterm integration.

Our candidate counterterm $K_{ij}$ is structurally similar to, and as simple as, 
the {\it FKS} counterterm for sector $ij$, however it has the advantage of being 
defined without any explicit parametrisation of the soft and collinear limits. Its 
constituent building blocks are the universal soft and collinear NLO kernels
which factorise from the radiative amplitude in the singular limits. We write
\beq
  \Si \, \Rl \left( \{ k \} \right) & = & - \, \Norm \,
  \sum_{\substack{l \neq i\\ m \neq i}} \, \mc I_{l m}^{(i)} \, 
  {\Bn}_{l m} \! \left( \{ k \}_{\slashed i} \right) \, ,
\label{eq:SiR} \\ [3pt]
  \Cj \, \Rl \left( \{ k \} \right) & = & \frac{\Norm}{s_{ij}} \, \Big[
  P_{ij} \, \Bn \! \left( \{ k \}_{\slashed i \slashed j}, k \right) + \, 
  Q_{ij}^{\mu \nu} \, 
  {\Bn}_{\mu \nu} \! \left( \{ k \}_{\slashed i \slashed j}, k \right) \Big] \nnb \\
  & \equiv & 
  \frac{\Norm}{s_{ij}} \, P^{\mu\nu}_{ij} \, 
  {\Bn}_{\mu \nu} \! \left( \{ k \}_{\slashed i \slashed j}, k \right) \, ,
\label{eq:CijR1} \\ [3pt]
  \Si \, \Cj \, \Rl \left( \{ k \} \right) & = & \frac{\Norm}{s_{ij}} \,\, 
  \Si \,  P_{ij} \, \Bn \! \left( \{ k \}_{\slashed i \slashed j}, k \right)
  \, = \, 2 \, \Norm \, C_{f_j} \, \mc I_{jr}^{(i)} \, 
  {\Bn} \! \left( \{ k \}_{\slashed i} \right) \, ,
\label{eq:SiCijR1}
\eeq
where we introduced several notations. Specifically, the prefactor $\Norm$ is 
defined as
\beq
  \Norm \, = \, 8 \pi \as \left( \frac{\mu^2 e^{\gam_E}}{4 \pi} \right)^{\eps} \, ,
\label{norm1}
\eeq
where $\mu$ is the renormalisation scale and $\gamma_E$ the Euler-Mascheroni 
constant; $\{k\}$ is the set of the $\npo$ final-state momenta in the 
radiative amplitude, while $\{k\}_{\slashed i}$ is the set of $n$ momenta 
obtained from $\{k\}$ by removing $k_i$; when a function takes the argument 
$( \{k\}_{\slashed i \slashed j}, k)$, it depends on the set of $n$ momenta 
obtained from $\{k\}$ by removing $k_i$ and $k_j$, and inserting their 
sum $k = k_i + k_j$; finally, $\Bn$ is the Born-level squared matrix element 
defined in \eq{pertsigNLOparts}, while
\beq
B_{lm} \, = \, {\cal A}_n^{(0)*} \, ({\bf T}_l \cdot {\bf T}_m) \, {\cal A}_n^{(0)}
\eeq
is the colour-connected Born-level squared matrix element, with ${\bf T}_a$ 
colour generators, and $B_{\mu\nu}$ is the spin-connected Born-level squared 
matrix element, obtained by stripping the spin polarisation vectors of the particle
with momentum $k$ from the Born matrix element and from its complex conjugate.

The NLO soft and collinear kernels are of course well known. In our 
notation, the eikonal kernel $\mc I_{lm}^{(i)}$, relevant for soft-gluon 
emissions, is given by
\beq
  \mc I_{l m}^{(i)} \, = \, \delta_{f_i g} \, \frac{s_{l m}}{s_{i l} \, s_{i m}} \, ,
\label{eq:eikkern}
\eeq
where $f_i$ indicates the flavour  of parton $i$, so that $\delta_{f_i g} = 1$ 
if parton $i$ is a gluon, and $\delta_{f_i g} = 0$ otherwise. In order to write 
the collinear kernels, we begin by introducing a Sudakov parametrisation for 
the momenta $k_i^\mu$ and $k_j^\mu$, as they become collinear. We 
introduce a massless vector $\bar k^\mu$, defining the collinear direction, 
using
\beq
  k^\mu \, \equiv \, k_i^\mu + k_j^\mu \, , \qquad \qquad 
  \bar k^\mu \, \equiv \, k^\mu - \frac{s_{i j}}{s_{i r} + s_{j r}} k_r^\mu \, ,
\label{colldir}
\eeq
where $k^2 = 2 \, k_i \cdot k_j = s_{i j}$, and $k_r$ is a massless reference 
vector (for example one of the on-shell momenta of the set $\{k\}$, with $r 
\neq i, j$), so that $\bar k^2 = 0$. We now write a Sudakov parametrisation 
of $k_a$ ($a = i, j$), as
\beq
  k_a^\mu \, = \, x_a \, \bar k^\mu + \kt_a^\mu - 
  \frac{1}{x_a} \, \frac{\kt_a^2}{2 \, k \dt k_r} \, k_r^\mu \, , 
\label{sudai}
\eeq
where we defined the transverse momenta $\kt_a^\mu$ with respect to the 
collinear direction $\bar k$, and the longitudinal momentum fractions $x_a$ 
along $\bar k$, as
\beq
  \kt_a^\mu & = & k_a^\mu - x_a \,k^\mu - 
  \left( \frac{k \dt k_a}{k^2} - x_a \! \right) \frac{k^2}{k \dt k_r} \, k_r^\mu \, , 
  \qquad \qquad
  \kt_i^\mu + \kt_j^\mu = 0 \, ,
  \nnb \\ [3pt]
  x_a & = & \frac{k_a \dt k_r}{k \dt k_r} \, = \, 
  \frac{s_{ar}}{s_{ir} + s_{jr}} \, , \qquad \qquad
  x_i + x_j = 1 \, .
\label{translong}
\eeq
The transverse momenta $\kt_a$, for $a = i, j$, satisfy
\beq
  \kt_a \cdot \bar k \, = \, \kt_a \cdot k_r \, = \, 0 \, .
\label{transetrans}
\eeq
We can now write the spin-averaged Altarelli-Parisi kernels $P_{ij}$, in a
flavour-symmetric notation, as
\beq
  \hspace{-5mm}  P_{ij} \, = \, P_{ij} \left( x_i, x_j \right) & = & 
  \delta_{f_i g} \delta_{f_j g} \, 2 \, C_A \left( \frac{x_i}{x_j} + 
  \frac{x_j}{x_i} + x_i x_j \right) + \delta_{ \{f_i f_j\} \{q \bar q\} } \, T_R
  \left( 1 - \frac{2 x_i x_j}{1 - \eps} \right) \nnb \\
  && \hspace{-3mm} + \, \delta_{f_i \{q, \bar q\}} \delta_{f_j g} \, C_F 
  \left( \frac{1 + x_i^2}{x_j} - \eps x_j \right) + 
  \delta_{f_i g} \delta_{f_j \{q, \bar q\}} \, C_F
  \left( \frac{1 + x_j^2}{x_i} - \eps x_i \right) \, ,
\label{eq:APkernels1}
\eeq
where we defined the flavour delta functions $\delta_{f \{q,\bar q\}} = \delta_{fq} +
\delta_{f \bar q}$, and $\delta_{\{f_i f_j\} \{q \bar q\}} = \delta_{f_i q} \delta_{f_j 
\bar q} + \delta_{f_i \bar q} \delta_{f_i q}$. In the following we will use interchangeably 
the notations $P_{ij}$, $P_{ij} (x_i,x_j)$, or $P_{ij}(s_{ir},s_{jr})$ to denote the 
collinear kernels of \eq{eq:APkernels1}, and similarly for the azimuthal 
kernels $Q^{\mu\nu}_{ij}$ and for $P^{\mu\nu}_{ij}$. The Casimir eigenvalues 
relevant for the SU($N_c$) gauge group are $C_F = (N_c^2 - 1)/(2 N_c)$ and
$C_A = N_c$, consistent with the normalisation $T_R = 1/2$. The azimuthal 
kernels $Q_{ij}^{\mu\nu}$ can be written as
\beq
  Q_{ij}^{\mu\nu} \, = \, Q_{ij}^{\mu\nu} (x_i, x_j) & = & Q_{ij} \left[ - \, g^{\mu\nu} 
  + (d - 2) \, \frac{\kt_i^\mu \kt_i^\nu}{\kt_i^2} 
  \right] \, ,\nnb \\ [5pt]
  Q_{ij} \, = \,  Q_{ij} (x_i, x_j) & = & - \, \delta_{f_ig} \, \delta_{f_jg} \, 2 \, C_A \, 
  x_i x_j + \delta_{\{f_i f_j\}\{q \bar q\}} \, T_R \, \frac{2 x_i x_j}{1 - \eps} \, .
\label{eq:APkernels2}
\eeq
We note that the presence of the azimuthal kernels $Q_{ij}^{\mu\nu}$ is necessary 
in order to achieve a \emph{local} subtraction of phase-space singularities. The 
collinear kernels satisfy the symmetry properties $P_{ij} = P_{ji}$, $Q_{ij} = Q_{ji}$. 

The final ingredient is the soft-collinear kernel for sector $ij$, which can be obtained 
by acting with the soft projector $\Si$ on the collinear kernel $P_{ij}$ (indeed, 
$Q_{ij}^{\mu\nu}$ is soft-finite). As detailed in \appn{app:commutationNLO}, 
one gets
\beq
  \Si \, P_{ij} \, = \, \delta_{f_i g} \, 2 \, C_{f_j} \frac{x_j}{x_i} \, = \, 
  \delta_{f_i g} \, 2 \, C_{f_j} \frac{s_{jr}}{s_{ir}} \ ,
  \qquad\Longrightarrow\qquad 
  \frac{\Si \,  P_{ij}}{s_{ij}}  \, = \, 2 \, C_{f_j} \, \mc I_{jr}^{(i)} \,  ,
\label{softcollker}
\eeq
where $C_{f_j} = C_A \, \delta_{f_jg} + C_F \, \delta_{f_j\{q\bar q\}}$. Importantly, 
the same soft-collinear kernel is obtained also by taking the collinear limit of 
\eq{eq:eikkern}: in other words, the two limits {\it commute}, as discussed in
detail in \appn{app:commutationNLO}. Subtracting from the collinear kernels
their soft limits, one gets the hard-collinear kernels
\beq
  P^{\hc}_{ij} \, = \, P^{\hc}_{ij} (x_i,x_j) & \equiv & P_{ij} - \delta_{f_i g} 
  C_{f_j}\frac{2 \, x_j}{x_i} - \delta_{f_j g} C_{f_i} \frac{2 \, x_i}{x_j}
  \nnb\\
  & = &
  \delta_{f_i g} \delta_{f_j g} \, 2 \, C_A \, x_i x_j + \delta_{ \{f_i f_j\} \{q \bar q\} } \, T_R
  \left( 1 - \frac{2 x_i x_j}{1 - \eps} \right) \nnb \\[5pt]
  && \hspace{-3mm} + \,
  \delta_{f_i \{q, \bar q\}} \delta_{f_j g} \,
  C_F (1-\eps) \, x_j + 
  \delta_{f_i g} \delta_{f_j \{q, \bar q\}} \,
  C_F (1-\eps) \, x_i \, .
\eeq
Although the candidate counterterm $K_{ij}$ defined above contains all phase-space 
singularities of the product $R \, \mc W_{ij}$, with no double counting, the kinematic
dependences on the right-hand sides of Eqs.~(\ref{eq:SiR}), (\ref{eq:CijR1}) and 
(\ref{eq:SiCijR1}) are not yet suited for a proper subtraction algorithm. Indeed, 
$\{k\}_{\slashed i}$ is a set of $n$ momenta that do not satisfy $n$-body momentum 
conservation away from the exact $\Si$ limit, and, similarly, in the set $(\{k\}_{\slashed i
\slashed j}, k)$ momentum $ k = k_i + k _j$ is off-shell away from the exact $\Cj$ 
limit. The Born-level squared amplitudes $\Bn$ appearing in the counterterm 
must instead feature valid ({\it i.e.}~on-shell and momentum conserving) $n$-body 
kinematics for all choices of the $\npo$ momenta in the radiative amplitude. 
A kinematic mapping is thus necessary, in order to factorise the $(\npo)$-body 
phase space into the product of Born ($n$-body) and radiation phase spaces, 
thereby allowing one to integrate the counterterms only in the latter.

Since the kernels in Eqs.~(\ref{eq:SiR})-(\ref{eq:SiCijR1}) are built in terms of 
Mandelstam invariants, and have not yet been parametrised at this stage, 
there is still full freedom to choose the most appropriate kinematic mapping 
in order to maximally simplify the analytic integrations to follow. In particular, 
at variance with what done in the {\it FKS} algorithm, in any given sector one 
can employ different mappings for different singular limits, or even for different 
contributions to \emph{the same} singular limit. In order to take advantage of 
this freedom, we introduce now a generic Catani-Seymour final-state mapping 
and parametrisation~\cite{Catani:1996vz}, as follows. Let $k_a$ and $k_b$
be two final-state on-shell momenta, and let $k_c$ be the on-shell momentum 
of another (massless) parton, with $c \neq a,b$. Now one can construct an 
on-shell, momentum conserving $n$-tuple of massless momenta 
$\{\bar k \}^{(abc)}$ as
\beq
  \kkl{abc} \, = \, \left\{ \kk{m}{abc} \right\}_{m \neq a} \, , && 
  \qquad
  \kk{i}{abc} \, = \, k_i,
  \quad
  \mbox{if } i \neq a,b,c, \nonumber \\ [3pt]
  \kk{b}{abc} \, = \, k_{a} + k_{b} - \frac{s_{ab}}{s_{ac} + s_{bc}} \, k_{c} \, , &&
  \qquad
  \kk{c}{abc} \, = \, \frac{s_{abc}}{s_{ac} + s_{bc}} \, k_{c} \, ,
\label{eq:CSmap}
\eeq
where $s_{abc} = s_{ab} + s_{ac} + s_{bc}$, and in particular the condition
\beq
  \bar k_{b}^{(abc)} + \bar k_c^{(abc)} \, = \, k_a + k_b + k_c
\label{momconsCS}
\eeq
ensures momentum conservation. Note that the collection of the $n$ light-like
momenta $\{\bar k\}^{(abc)}$ can also be expressed as 
\beq
  \{ \bar k \}^{(abc)} & = & \left\{ \{ k \}_{\slashed a \slashed b \slashed c}, \, 
  \bar k_{b}^{(abc)}, \, \bar k_c^{(abc)} \right\} \, .
\label{precset}
\eeq
Next, we select different values of $a,b,c$ in different sectors and limits. Consistently 
with the general structure of factorised virtual amplitudes~\cite{Magnea:2018ebr}, we 
treat separately the soft and the hard-collinear limits. For the hard-collinear kernel 
in sector $ij$, $(\Cj - \Si \, \Cj) \,\Rl \, \mc W_{ij}$, we choose to assign the labels 
$a$, $b$, and $c$ of \eq{eq:CSmap} as $a = i$,~$b = j$, and $c = r$: partons 
$i$ and $j$ specify the collinear sector, while parton $r$, introduced in \eq{colldir}, 
is the `spectator'. For the soft kernel, $\Si \, \Rl \, \mc W_{ij}$, we choose to 
map differently \emph{each term} in the sum over $l,m$ in \eq{eq:SiR}, with 
assignments $a = i$,~$b = l$, and $c = m$. We then define the local counterterm 
as
\beq
  \hspace{-5mm} \overline{K}
  & = & \sum_{i} \Big[ \sum_{j \neq i} \Si \, \W{ij} \Big] \, \bSi \, \Rl + 
  \sum_{i, \, j > i} \Big[ \Cj \big( \W{ij} + \W{ji} \big) \Big] \bCj \, \Rl
  - \sum_{i,j \neq i} \! \Big[ \Si \, \Cj \, \W{ij} \Big] \, \bSi \, \bCj \, \Rl \, ,
  \label{eq:NLOKsummedfinal}
\eeq
where the barred projectors select soft and collinear limits, and assign the 
appropriate set of on-shell momenta to the kernels. Explicitly
\beq
  \bSi \, \Rl \left( \{k\} \right) & = & - \, \Norm \sum_{\substack{l \neq i\\ m \neq i}} 
  \mc I_{l m}^{(i)} \,\, {\Bn}_{l m} \!\left( \{\bar k\}^{(ilm)} \right) \, ,
\label{eq:SiRdef} \\ [3pt]
  \bCj \, \Rl \left( \{k\} \right) & = & \frac{\Norm}{s_{ij}} \, P^{\mu\nu}_{ij} \, 
  {\Bn}_{\mu\nu} \! \left( \{ \bar k \}^{(ijr)} \right) \, ,
\label{eq:CijR1def} \\ [4pt]
  \bSi \, \bCj \, \Rl \left( \{k\} \right) & = & 2 \, \Norm \, C_{f_j} \, \mc I_{jr}^{(i)} \, 
  \Bn \! \left( \{\bar k\}^{(ijr)} \right) \, ,
\label{eq:SiCijR1def}
\eeq
where we stress that $r\neq i,j$ can be chosen differently for different $ij$ pairs, 
with the constraint that the same $r$ should be chosen for all permutations of $ij$.
The expression in \eq{eq:NLOKsummedfinal} can be rewritten in terms of a sum 
over sectors of local counterterms $\overline K_{ij}$, each containing all the 
singularities of the product $R\,\W{ij}$:
\beq
  \overline K \, = \, \sum_{i, j \neq i} \overline K_{ij} \, , 
  \qquad \qquad \overline{K}_{ij} \, = \,
  \left( \bSi + \bCj - \bSi \, \bCj \right) \Rl \, \mc W_{ij} \, ,
\label{eq:NLOKdef}
\eeq
where it is understood that the action of barred projectors on sector functions is 
the same as that of un-barred ones, namely $\bSi \, \mc W_{ab} = \Si \, \mc 
W_{ab}$, and $\bCj \, \mc W_{ab} = \Cj \, \mc W_{ab}$. To obtain \eq{eq:NLOKdef} 
we have used the symmetry under exchange $i\lra j$ in our definition of $\bCj\,\Rl$.

%%%%%%%%%%%%%%%%

\subsection{Counterterm integration}
\label{sec:NLOcntint}

The counterterm defined in \eq{eq:NLOKdef} is a sum of terms, each 
factorised into a matrix element with Born-level kinematics, multiplying 
a kernel with real-radiation kinematics. The analytic integration of the 
latter in the radiation phase space proceeds by first summing over all 
sectors, as done in {\it FKS}. This operation matches the fact that the 
integrated counterterm must eventually cancel the singularities of the 
virtual contribution, which obviously is not split into sectors.

Upon summation over sectors, the integrand becomes independent of 
sector functions. In fact
\beq
  \overline{K}
  & = & \sum_{i} \, \bSi \, \Rl + \sum_{i, \, j > i} 
  \bCj \!\left( 1 - \bSi - \bSj \right) \Rl \, .
\label{eq:Ksymm}
\eeq
In the soft term we have considered that the kinematic mapping is 
$j$-independent, and performed the sum over $j$, exploiting the soft 
sum rule in \eq{eq:sec_prop4}; in the hard-collinear contribution we have 
used the symmetry of the kinematic mapping and of the collinear operator 
$\bCj$ under the interchange $i\lra j$, exploited the collinear sum rule in  
\eq{eq:sec_prop4}, and the fact that $\Si \, \Cj \, \mc W_{ij} = \Sj \, \Cj \, 
\mc W_{ji} = 1$ (see \eq{eq:commutSfun1} and \eq{eq:commutSfun2}). 
The form of the counterterm in \eq{eq:Ksymm} is now suitable for analytic 
phase-space integration.

We start by introducing the Catani-Seymour parameters 
\beq
  y \, = \, \frac{s_{ab}}{s_{abc}} \, , \qquad 
  z \, = \, \frac{s_{ac}}{s_{ac} + s_{bc}} \, , 
\label{eq:CSparam}
\eeq
which satisfy
\beq
  s_{ab} \, = \, y \, s_{abc} \, , \qquad
  s_{ac} \, = \, z (1 - y) \, s_{abc} \, , \qquad
  s_{bc} \, = \, (1 - z)(1 - y) \, s_{abc} \, ,
\eeq
so that $0 \leq y \leq 1$ and 
$0 \leq z \leq 1$. We use these variables to parametrise the $(\npo)$-body 
phase space, consistently with the mappings in \eq{eq:CSmap}, as 
\beq
  d \Phi_\npo & = & d \Phi_n^{(abc)} \, d \Phi_{\rm rad}^{(abc)} \, ,
  \qquad \qquad d \Phi_{\rm rad}^{(abc)} \, \equiv  \, 
  d \Phi_{\rm rad} \left( \bar{s}_{bc}^{(abc)}; y, z, \phi \right) \, ,
\eeq
leading to the explicit expression
\beq
  \hspace{-5mm} \int d \Phi_{\rm rad} \left(s; y, z, \phi \right)
  & \equiv & N (\eps) \, s^{1 - \eps} \!
  \int_0^\pi \!\! d \phi \, \sin^{- 2 \eps} \! \phi \int_0^1 \!\! dy 
  \int_0^1 \!\! dz
  \Big[ y (1 - y)^2 \, z (1 - z) \Big]^{- \eps} \! (1 - y) \, ,
\label{parphsp}
\eeq
where $d \Phi_n^{(abc)}$ is the $n$-body phase space for partons with 
momenta $\{\bar k\}^{(abc)}$, $\phi$ is the azimuthal angle between 
$\vec k_a$ and an arbitrary three-momentum (other than $\vec k_b, \vec 
k_c$), taken as reference direction, and we have set
\beq
  N(\eps) \, \equiv \, \frac{(4\pi)^{\eps - 2}}{\sqrt \pi \, \Gamma (1/2 - \eps)} \, , 
  \qquad\qquad
  \bar s^{(abc)}_{bc} \, \equiv \, 2 \, \bar k_{b}^{(abc)} \cdot \bar k_c^{(abc)} 
  \, = \, s_{abc} \, .
\label{normevar}
\eeq
We first consider the integral $I^\hc$ of the hard-collinear counterterm 
\beq
  \overline{K}^\hc \, = \, \sum_{i, \, j > i}  \bCj \, \left( 1 \, - \, \bSi \, - \, 
  \bSj \right) \, \Rl \, = \, \sum_{i, \, j > i} \frac{\Norm}{s_{ij}} \, P^{\hc\,\mu\nu}_{ij} \, 
  {\Bn}_{\mu\nu} \! \left( \{ \bar k \}^{(ijr)} \right) \, ,
\label{barKhc}
\eeq
where
\beq
P^{\hc\,\mu\nu}_{ij}
\Bn_{\mu\nu}\! \left( \{ \bar k \}^{(ijr)} \right)
\, = \,
P^{\hc}_{ij}\,\Bn\! \left( \{ \bar k \}^{(ijr)} \right)
+
Q^{\mu\nu}_{ij}\,\Bn_{\mu\nu}\! \left( \{ \bar k \}^{(ijr)} \right)
\, .
\eeq 
Each term in the double sum in $\overline{K}^\hc$ is parametrised assigning 
labels $a = i$,~$b = j$, and $c = r$, as detailed below \eq{momconsCS}. 
We have
\beq
  I^\hc & = & \frac{\varsi_\npo}{\varsi_n} \sum_{i, \, j > i} 
  \int d \Phi_{\rm rad}^{(ijr)} \,\, \bCj \, \left( 1 - \bSi - \bSj \right) \, 
  \Rl \left( \{k\} \right) \, ,
\label{eq:IHC}
\eeq
where $\varsi_k$ indicates the symmetry factor associated to the $k$-body 
final state. We note that the integral does not receive any contribution from the 
azimuthal kernels $Q_{ij}^{\mu\nu}$, as the latter integrate to zero in the radiation 
phase space. In our chosen parametrisation, the variable $z$ coincides with the 
collinear fraction $x_i$ defined in \eq{translong}, while $s_{ij} = y \, \bar s^{(ijr)}_{jr}$. 
The analytic integration of the counterterm is therefore straightforward, and can 
be carried out exactly to all orders in $\eps$. By defining
\beq
\label{eq:Jijhc}
  J^\hc_{ij}(s,\eps) & \equiv & \frac1s \int d \Phi_{\rm rad}(s;y,z,\phi) \,
  \frac{P_{ij}^{\hc}(z,1-z)}{y} \nnb \\
  & = & - \, \frac{(4\pi)^{\eps - 2}}{s^\eps} \, \frac{\Gam(1 - \eps) 
  \Gam(2 - \eps)}{\eps \, \Gam(2 - 3 \eps)} 
  \label{Ihc1} \\
  & & \times \, \Bigg[ \frac{C_A}{3 - 2\eps} \, \delta_{f_ig} \delta_{f_jg}
  + \frac{C_F}{2} \left(\delta_{f_i \{q, \bar q\}} \delta_{f_j g} + \delta_{f_j \{q, \bar q\}}
  \delta_{f_ig}\right) + \frac{2\,T_R}{3 - 2 \eps} \, \delta_{\{f_i f_j\} \{q \bar{q}\} } \, 
  \Bigg] \, , \nnb
\eeq
one finds
\beq
  I^{\rm \, hc} & = & \Norm \, \frac{\varsi_\npo}{\varsi_n} \sum_{i, \, j > i} \, 
  J^\hc_{ij} \left( \bar s_{jr}^{(ijr)}, \eps \right) \,\, \Bn \! \left( \{\bar k\}^{(ijr)} \right)
\label{inthcout} \\ [4pt]
  & = & - \, \frac{\as}{2 \pi} \left( \frac{\mu^2}{s} \right)^{\eps}
  \sum_{p} \, \Bn \! \left( \{\bar k\}^{(ijr)} \right) \Bigg[ \delta_{f_p g} \,
  \frac{C_A + 4 \, T_R N_f}{6} \left( \frac{1}{\eps} + \frac{8}{3} - \ln \bar\eta_{pr} \right)
  \nnb \\
  & & \hspace{50mm} + \, \, \delta_{f_p \{q, \bar q\}} \, \frac {C_F}{2}
  \left( \frac{1}{\eps} + 2 - \ln \bar\eta_{pr} \right) \Bigg]
  \, + \, \mc O(\eps) \, , \nnb
\eeq
where in the last step we replaced the sum over $i, \, j$ with a sum over `parent' 
partons $p$ (which has absorbed the $\varsi_\npo/\varsi_n$ symmetry factor), 
carrying momentum $\bar k_{j}^{(ijr)}$ (see \eq{eq:CSmap}), we included a $1/2$ 
Bose-symmetry factor in the $C_A$ term, accounting for gluon indistinguishability, 
and we considered $N_f$ light $q\bar q$ pairs. The invariant $\bar\eta_{pr}$ is 
defined as $\bar \eta_{pr} = \bar s^{(ijr)}_{jr}/s = s_{ijr}/s$, with $r\neq p$.
Notice that the result contains only a single $\ep$ pole, consistently with the
fact that soft singularities are excluded.

Next we turn to the integral $I^\so$ of the soft counterterm 
\beq
  \overline K^\so \, = \, \sum_{i} \bSi \, \Rl \, .
\label{soco1}
\eeq
We parametrise it by assigning different labels to each term in the eikonal 
sum, with $a = i$,~$b = l$ and $c = m$, as detailed below \eq{momconsCS}, 
obtaining
\beq
  I^\so & = & \frac{\varsi_\npo}{\varsi_n} \, \sum_i \int d \Phi_{\rm rad} \, \bSi \, 
  \Rl \left( \{k\} \right) \nnb \\
  & = & - \, \Norm \, \frac{\varsi_\npo}{\varsi_n} \, \sum_i 
  \sum_{\substack{l \neq i\\m \neq i}} \,
  {\Bn}_{lm} \! \left( \{\bar k\}^{(ilm)} \right) 
  \int d \Phi_{\rm rad}^{(ilm)} \, \mc I_{lm}^{(i)} \, .
\label{eq:IS}
\eeq
In our chosen parametrisation $s_{lm}/s_{im} = (1 - z)/z$, and $s_{il} = y \, 
\bar s^{(ilm)}_{lm}$: the soft counterterm can then be analytically integrated, 
once again to all orders in $\eps$. By defining, for each term of the eikonal sum,
\beq
\label{eq:Jsoft}
  J^\so(s,\eps) & \equiv & \frac{1}{s} \int d \Phi_{\rm rad} \left(s; y, z, \phi \right)
  \frac{1 - z}{yz} \, = \, \frac{(4 \pi)^{\eps - 2}}{s^\eps} \,
  \frac{\Gam(1 - \eps) \Gam(2 - \eps)}{\eps^2 \, \Gam(2 - 3 \eps)} \, ,
\label{Iseach}
\eeq
we get the simple result
\beq
  I^\so & = & - \, \Norm \, \frac{\varsi_\npo}{\varsi_n} \sum_i \, \delta_{f_i g} 
  \sum_{\substack{l \neq i\\ m \neq i}} \, J^\so \left( \bar s_{lm}^{(ilm)}, \eps \right) \,\, 
  {\Bn}_{lm} \! \left( \{\bar k\}^{(ilm)} \right) \nnb \\ [5pt]
  & = & \frac{\as}{2 \pi} \left( \frac{\mu^2}{s} \right)^{\eps} \,
  \Bigg[ \sum_l \, C_{f_l} \, \Bn \! \left( \{\bar k\} \right) \, \Bigg(
  \frac{1}{\eps^2} + \frac{2}{\eps} \, + \, 6 - \frac{7}{2} \, \zeta_2
  \Bigg) \nnb \\
  && \quad + \,\, \sum_{l, \, m\neq l} \, {\Bn}_{lm} \! \left( \{\bar k\} \right) \,
  \ln \bar\eta_{lm} \, \Bigg( \frac{1}{\eps} + 2 - \frac{1}{2} \ln \bar\eta_{lm} 
  \Bigg) \Bigg] \, + \, \mc O(\eps) \, ,
\label{eq:IScntcorr}
\eeq
where in the second step we have remapped all identical soft-gluon contributions 
on the same Born-level kinematic configuration $\{\bar k\}$, and the sum $\sum_i \, 
\delta_{f_i g}$ has absorbed the symmetry factor $\varsi_\npo/\varsi_n$. Note that 
\eq{eq:IScntcorr} correctly features a double $\ep$ pole, coming from soft-collinear 
configurations.

We can finally combine soft and hard-collinear integrated counterterms, obtaining, 
up to $\mc O(\eps)$ corrections,
\beq
  I \left( \{\bar k\} \right) & = & I^\so \left( \{\bar k\} \right) + 
  I^\hc \left( \{\bar k\} \right) \nnb \\
  & = & \frac{\as}{2\pi} \left( \frac{\mu^2}{s} \right)^{\eps} \,
  \Bigg\{ \Bigg[ \, \Bn \! \left( \{\bar k\} \right) \sum_{k} \Bigg(
  \frac{C_{f_k}}{\eps^2} + \frac{\gamma_k}{\eps} \Bigg) + 
  \sum_{k,\, l\neq k} \, {\Bn}_{kl} \! \left( \{\bar k\} \right) \, \frac{1}{\eps} \ln \bar\eta_{kl} 
  \Bigg] \nnb \\
  && \hspace{1cm} + \, \Bigg[ \, \Bn \! \left( \{\bar k\} \right) \sum_{k} \,
  \Bigg( \delta_{f_k g} \, \frac{C_A + 4\,T_R\,N_f}{6}
  \left( \ln \bar\eta_{kr} - \frac{8}{3} \right) \nnb \\
  && \hspace{1cm} + \, \, \delta_{f_k g} \, C_A
  \left( 6 - \frac72 \zeta_2 \right) \, + \, \delta_{f_k \{q,\bar q\}} \frac {C_F}{2}
  \big( 10 - 7 \zeta_2 + \ln \bar\eta_{kr} \big) \Bigg) \nnb \\ 
  && \hspace{1cm} + \, \sum_{k,\, l\neq k} \, {\Bn}_{kl} \left( \{\bar k\} \right) \, 
  \ln \bar \eta_{kl} \, 
  \left( 2 - \frac{1}{2} \ln \bar\eta_{kl} \right) \Bigg] \Bigg\} \, ,
\label{eq:intcntNLO}
\eeq
where we introduced the spin-dependent one-loop collinear anomalous dimension
\beq
  \gamma_k \, = \,  \delta_{f_k g} \, \frac{11 C_A - 4 \,T_R N_f}{6} \, + \, 
  \delta_{f_k \{q,\bar q\}} \, \frac32 C_F \, .
\label{collgam1}
\eeq
The integrated counterterm in \eq{eq:intcntNLO} successfully reproduces the pole 
structure of the virtual NLO contribution (see for example~\cite{Catani:1998bh}), 
which provides a check of validity of the subtraction method. Moreover, we note 
the simplicity of the integrated counterterms to all orders in $\eps$, which is a direct 
consequence of having optimally adapted term by term the kinematic mapping and 
parametrisation.

We conclude this Section with three considerations on the structure of the counterterm.
First, the strong coupling $\as$ has been treated as a constant throughout the 
computation. A dynamical scale for the coupling can simply be reinstated in the 
counterterm by evaluating it with the Born-level kinematics $\{\bar k\}$. Second, in 
the counterterm definition in \eq{eq:NLOKsummedfinal} we have chosen to apply
projectors $\bSi$ and $\bCj$ only on the product $\Rl \, \mc W_{ij}$, while treating
exactly the phase-space measure $d \Phi_{\rm rad}$. In other words, the counterterm
phase space is exact, and coincides with that of the real-radiation matrix element.
We stress that this feature is not crucial to our method: one could as well consider 
approximate phase-space measures $d \widehat{\Phi}_{\rm rad}$, provided they 
correctly reproduce the exact $d \Phi_{\rm rad}$ in the singular limits. In the 
massless final-state case detailed in this article, as evident from the above 
calculation, no computational advantage would result from such an approximation, 
however the latter may become relevant in more complicated cases. Analogously, 
restrictions on the counterterm phase space could be applied in order to improve
the convergence of a numerical implementation. We leave these possibilities open
for future studies.

Third, \eq{eq:NLOKdef} and \eq{eq:Ksymm} are analytically equivalent, but they 
underpin different philosophies in the implementation of the subtraction scheme.
In \eq{eq:NLOKdef}, which is our preferred choice, subtraction is seen as the
incoherent sum of terms, each of which features a minimal singularity structure
and is separately optimisable, in the same spirit of the \emph{FKS} method but,
we believe, featuring enhanced flexibility. \eq{eq:Ksymm}, which in what we 
have presented is employed only for analytic integration, represents a single 
local subtraction term containing \emph{all} singularities of the real matrix element, 
hence it has the same essence as \emph{CS} subtraction, but with much simpler 
counterterms. Our method at NLO represents thus a bridge between these two 
long-known subtraction methods, aiming at retaining the virtues of both, and 
not being limited by the mutual suboptimal features.

%%%%%%%%%%%%%%%%%%%%%%%%%%%%%%%%%

\section{Local analytic sector subtraction at NNLO}
\label{NNLO}

%%%%%%%%%%%%%%%%

\subsection{Generalities}
\label{sec:NNLOgen}

The NNLO contribution to the differential cross section with respect to a generic 
IR-safe observable $X$ can be schematically written as
\beq
  \frac{d \sig_\NNLO}{dX} \, = \, \int d \Phi_n \, \VV \, \delta_n(X) +
  \int d \Phi_\npo \, \RVl \,\delta_\npo(X) + \int d \Phi_\npt \, \RR \,\delta_\npt(X) \, ,
\label{eq:NNLO_struct}
\eeq
where $\RR$, $\VV$, and $\RVl$ are the double-real, the UV-renormalised 
double-virtual, and the UV-renormalised real-virtual corrections, respectively, 
with
\beq
  && \RR \, = \, \left| {\cal A}_\npt^{(0)} \right|^2 \, , \qquad 
  \RVl \, = \, 2 \, {\bf Re} \left[ {\cal A}_\npo^{(0) *} \, {\cal A}_\npo^{(1)} \right] \, , 
  \nnb \\
  && \hspace{8mm} \VV \, = \, \left| {\cal A}_n^{(1)} \right|^2 + \, 
  2 \, {\bf Re} \left[ {\cal A}_n^{(0) *} \, {\cal A}_n^{(2)} \right] \, .
\label{rrrvvv}
\eeq
In dimensional regularisation, $\VV$ features up to a quadruple IR pole in 
$\eps$; $\RR$ is finite in $d = 4$, but it is affected by up to four singularities 
in the double-radiation phase space, stemming from configurations that 
feature up to two soft and/or collinear emissions; $\RVl$ has up to a double 
IR pole in $\eps$, originating from its one-loop nature, on top of a double 
singularity in the single-radiation phase space. The sum of these three 
contributions is finite due to the IR safety of $X$ and to the KLN theorem. 
It is however clear that the difficulty of evaluating and integrating complete
radiative matrix elements in arbitrary dimension at NNLO is significantly more
severe  than at the NLO, hence the necessity of a subtraction procedure.

Subtraction at NNLO amounts to modifying~\eq{eq:NNLO_struct} by adding 
and subtracting three sets of counterterms: single-unresolved, double-unresolved, 
and real-virtual, which we write as
\beq
  \int d \widehat \Phi_\npt \, \overline{K}^{\one} \, \delta_\npo(X) \, , \quad \,\,
  \int d \widehat \Phi_\npt \, \left( \overline{K}^{\two} + \overline{K}^{\otwo} \right) 
  \, \delta_n(X) \, , \quad \,\,
  \int  d \widehat \Phi_\npo \, \overline{K}^{\RV} \, \delta_n (X) \, ,
\label{eq:defNNLOcnt}
\eeq
and which can be characterised as follows. The single-unresolved counterterm 
$d \widehat \Phi_\npt \, \overline{K}^{\one}$ features the subset of phase-space 
singularities of $d\Phi_\npt \,\RR$ which correspond to configurations where only 
one parton becomes unresolved, analogously to what happens at NLO. The sum 
$d \widehat \Phi_\npt \, \big( \overline{K}^{\two} + \overline{K}^{\otwo} \big)$ contains 
all singularities stemming from kinematic configurations where exactly two partons 
become unresolved. At NNLO, this exhausts all possible phase-space singularities. 
We note that the Dirac delta functions associated with these two counterterms 
mirror their physical meaning, with $\delta_\npo (X)$ associated with 
$\overline{K}^{\one}$, and $\delta_n (X)$ with $(\overline{K}^{\two} + 
\overline{K}^{\otwo})$. The distinction between $\overline{K}^{\two}$ and 
$\overline{K}^{\otwo}$ will be described in detail in \secn{sec:NNLOcntdef}.
The third subtraction term, $d \widehat\Phi_\npo \, \overline{K}^{\RV}$ cancels 
the phase-space singularities of $d \Phi_\npo \, \RVl$.

Denoting the corresponding phase-space-integrated counterterms with
\beq
  I^{\one} \, = \,  \int d \widehat \Phi_{{\rm rad}, 1} \, \overline{K}^{\one} \, , 
  \qquad &&
  I^{\two} \, = \, \int d \widehat \Phi_{{\rm rad}, 2} \, \overline{K}^{\two} \, ,
  \nnb \\ [3pt]
  I^{\otwo} \, = \, \int d \widehat \Phi_{{\rm rad}, 1} \, \overline{K}^{\otwo} \, , 
  \qquad && 
  I^{\RV} \, = \, \int d \widehat \Phi_{\rm rad} \, \overline{K}^{\RV} \, ,
\label{eq:int_count_NNLO}
\eeq
where $d \widehat \Phi_{{\rm rad}, 1} = d \widehat \Phi_{\npt}/d \widehat 
\Phi_\npo$,~ $d \widehat \Phi_{{\rm rad}, 2} = d \widehat \Phi_{\npt}/d \Phi_n$, 
and $d \widehat \Phi_{\rm rad} = d \widehat \Phi_{\npo}/d \Phi_n$, the 
subtracted NNLO cross section can be identically rewritten as
\begin{eqnarray}
  \frac{d \sig_\NNLO}{dX} & = & \int d \Phi_n \, 
  \left( \VV + I^{\two} +I^{\RV} \right) \, \delta_n (X) 
\label{eq:NNLO_structure} \\
  && + \int \bigg[ \left( d \Phi_\npo \, \RVl + d \widehat \Phi_\npo \, I^{\one} \right) \,
  \delta_\npo(X) \, - \, d \widehat \Phi_\npo \, \left( \overline{K}^{\RV} - I^{\otwo} \right) \,
  \delta_n(X) \bigg] \nnb \\
  && + \int \bigg[ d \Phi_\npt \, \RR \, \delta_\npt(X) - d \widehat \Phi_\npt \, 
  \overline{K}^{\one} \, \delta_\npo(X) - d \widehat \Phi_\npt \, \left( 
  \overline{K}^{\two} + \overline{K}^{\otwo} \right) \, \delta_n(X) \bigg] \, . \nnb
\end{eqnarray}
In the third line of \eq{eq:NNLO_structure}, all terms are separately finite in 
$d=4$, and their sum is finite in the double-radiation phase space, making this 
contribution fully regular and integrable numerically. In the second line, $I^{\one}$ 
features the same poles in $\eps$ as $\RVl$, up to a sign, so that their sum 
is finite in $d = 4$. The counterterm $\overline{K}^{\RV}$ locally subtracts the 
phase-space singularities of $\RVl$; it contains however explicit poles in $\eps$, 
and the local counterterm $\overline{K}^{\otwo}$ is such that the integral
$I^{\otwo}$ cancels those poles; furthermore, the finite sum 
$\RVl + I^{\one}$ features phase space singularities, and these must be 
cancelled by the finite sum $\overline{K}^{\RV} - I^{\otwo}$. In total, the sum 
of the four terms in the second line of \eq{eq:NNLO_structure} is both finite 
in $d = 4$ and integrable in the single-radiation phase space, making this 
contribution numerically tractable. Finally, in the first line of \eq{eq:NNLO_structure}, 
the sum $I^{\two} + I^{\RV}$ features the same poles in $\eps$ as $\VV$, up 
to a sign, making the Born-like contribution finite and integrable.

%%%%%%%%%%%%%%%%

\subsection{Sector functions}
\label{sec:NNLOsecfun}

As in the NLO case, we start by partitioning the phase space in sectors, 
each of which selects the singularities stemming from an identified subset 
of partons. We thus introduce sector functions $\W{abcd}$, with as many 
indices as the maximum number of partons that can simultaneously be 
involved in an NNLO-singular configuration. We reserve the first two 
indices for singularities of single-unresolved type, implying that  $b$, $c$, 
and $d$ differ from $a$. As far as double-unresolved configurations are 
concerned, in particular those of collinear nature, they can involve three 
or four different partons, hence either indices $b$, $c$, and $d$ are all 
different, or two of them are equal. Without loss of generality we choose 
the third and the fourth indices to be always different, so that the allowed 
combinations of indices, that we refer to as \emph{topologies}, are
\beq
  \W{ijjk}\, , \qquad \,\, \W{ijkj}\, , \qquad \,\, \W{ijkl} \, ,
  \qquad \qquad \,\,
  i,j,k,l \quad \mbox{all different} \, .
\label{topo}
\eeq
Since the sector functions must add up to $1$, in order to represent a unitary 
partition of phase space, they can be defined as ratios of the type
\beq
  \W{abcd} \, = \,  \frac{\sigma_{abcd}}{\sigma} \, , \qquad
  \sigma \, = \, \!\!
  \sum_{\substack{a, \, b \neq a}} \sum_{\substack{c \neq a \\ d \neq a, c}}
  \sigma_{abcd}
  \qquad \Longrightarrow \quad
  \sum_{\substack{a, \, b \neq a}} \sum_{\substack{c \neq a \\ d \neq a, c}}
  \W{abcd} \, = \, 1 \, .
\label{eq:sec_fun_NNLO}
\eeq
There is a certain freedom in the definition of $\sigma_{abcd}$. Analogously to 
the NLO case, we design them in such a way as to minimise the number of IR 
limits that contribute to a given sector. In addition, at NNLO there is another 
property to be required, new with respect to NLO, and related to the fact that 
the integrated single-unresolved counterterm $I^{\one}$ must be combined 
with the real-virtual contribution, to cancel its explicit poles in $\eps$, as detailed 
in \secn{sec:NNLOgen}. Since $\RVl$, as any term with $(\npo)$-body kinematics, 
is split into NLO-type sectors, the same must be true for $I^{\one}$. This implies that, 
roughly speaking, sector functions with four indices must factorise sector functions 
with two indices in the single-unresolved limits, in order for the cancellation of 
poles to take place NLO-sector by NLO-sector. 

A possible expression for $\sigma_{abcd}$ with the required properties is
\beq
  \sigma_{abcd} \, = \,  \frac{1}{(e_a)^{\alpha} \, (w_{ab})^{\beta}}
  \frac{1}{(e_c \, + \, \delta_{bc} \, e_a) \, w_{cd}} \, , \qquad \qquad
  \alpha \, > \, \beta \, > \, 1 \, .
\label{eq:sigmadef}
\eeq
With the sector functions defined in \eq{eq:sec_fun_NNLO} and \eq{eq:sigmadef}, 
the list of singular limits acting non-trivially in each NNLO sector includes the 
single-unresolved projectors ${\bf S}_a$ and $\bC{ab}$, already considered at 
NLO, as well as the following double-unresolved limits:
\beq
  \bS{ab} \, : && e_{a}, \, e_b  \, \to  \, 0 \, , \quad
  e_a/e_b \, \to \, \mbox{constant} 
  \nnb \\
  &&
  \mbox{(uniform double-soft configuration of partons ($a, b$))} \, ,
  \nnb \\
  \bC{abc} \, : && w_{ab}, \, w_{ac}, \, w_{bc} \, \to  \, 0 \, , \quad
  w_{ab}/w_{ac}, \, w_{ab}/w_{bc}, \, w_{ac}/w_{bc} \, \to  \, \mbox{constant}
  \nnb \\
  &&
  \mbox{(uniform double-collinear configuration of partons $(a, b, c)$)} \, , 
  \nnb \\
  \bC{abcd} \, : && w_{ab}, \, w_{cd} \, \to  \, 0 \, , \quad
  w_{ab}/w_{cd} \, \to \,  \mbox{constant}
  \nnb \\
  &&
  \mbox{(uniform double-collinear configuration of partons $(a, b)$ and 
  $(c, d)$)} \, ,
  \nnb \\
  \bSC{abc} \, : && e_a, \, w_{bc} \, \to  \, 0 \, , \quad
  e_a/w_{bc} \, \to  \, 0 \, , \qquad
  \bSC{abc}(f) \, = \, \bC{bc} \, \big[ \bS{a}(f) \big] \, ,
  \nnb \\
  &&
  \mbox{(ordered soft (first) and collinear configuration of partons $a$ 
  and $(b, c)$)} \, ,
  \nnb  \\
  \bCS{abc} \, : && w_{ab}, \, e_c \, \to  \, 0 \, , \quad
  w_{ab}/e_c \, \to \, 0 \, , \qquad \bCS{abc}(f) = \bS{c} \, \big[ \bC{ab}(f) \big] \, ,
  \nnb \\
  &&
  \mbox{(ordered collinear (first) and soft configuration of partons $(a, b)$ 
  and $c$)} \, .
\label{eq:2unreslim}
\eeq
Notice that only the first two limits of the list (\ref{eq:2unreslim}) are genuinely 
double-unresolved\footnote{In the literature the configuration $\bC{abc}$ 
is sometimes referred to as \emph{triple-collinear}. We call it \emph{double-collinear},
following~\cite{Frixione:2004is}, in order to consistently specify the type of 
configuration as being double-unresolved, rather than indicating the number 
of partons that become collinear.}, namely they cannot be reduced to compositions 
of single-unresolved limits when acting on the double-real matrix elements; the 
remaining three configurations are compositions of single-unresolved limits 
when acting on matrix elements, but not when they are applied to the sector 
functions in \eq{eq:sec_fun_NNLO}, therefore they have to be introduced as 
independent limits. In \appn{app:SfunNNLOprop} we show that, among the 
single- and double-unresolved limits that we are considering, only a subset
give a non-zero contribution in the various topologies. They are 
\beq
  \W{ijjk} &:& \quad \Si \, , ~~ \Cj \, , ~~ \SSj \, , ~~ \CCj \, , ~~ \bSC{ijk} \, ;
  \nnb \\ 
  \W{ijkj} &:& \quad \Si \, , ~~ \Cj \, , ~~ \SSk \, , ~~ \CCj \, , ~~ 
  \bSC{ijk} \, , ~~ \bCS{ijk} \, ;
  \nnb \\
  \W{ijkl} &:& \quad \Si \, , ~~ \Cj \, , ~~ \SSk \, , ~~ \bC{ijkl} \, , ~~ 
  \bSC{ikl} \, , ~~ \bCS{ijk} \, .
\label{eq:NNLOprop2}
\eeq
In \appn{app:SfunNNLOprop} we also show that all the limits reported in 
\eq{eq:NNLOprop2} commute when acting on the sector functions, and that the
combinations of these limits exhaust all possible single- and double-unresolved 
configurations in each sector. We stress that this structure depends 
on our choice of sector functions; with other functions the surviving limits would
in general be different.

It is now necessary to study the properties of the sector functions defined in 
\eq{eq:sec_fun_NNLO} and \eq{eq:sigmadef} under the action of single-unresolved 
limits. As noted above, in these configurations the NNLO sector functions must 
factorise into products of NLO-type sector functions. To this end, let us define
\beq
  \sigma_{ab}^{(\alpha \beta)} \, = \, \frac{1}{(e_a)^{\alpha}(w_{ab})^{\beta}} \, ,
  \qquad \qquad
  \Wab{ij} \, = \, \frac{\sigma_{ij}^{(\alpha \beta)}}{\sum \limits_{a, \, b \neq a}
  \sigma_{ab}^{(\alpha\beta)}} \, ,
\label{eq:Wab_def}
\eeq
so that the NLO sector functions in \eq{eq:sfunNLO} are given by $\mc W_{ij} =
\mc W_{ij}^{(11)}$, and similarly $\sigma_{ab} = \sigma_{ab}^{(11)}$. One easily 
verifies that the functions $\Wab{ij}$ satisfy all the requirements that must apply 
to NLO sector functions.  It is now
straightforward to verify that the NNLO sector functions defined in 
\eq{eq:sec_fun_NNLO} and \eq{eq:sigmadef} satisfy
\beq
  \bS{i}  \, \W{ijjk} \, = \, \W{jk}  \, \bS{i}  \, \Wab{ij}  \, , ~~
  \bC{ij}  \, \W{ijjk} & = & \W{[ij]k}  \, \bC{ij} \,  \Wab{ij}  \, , ~~ \,
  \bS{i} \, \bC{ij} \,  \W{ijjk} = \W{jk}  \, \bS{i} \, \bC{ij}  \, \Wab{ij} \,  ,
  \nnb \\ 
  \bS{i}  \, \W{ijkj} \, =  \, \W{kj}  \, \bS{i}  \, \Wab{ij}  \, ,  ~~
  \bC{ij}  \, \W{ijkj} & = & \W{k[ij]}  \, \bC{ij}  \, \Wab{ij}  \, , ~~
  \bS{i} \, \bC{ij} \,  \W{ijkj} \, = \, \W{kj}  \, \bS{i} \, \bC{ij} \,  \Wab{ij}  \, ,
  \nnb \\ 
  \bS{i}  \, \W{ijkl} \, = \,  \W{kl} \, \bS{i}  \, \Wab{ij} \, , \hspace{3.3mm}
  \bC{ij} \,  \W{ijkl} & = & \W{kl}  \, \bC{ij}  \, \Wab{ij}  \, , \hspace{5.7mm}
  \bS{i} \, \bC{ij}  \, \W{ijkl} \, =  \,  \W{kl}  \, \bS{i} \, \bC{ij}  \, \Wab{ij}  \, ,
  \nnb \\
\label{eq:propNNLOfact}
\eeq
where $\W{[ab]c}$ is the NLO sector function defined in the $(\npo)$-particle 
phase space with respect to the parent parton $[ab]$ of the collinear pair $(a,b)$. 

Finally, the NNLO sector functions satisfy sum rules analogous to the NLO ones in 
\eq{eq:sec_prop4}, and which stem from their definition 
in \eq{eq:sec_fun_NNLO}. One may verify that
\beq
  && \bS{ik} \, \Big( \sum_{b \neq i} \sum_{d \neq i,k} \, \W{ibkd}
  + \sum_{b \neq k} \sum_{d \neq k,i} \, \W{kbid} \Big) \, = \, 1 \, , 
\label{srule1} \\ [7pt]
  && \bC{ijk} \!\!\!\!\! \sum_{abc \, \in \, \pi(ijk)}
  \!\!\!\!\!\! \big( \W{abbc} + \W{abcb} \big) \, = \, 1 \, ,
  \hspace{7mm}
  \bC{ijkl} \!\!\!\!\! \sum_{\substack{ab \, \in \, \pi(ij) \\ 
  cd \, \in \, \pi(kl)}} \!\!\!\!\!\!
  \big( \W{abcd} + \W{cdab} \big) \, = \, 1 \, , 
\label{srule2} \\
  && \bSC{ikl} \, \sum_{b \neq i} \big( \W{ibkl} + \W{iblk} \big) \, = \, 1 \, ,
  \hspace{9mm}
  \bCS{ijk} \, \Big( \sum_{d \neq i,k} \W{ijkd} + 
  \sum_{d \neq j,k} \W{jikd} \Big) \, = \, 1 \, ,
\label{srule3}
\eeq
where by $\pi(ijk)$ we denote the set $\{ijk, \, ikj, \, jik, \, jki, \, kij, \, kji \}$. 
Sum rules for composite double-unresolved limits, that follow from those 
reported in Eqs.~(\ref{srule1})-(\ref{srule3}), will be further detailed in 
\secn{sec:2unrescntint}, where we describe the structure of the 
double-unresolved counterterm. We stress that the properties in 
Eqs.~(\ref{srule1})-(\ref{srule3}), in full analogy with the NLO case, 
allow one to perform sums over all the sectors that share a given set 
of double-unresolved singular limits, eliminating the corresponding sector 
functions prior to countertem integration. This feature, distinctive of our 
method at NNLO, is crucial for the feasibility of the analytic integration 
of counterterms.

%%%%%%%%%%%%%%%%

\subsection{Definition of local counterterms}
\label{sec:NNLOcntdef}

As reported in \eq{eq:NNLOprop2}, a limited number of products of IR projectors 
is sufficient to collect all singular configurations of the double-real matrix elements 
in each sector. By subtracting these products from the matrix element, one gets, 
for the different topologies, the finite expressions
\beq
  \RR_{ijjk}^{\, \sub} & = & \big( 1 - \Si \big) \big( 1 - \Cj \big) \big( 1 - \SSj \big) 
  \big( 1 - \CCj \big) \big( 1 - \SCj \big) \, \RR \, \W{ijjk}
  \nnb \\
  & \equiv & 
  \left( 1 - \bL{\one}{ij} \right) \left( 1 - \bL{\two}{ijjk} - \bL{\twosc}{ijjk} \right) \, 
  \RR \, \W{ijjk} \, ,
  \nnb\\ [7pt] 
  \RR_{ijkj}^{\, \sub} & = & \big( 1- \Si \big) \big( 1 - \Cj \big) \big( 1 - \SSk \big) 
  \big( 1 - \CCj \big) \big( 1 - \SCj \big) \big( 1 - \bCS{ijk} \big) \, \RR \, \W{ijkj} 
  \nnb\\
  & \equiv & 
  \left( 1 - \bL{\one}{ij} \right) \left( 1 - \bL{\two}{ijkj} - \bL{\twosc}{ijkj} \right) \, 
  \RR \, \W{ijkj} \, , 
  \nnb \\ [7pt]
  \RR_{ijkl}^{\, \sub} & = & \big( 1 - \Si \big) \big( 1 - \Cj \big) \big( 1 - \SSk \big)
  \big( 1 - \CCl \big) \big( 1 - \SCl \big) \big( 1 - \bCS{ijk} \big) \, \RR \; \W{ijkl}
  \nnb\\
  & \equiv & 
  \left( 1 - \bL{\one}{ij} \right) \left( 1 - \bL{\two}{ijkl} - \bL{\twosc}{ijkl} \right) \, 
  \RR \, \W{ijkl} \, ,
\label{formdefctnnlo}
\eeq
where we separated the action of the single-unresolved limits $\bL{\one}{ij}$, 
defined in \eq{eq:NLOK}, from that of the double-unresolved ones $\bL{\two}{T} 
+ \bL{\twosc}{T}$, defined for the various topologies $T \, = \, ijjk, \, ijkj, \, ijkl$
by the expressions
\beq 
  \bL{\two}{ijjk} & = &  \SSj + \CCj \big( 1 - \SSj \big) \, , \qquad \,\,
  \bL{\twosc}{ijjk} \, = \,  \SCj \, \big( 1 - \SSj \big) \big( 1 - \CCj \big) \, , \\
  \bL{\two}{ijkj} & = &  \SSk + \CCj \big( 1 - \SSk \big) \, , \qquad \,
  \bL{\twosc}{ijkj} \, = \,  \Big[ \SCj + \bCS{ijk} \big( 1 - \SCj \big) \Big] 
  \big( 1 - \SSk \big) \big( 1 - \CCj \big) \, , \nnb \\
  \bL{\two}{ijkl} & = &  \SSk + \CCl \big( 1 - \SSk \big) \, , \qquad \! \hspace{1pt}
  \bL{\twosc}{ijkl} \, = \,  \Big[ \SCl + \bCS{ijk} \big( 1 - \SCl \big) \Big] 
  \big( 1 - \SSk \big) \big( 1 - \CCl \big) \, . \nnb
\eeq
The order with which the various operators are applied to matrix elements 
is irrelevant, as all limits commute. In \appn{app:SfunNNLOprop} we show 
that this property is also respected by the sector functions defined in 
\eq{eq:sec_fun_NNLO}.  Candidate double-real local counterterms for 
the various topologies $T$ can thus be defined, in analogy with 
\eq{eq:NLOK}, as
\beq
  K^{\one}_{T} + K^{\otwo}_{T} + K^{\two}_{T} & = &  
  \RR \, \W{T} - \RR_{\, T}^{\sub} \\ & = &
  \left[ \bL{\one}{ij} + \left( \bL{\two}{T} + \bL{\twosc}{T} \right) - 
  \bL{\one}{ij} \left( \bL{\two}{T} + \bL{\twosc}{T} \right)
  \right] \RR \, \W{T} \, . \nnb
\label{eq:cnt_NNLO_loc_sum}
\eeq
The different contributions are naturally split according to their kinematics. All 
terms containing only single-unresolved limits are assigned to $K^{\one}$, the 
single-unresolved counterterm; terms containing only double-unresolved limits 
are assigned to $K^{\two}$, which we refer to as \emph{pure} double-unresolved 
counterterm; all remaining terms, containing overlaps of single- and double-unresolved 
limits, while still featuring double-unresolved kinematics, are assigned to $K^{\otwo}$, 
which we refer to as \emph{mixed} double-unresolved counterterm. A direct
characterisation of mixed double-unresolved counterterms in terms of factorisation
kernels will be discussed in Ref.~\cite{Magnea:2018ebr}. We write therefore, for
each topology $T$,
\beq
  K^{\one}_T & = & \bL{\one}{ij} \, \RR \, \W{T} \, , 
\label{eq:cnt_NNLO_loc1} \\ [3pt]
  K^{\two}_T & = & \left( \bL{\two}{T} + \bL{\twosc}{T} \right) \, \RR \, \W{T} 	\, , 
\label{eq:cnt_NNLO_loc2} \\ [3pt]
  K^{\otwo}_T & = & - \, \bL{\one}{ij} \, \left( \bL{\two}{T} + \bL{\twosc}{T} \right) \,
  \RR \, \W{T} \, .
\label{eq:cnt_NNLO_loc12}
\eeq
The definitions in Eqs.~(\ref{eq:cnt_NNLO_loc1})-(\ref{eq:cnt_NNLO_loc12}) are 
very intuitive and compact. First, notice that the candidate single-unresolved counterterm 
has the very same structure as the NLO counterterm, as one can deduce by
comparing \eq{eq:cnt_NNLO_loc1} with \eq{eq:NLOK}. This correspondence 
is strict: indeed, if one imagines removing from a given process all $n$-body 
contributions, for instance by means of phase-space cuts, the original NNLO 
computation reduces to the NLO computation for the process with $n+1$ 
particles at Born level, with $\RR$ playing the role of single-real correction, 
and $\RVl$ that of virtual contribution; in this scenario, $K^{\one}$ becomes 
\emph{exactly} the candidate NLO local counterterm. As for the double-unresolved 
contributions, $K^{\two}$ is to be integrated in $d \widehat\Phi_{\rm rad,2}$, 
giving rise to up to four poles in $\eps$, multiplied by Born-like matrix elements, 
analogously to $\VV$; the single-unresolved structure in $K^{\otwo}$, on the 
other hand, makes it suitable for integration in $d \widehat \Phi_{\rm rad,1}$;
once this is achieved, its double-unresolved projectors naturally become 
single-unresolved projectors for the parent parton which originated the first
splitting, thus reproducing the structure of $K^{\RV}$. This is necessary, since 
the integral of $K^{\otwo}$ must compensate the explicit poles in $\eps$ of $K^{\RV}$. 
This cancellation also relies on the factorisation properties of sector functions, 
presented in \eq{eq:propNNLOfact}, as will be further detailed below.

The double-unresolved kernels appearing in the counterterm definitions of 
Eqs.~(\ref{eq:cnt_NNLO_loc1})-(\ref{eq:cnt_NNLO_loc12}) can be derived 
from soft and collinear limits of scattering amplitudes, which are universal, 
and for the massless case relevant to this article they were computed in 
Refs.~\cite{Catani:1998nv,Catani:1999ss}. General expressions for the 
kernels can also be derived starting from the factorisation of soft and collinear 
poles in virtual corrections to fixed-angle scattering amplitudes, as will
be discussed in detail in Ref.~\cite{Magnea:2018ebr}. Here we just write 
symbolically
\beq
  \bS{ij} \RR \big( \{k\} \big) & = & \frac{\Norm^{\,2}}2
  \sum_{\substack{c \neq i,j \\ d \neq i,j}} \bigg[ 
  \sum_{\substack{e\neq i,j\\f\neq i,j}} \mc I_{cd}^{(i)} \, \mc I_{ef}^{(j)} \,
  {\Bn}_{cdef} \! \left( \{k\}_{\slashed i \slashed j} \right) +
   \mc I_{cd}^{(ij)} \, {\Bn}_{cd} \!
  \left( \{k\}_{\slashed i\slashed j} \right) \bigg] \, ,
\label{eq:NNLO2unreslimitsSS} \\ [3pt]
  \bC{ijk} \RR \big( \{k\} \big) & = & \frac{\Norm^{\,2}}{s_{ijk}^2} \,
  \Big[ P_{ijk} \, \Bn \! \left( \{k\}_{\slashed i \slashed j \slashed k}, k \right) 
  \,+ \, Q_{ijk}^{\mu \nu} \, {\Bn}_{\mu \nu} \!
  \left( \{k\}_{\slashed i \slashed j \slashed k}, k \right) \Big] \nnb \\
  & \equiv & \frac{\Norm^{\,2}}{s_{ijk}^2} \, P_{ijk}^{\mu\nu} \, {\Bn}_{\mu\nu}
  \! \left( \{k\}_{\slashed i \slashed j \slashed k}, k \right) \, ,
\label{eq:NNLO2unreslimitsCC} \\ [3pt]
  \bC{ijkl} \RR \big( \{k\} \big) & = &
  \frac{\Norm^{\,2}}{s_{ij} s_{kl}} \, P^{\mu\nu}_{ij} \, P^{\rho\sigma}_{kl} \, 
  {\Bn}_{\mu\nu\rho\sigma} \! \left( \{k\}_{\slashed i \slashed j \slashed k \slashed l},
  k_{ij}, k_{kl} \right) \, ,
\label{eq:NNLO2unreslimitsCCijkl} \\ [3pt]
  \bSC{ijk} \RR \big( \{k\} \big) & = & - \, 
  \frac{\Norm^{\, 2}}{s_{jk}} \, P^{\mu\nu}_{jk} \sum_{c,d \neq i} \mc I^{(i)}_{cd} \, 
  {\Bn}^{cd}_{\mu\nu} \! \left( \{k\}_{\slashed i \slashed j \slashed k}, k_{jk} \right) 
  \, = \, \bCS{jki} \, \RR \big( \{k\} \big) \, .
\label{eq:NNLO2unreslimitsSCijk}
\eeq
In the equations above, and in the following, the sum over indices $c$
and $d$ is understood to run over the partons that are present at Born level.
In the double-soft limit, ${\Bn}_{cdef}$ is the doubly-colour-connected Born 
matrix element, defined for instance in Eq.~(113) of \cite{Catani:1999ss}; 
the eikonal kernels $\mc I_{ab}^{(i)}$ have been defined in \eq{eq:eikkern}, 
while the kernels $\mc I_{cd}^{(ij)}$ are defined in Eqs.~(96) and (110) 
of~\cite{Catani:1999ss}\footnote{According to our conventions, $\mc I_{cd}^{(ij)}$ 
  corresponds to Eq.~(96) of \cite{Catani:1999ss}, multiplied times $T_R/2$ in the 
  $q\bar q$ case, while it corresponds to Eq.~(110) of \cite{Catani:1999ss}, multiplied 
  times $- C_A/2$ in the $gg$ case. Furthermore, in order to get $\mc I_{cd}^{(ij)}$, 
  one should replace $q_1$ with $k_i$, $q_2$ with $k_j$, $p_i$ with $k_c$, and 
  $p_j$ with $k_d$.}.
In the non-factorisable double-collinear limit $\bC{ijk}$, the set of momenta 
$(\{k\}_{\slashed i\slashed j\slashed k},k)$ refers to a set of $n$ partons obtained 
from $\{k\}$ by removing $k_i$, $k_j$, and $k_k$, and inserting their sum 
$k = k_i \, + \, k_j \, + \, k_k$. The expressions for the double-collinear spin-averaged 
kernels $P_{ijk}$ and for the azimuthal kernels $Q_{ijk}^{\mu\nu}$, all symmetric 
under permutations\footnote{Symmetry under permutations of $i$, $j$, and $k$ 
  does {\emph not} mean symmetry under flavour exchange, but only that kernels 
  and flavour Kronecker delta symbols combine in a symmetric way: this is 
  analogous to what happens in the case of a $q \to qg$ collinear splitting at 
  NLO in \eq{eq:APkernels1}. } of $i$, $j$, and $k$, can be easily extracted from 
\cite{Catani:1998nv,Catani:1999ss}, but the expressions are long and therefore 
will not be reproduced here. We note however that $Q_{ijk}^{\mu\nu}$ can always 
be cast in the form 
\beq
  Q_{ijk}^{\mu\nu} \, = \, \sum_{a = i,j,k} Q^{(a)}_{ijk}
  \left[ - \,g^{\mu\nu} + (d - 2) \, \frac{\kt_{a}^\mu \kt_{a}^\nu}{\kt_{a}^2} 
  \right] \, ,
\label{eq:QNNLO}
\eeq  
where, in analogy with \eq{translong},
\beq
  \kt_{a}^\mu & = & k_a^\mu - z_a \, k^\mu - \left( \frac{k \dt k_a}{k^2} - z_a \right) 
  \frac{k^2}{k \dt k_r} \, k_r^\mu \, , \qquad \quad
  \kt_{i}^\mu + \kt_{j}^\mu + \kt_{k}^\mu \, = \, 0 \, ,\nnb \\
  z_a & = & \frac{k_a \dt k_r}{k \dt k_r} \, = \, \frac{s_{ar}}{s_{ir} + s_{jr} + s_{kr}} \, , 
  \qquad \qquad z_i + z_j + z_k \, = \, 1 \, ,
\label{eq:NNLOcollquantities}
\eeq
and $k_r^\mu$ is a light-like vector which specifies how the collinear limit 
is approached. The Lorentz structure in \eq{eq:QNNLO}, identical to the NLO
one in \eq{eq:APkernels2}, is such that the radiation-phase-space integral
of the double-collinear azimuthal terms vanishes identically. Hence, once 
more, the analytic integration of the counterterms involves only 
spin-averaged kernels. The factorisable double-collinear limit $\bC{ijkl}$ features 
the doubly-spin-correlated Born matrix element ${\Bn}_{\mu\nu\rho\sigma}$, 
with a kinematics obtained from $\{k\}$ removing $k_i$, $k_j$, $k_k$, and 
$k_l$, and inserting the sums $k_{ij} \, = \, k_i \, + \, k_j$, and $k_{kl} \, = \, 
k_k \, + \, k_l$; the corresponding kernel is defined as
\beq
  P^{\mu\nu}_{ij} \, P^{\rho\sigma}_{kl} \, {\Bn}_{\mu\nu\rho\sigma} \, = \, 
  P_{ij} \, P_{kl} \, {\Bn} \, + \, Q^{\mu\nu}_{ij} \, P_{kl} \, {\Bn}_{\mu\nu} \, + \, 
  P_{ij} \, Q^{\rho\sigma}_{kl} \, {\Bn}_{\rho\sigma} \, + \, Q^{\mu\nu}_{ij} \, 
  Q^{\rho\sigma}_{kl} \, {\Bn}_{\mu\nu\rho\sigma} \, .
\label{docoke}
\eeq
Finally, the soft-collinear limit $\bSC{ijk}$ features a colour- and spin-correlated 
Born contribution ${\Bn}^{cd}_{\mu\nu}$, obtained from the colour-correlated 
Born matrix element ${\Bn}^{cd}$ by stripping external spin polarisation vectors.

We now note that, while Eqs.~(\ref{eq:cnt_NNLO_loc1})-(\ref{eq:cnt_NNLO_loc12}) 
are quite natural, they contain a certain degree of redundancy. In fact, the 
double-real matrix element $\RR$ can feature at most four phase-space 
singularities, hence not all of the projectors relevant to a given topology, listed 
in \eq{eq:NNLOprop2}, carry independent information on its singularity structure. 
These redundancies can be eliminated by exploiting the idempotency of projection 
operators: for instance, once $\bSC{icd}$ has been applied to the double-real 
matrix element, further action on the latter by $\Si$ does not produce any 
effect, and analogously if the limit $\bC{ij}$ is applied after the action of 
$\bCS{ijk}$. This ultimately stems from the factorisable nature of $\bSC{icd} \, 
\RR$, and of $\bCS{ijk}\, \RR$, namely
\beq
  \bSC{icd} \, \RR & = & \Si \, \bC{cd} \, \RR \, = \, \bC{cd} \, \Si \, \RR \, ,
  \nnb \\
  \bCS{ijk} \, \RR & = & \bC{ij} \, \bS{k} \, \RR \, = \, \bS{k} \, \bC{ij} \, \RR \, .
\label{scfactor}
\eeq
Even if this factorisation property does \emph{not} hold when the $\bSC{icd}$ 
and $\bCS{ijk}$ limits are applied to the sector functions of \eq{eq:sec_fun_NNLO}, 
the commutation relations discussed in \appn{app:SfunNNLOprop} are sufficient 
to prove that\footnote{Also the limit $\bC{ijkl}$ has a factorisable nature when 
applied on the double-real matrix element,
\beq
  \bC{ijkl} \, \RR \, = \, \Cj \, \Cl \, \RR \, = \,  \Cl \, \Cj \, \RR \, , \nnb
\label{footeq}
\eeq
however, in this case, the relevant commutation relations are not sufficient to 
obtain the analogue of \eq{eq:NNLOredundancy}.}
\beq
  \Si \,\, \bSC{icd} \, \RR \, \W{ibcd} \, = \, \bSC{icd} \, \RR \, \W{ibcd} \, , \nnb \\
  \Cj \,\, \bCS{ijk} \, \RR \, \W{ijkd} \, = \, \bCS{ijk} \, \RR \, \W{ijkd} \, .
\label{eq:NNLOredundancy}
\eeq
As a consequence of \eq{eq:NNLOredundancy}, the candidate mixed double-unresolved 
counterterm $K^{\otwo}$ simplifies to
\beq
  K^{\otwo}_{T} & = & - \, \left( \bL{\one}{ij} \, \bL{\two}{T} + \bL{\twosc}{T} \right) \,
  \RR \, \W{T} \, , 
\label{eq:K12_loc}
\eeq
and the sum of $K_T^{\otwo} + K_T^{\two}$ becomes
\beq
  K^{\otwo}_{T} + K^{\two}_{T}  & = & \left( 1 - \bL{\one}{ij} \right) \, 
  \bL{\two}{T} \, \RR \, \W{T} \, , 
\label{simpk2k12}
\eeq
free of any contribution from the limit $\bL{\twosc}{T}$. A similar simplification 
occurs in the definition of $\bL{\twosc}{ijkj}$ and $\bL{\twosc}{ijkl}$, where 
one can exploit the relations
\beq
  \SCj \,\, \bCS{ijk} \, \left( 1 - \SSk \right) \, = \, \SCl \,\, \bCS{ijk} \, 
  \left( 1 - \SSk \right) \, = \, 0 \, ,
\label{simpsccs}
\eeq
valid both on matrix elements and on sector functions, to rewrite
\beq
  \bL{\twosc}{ijkj} \, = \,  \left( \SCj + \bCS{ijk} \right) \left( 1 - \SSk \right)
  \left( 1 - \CCj \right) \, , \nnb \\
  \bL{\twosc}{ijkl} \, = \,  \left( \SCl + \bCS{ijk} \right) \left( 1 - \SSk \right)
  \left( 1 - \CCl \right) \, .
\label{lscsimp}
\eeq
After the simplifications just discussed, we are finally in a position to write down 
the definition of the candidate local counterterms for all 
contributing topologies $T = ijjk, \, ijkj, \, ijkl$:
\beq
\label{eq:NNLOcnt_final}
   K^{\one}_{T}  & = & \Big[ \, \Si + \Cj \, \left( 1 - \Si \right) \Big]
  \RR \, \W{T} \, , \nnb \\ [3pt]
   K^{\two}_{ijjk}  & = & \Big[ \, \bS{ij} + \bC{ijk} \, \left( 1 - \bS{ij} \right) + \,
  \bSC{ijk} \left( 1 - \bS{ij} \right) \left( 1 - \bC{ijk} \right) \Big] 
  \RR \, \W{ijjk} \, , \nnb \\ [3pt]
   K^{\two}_{ijkj}  & = & \Big[ \, \bS{ik} + \bC{ijk} \, \left( 1 - \bS{ik} \right) + 
  \left( \, \bSC{ijk} + \bCS{ijk} \right) \left( 1 - \bS{ik} \right) \left( 1 - \bC{ijk} \right)
  \Big] \RR \, \W{ijkj} \, , \nnb \\ [3pt]
   K^{\two}_{ijkl}  & = & \Big[ \, \bS{ik} + \bC{ijkl} \, \left( 1 - \bS{ik} \right) + 
  \left( \, \bSC{ikl} + \bCS{ijk} \right) \left( 1 - \bS{ik} \right) \left( 1 - \bC{ijkl} \right)
  \Big] \RR \, \W{ijkl} \, , \nnb \\ [3pt]
   K^{\otwo}_{ijjk} & = &  - \, \Big\{ \Big[ \, \Si + \Cj \, \left( 1 - \Si \right)
  \Big] \Big[ \, \bS{ij} + \bC{ijk} \, \left( 1 - \bS{ij} \right) \Big]
  \nnb\\
  && \hspace{32mm} + \,\,
  \bSC{ijk} \, \left( 1 - \bS{ij} \right) \left( 1 - \bC{ijk} \right) \Big\} \RR \, \W{ijjk} \, ,
  \nnb \\ [3pt]
   K^{\otwo}_{ijkj} & =&  - \, \Big\{ \Big[ \, \Si + \Cj \, \left( 1 - \Si \right)
  \Big] \Big[ \, \bS{ik} + \bC{ijk} \, \left( 1 - \bS{ik} \right) \Big] \nnb \\
  && \hspace{32mm} + \,\,
  \left( \, \bSC{ijk} + \bCS{ijk} \right) \left( 1 - \bS{ik} \right)
  \left( 1 - \bC{ijk} \right) \Big\} \RR \, \W{ijkj} \, , \nnb \\ [3pt]
   K^{\otwo}_{ijkl} & = & - \, \Big\{ \Big[ \, \Si + \Cj \, \left( 1 - \Si \right)
  \Big] \Big[ \, \bS{ik} + \bC{ijkl} \, \left( 1 - \bS{ik} \right) \Big] \nnb \\
  && \hspace{32mm} + \,\,
  \left( \, \bSC{ikl} + \bCS{ijk} \right) \left( 1 - \bS{ik} \right) 
  \left( 1 - \bC{ijkl} \right) \Big\} \RR \, \W{ijkl} \, .
\eeq
The final step for the construction of the NNLO counterterms, analogously to 
what happens in the NLO case discussed in \secn{sec:NLOcntdef}, is to apply 
kinematic mappings to \eq{eq:NNLOcnt_final}. There is ample freedom in the 
choice of these mappings, and in principle different mappings can be employed 
for different kernels, or even for different contributions to the same kernel. The 
detailed definition of the kinematic mappings we employ for each counterterm 
is given in Sections~\ref{sec:1unrescntint} and \ref{sec:2unrescntint} where, 
as usual, all remapped quantities will be denoted with a bar. Finally, the
real-virtual counterterm has formally the same structure as the NLO  counterterm
of \eq{eq:NLOKdef}, with the replacement $\Rl \, \to \, \RVl$,  and will be sketched
in \secn{sec:RVcntint}.

%%%%%%%%%%%%%%%%

\subsection{Single-unresolved counterterm}
\label{sec:1unrescntint}

We start by separating the hard-collinear and the soft contributions to the 
candidate single-unresolved counterterm:
\beq
   K^{\one} & = &  K^{\onehc} +  K^{\ones} \, ,
\label{eq:K1hcps} \\
   K^{\onehc} & = & \sum_{i, \, j \neq i} \bC{ij} \, \left( 1 - \bS{i} \right) \, 
  \RR \sum_{k \neq i, j} \bigg( \W{ijjk} + \W{ijkj} + \sum_{l \neq i,j,k} \W{ijkl}
  \bigg) \, , 
\label{eq:K1hc} \\
   K^{\ones} & = & \sum_{i, \, j \neq i} \bS{i} \, \RR 
  \sum_{k \neq i,j} \bigg( \W{ijjk} + \W{ijkj} + \sum_{l \neq i,j,k} \W{ijkl} \bigg) \, .
\label{eq:K1so}
\eeq
Using the factorisation properties (\ref{eq:propNNLOfact}) we can proceed as
done at NLO. We define the appropriate counterterms with remapped kinematics,
where in this case barred projectors apply not only to matrix elements, but
also to sector functions:
\beq
  \overline K^{\onehc} & = & \sum_{i, \, j \neq i} 
  \sum_{\substack{k \neq i \\ l \neq i,k}} \bigg[ \Big( \bC{ij} \, \Wab{ij} \Big)
  \left( \bbC{ij} \,\RR \right) \, \bW{kl} - \Big( \bS{i} \, \bC{ij} \Wab{ij} \Big)
  \left( \bbS{i} \, \bbC{ij} \, \RR \right) \, \bW{kl} \bigg] \, ,
\label{eqK1hc} \nnb \\
  \overline K^{\ones} & = & \sum_{i, \, j \neq i} 
  \sum_{\substack{k \neq i \\ l \neq i, k}} \Big( \bS{i} \, \Wab{ij} \Big)
  \left( \bbS{i} \RR \right) \, \bW{kl} \, .
\label{eqK1so}
\eeq
The kinematic mapping of sector functions, once the integrated counterterm 
is considered, allows to factorise the structure of NLO sectors out of the radiation 
phase space, and integrate analytically only single-unresolved kernels. Explicitly
\beq
  \left( \bbS{i} \, \RR \right) \, \bW{kl} & \equiv & - \, \Norm \,
  \sum_{\substack{a\neq i\\b\neq i}} \, \mc I^{(i)}_{ab} \,
  {\Rl}_{ab} \Big( \kkl{iab} \Big) \, \bW{kl}^{(iab)} \, ,
\label{eq:K1Smapdef} \\
  \left( \bbC{ij} \, \RR \right) \, \bW{kl} & \equiv & \frac{\Norm}{s_{ij}} \, 
  P^{\mu\nu}_{ij} \, {\Rl}_{\mu\nu} \Big( \kkl{ijr} \Big) \, \bW{kl}^{(ijr)} \, , 
\label{eq:K1Cmapdef} \\
  \left( \bbS{i} \, \bbC{ij} \, \RR \right) \, \bW{kl} & \equiv & 2 \, \Norm \, C_{f_j} \, 
  \mc I_{jr}^{(i)} \Rl \Big( \kkl{ijr} \Big) \, \bW{kl}^{(ijr)} \, ,
\label{eq:K1SCmapdef}
\eeq
where $\Rl_{ab}$ and $\Rl_{\mu\nu}$ are the colour- and spin-correlated real 
matrix elements and
\beq
  \bW{kl}^{(abc)} & = & \frac{\bar\sigma_{kl}^{(abc)}}{\sum\limits_{i, \, j\neq i}
  \bar\sigma_{ij}^{(abc)}} \, , \qquad \qquad \bar\sigma_{ij}^{(abc)} \, = \, 
  \frac{1}{\bar e_i^{(abc)} \, \bar w_{ij}^{(abc)}} \, , \\
  \bar e_i^{(abc)} & = & \frac{\sk{\q i}{abc}}s \, ,  \qquad \qquad \qquad \,\,
  \bar w_{ij}^{(abc)} \, = \, \frac{s \, \sk{ij}{abc}}{\sk{\q i}{abc} \, \sk{\q j}{abc}} \, .
\eeq
In Eqs.~(\ref{eq:K1Cmapdef}) and (\ref{eq:K1SCmapdef}) the choice of $r\neq i,j$ is
as follows: if $k=j$, the same $r$ should be chosen for all permutations of $ijl$,
and analogously for the case $l=j$; if both $k\neq j$ and $l\neq j$,  the same $r$
should be chosen for all permutations in $\pi(\pi(ij)\,\pi(kl))$.

%%%%%%%%%

\subsubsection{Integration of the single-unresolved counterterm}
\label{intsuc}

As done at NLO, we now integrate the single-unresolved counterterm 
in its radiation phase space. We first get rid of the NLO sector functions 
$\Wab{ij}$ using their NLO sum rule, obtaining 
\beq
  \overline K^{\onehc} & = & \sum_{i, \, j > i} \sum_{\substack{k \neq i \\ l \neq i, k}}
  \Big[ \bbC{ij} \left( 1 - \bbS{i} - \bbS{j} \right) \RR \Big] \, \bW{kl} \, , \\
  \overline K^{\ones} & = & \sum_i \sum_{\substack{k \neq i \\ l \neq i, k}}
  \left( \bbS{i} \RR \right) \, \bW{kl} \, ,
\eeq
two expressions which are suitable for analytic integration. Indeed, the integral of 
$\overline K^{\onehc}$ in the single-unresolved radiation phase space 
$d \Phi_{\rm rad, 1}^{(abc)} \, = \, d \Phi_{\rm rad}^{(abc)}$ reads
\beq
  {\hspace{-5mm}} I^{\onehc} & = & \frac{\varsi_\npt}{\varsi_\npo} 
  \sum_{i, \, j>i} \, \sum_{\substack{k \neq i \\ l \neq i, k}} \bW{kl} 
  \int d \Phi_{\rm rad,1}^{(ijr)} \,\, 
  \bCj \, \left( 1 - \bSi - \bSj \right) \, \RR \left( \{k\} \right) \nnb \\
  & = &  \Norm \,\, \frac{\varsi_\npt}{\varsi_\npo} \sum_{i, \, j > i} \, 
  \sum_{\substack{k \neq i \\ l \neq i, k}} \, J^\hc_{ij} 
  \left( \bar s_{jr}^{(ijr)}, \eps \right) \, \, \Rl \Big( \{\bar k\}^{(ijr)} \Big) \, \bW{kl}^{(ijr)}  
\label{eq:IHC1totfin} \\
  & = &  - \, \frac{\as}{2 \pi} \left( \frac{\mu^2}{s} \right)^{\eps}
  \sum_{p} \, \sum_{k, \, l\neq k} \, \bW{kl}^{(ijr)} \, \Rl \Big( \{\bar k\}^{(ijr)} \Big) 
  \Bigg[ \delta_{f_p g} \, \frac{C_A + 4 \, T_R N_f}{6} \left( \frac{1}{\eps} + 
  \frac{8}{3} - \ln \bar\eta_{pr} \right) \nnb \\
  & & \hspace{62mm} + \, \delta_{f_p \{q, \bar q\}} \, \frac {C_F}{2}
  \left( \frac{1}{\eps} + 2 - \ln \bar\eta_{pr} \right) \Bigg]
  \, + \, \mc O(\eps) \, , \nnb
\eeq
fully analogous to its NLO counterpart in \eq{inthcout}. The integral of 
$\overline K^{\ones}$ similarly yields
\beq
\label{eq:Ionesoftintegrated}
  I^{\ones} & = & \frac{\varsi_\npt}{\varsi_\npo} \, \sum_{i} \,
  \sum_{\substack{k \neq i \\ l \neq i, k}} \bW{kl}
  \int d \Phi_{\rm rad, 1} \, \bSi \, \RR \left( \{k\} \right) \nnb \\
  & = & - \, \Norm \,\, \frac{\varsi_\npt}{\varsi_\npo} \, 
  \sum_i \, \delta_{f_i g} \sum_{\substack{k \neq i \\ l \neq i, k}}
  \sum_{\substack{a \neq i \\ b \neq i}} \, 
  J^\so \Big( \bar s_{ab}^{(iab)}, \eps \Big) \, 
  {\Rl}_{ab} \Big( \{ \bar k\}^{(iab)} \Big) \, \bW{kl}^{(iab)}
  \label{i1snnlo} \\ [5pt]
  & = & \frac{\as}{2 \pi} \left( \frac{\mu^2}{s} \right)^{\eps} \,
  \sum_{k,\, l\neq k} \, \bW{kl} \Bigg[ \sum_a \, C_{f_a} \, \Rl \! \left( \{\bar k\} \right) \, 
  \left( \frac{1}{\eps^2} + \frac{2}{\eps} \, + \, 6 - \frac{7}{2} \, \zeta_2
  \right) \nnb \\
  && \quad + \,\, \sum_{a,\,b\neq a} \, {\Rl}_{ab} \! \left( \{\bar k\} \right) \,
  \ln \bar\eta_{ab} \, \left( \frac{1}{\eps} + 2 - \frac{1}{2} \ln \bar\eta_{ab} \right)
  \Bigg] \, + \, \mc O(\eps) \, , \nnb
  \eeq
where, in the last step, all identical soft-gluon contributions have been remapped 
on the same real kinematics $\{\bar k\}$, and the sum $\sum_i \, \delta_{f_i g}$ 
has absorbed the symmetry factor $\varsi_\npt/\varsi_\npo$. The combination of 
hard-collinear and soft contributions is straightforward, as in the NLO case, 
yielding
\beq
  \hspace{-5mm} I^{\one}(\{\bar k\}) & = & I^{\ones} \left( \{\bar k\} \right) + 
  I^{\onehc} \left( \{\bar k\} \right) \, = \, \sum_{h, \, q \neq h} I^{\one}_{hq}
  (\{\bar k\}) \nnb \\ & = & \frac{\as}{2 \pi}
  \left(\frac{\mu^2}{s}\right)^{\eps} \sum_{h, \, q \neq h} \, \bW{hq} \,
  \Bigg\{ \Bigg[
  \Rl \! \left( \{\bar k\} \right) \sum_{a} \left( \frac{C_{f_a}}{\eps^2} + 
  \frac{\gamma_a}{\eps} \right) + \sum_{a, \, b\neq a} \, 
  {\Rl}_{ab} \! \left( \{\bar k\} \right) \,
  \frac{1}{\eps} \ln \bar\eta_{ab} \Bigg] \nnb \\
  && \hspace{6mm}
  + \Bigg[ \Rl \! \left( \{\bar k \} \right) \sum_{a} \, 
  \Bigg( \delta_{f_ag} \frac{C_A + 4 \, T_R \, N_f}6
  \left( \ln \bar\eta_{ar} - \frac{8}{3} \right)
  + \delta_{f_ag} C_A \left( 6 - \frac72 \zeta_2 \right)
\label{eq:intcnt1NNLO} \\
  && \hspace{6mm} + \,\, \delta_{f_a \{ q, \bar q\}} \frac {C_F}2
  \big( 10 - 7 \zeta_2 + \ln \bar\eta_{ar} \big)
  \Bigg) + \sum_{a, \, b\neq a} \, {\Rl}_{ab} \! \left( \{\bar k\} \right) \, 
  \ln \bar\eta_{ab} \, \left( 2 - \frac{1}{2} \ln \bar\eta_{ab} \right) 
  \Bigg] \Bigg\} \, , \nnb
\eeq
where indices $h$ and $q$ run over the NLO multiplicity, barred momenta
and invariants refer to NLO kinematics, and $r\neq a$. \eq{eq:intcnt1NNLO} 
exhibits the same poles in $\eps$ as the ones shown at NLO in \eq{eq:intcntNLO}, 
due to the single-unresolved nature of the involved projectors. Such poles 
are identical (up to a sign) to the ones of the real-virtual matrix element, thus 
showing the finiteness in $d = 4$ of the sum $\RVl + I^{\one}$. It is important to 
note, however, that in \eq{eq:intcnt1NNLO}, as well as in $\RVl$, the full 
structure of NLO sector functions $\bW{hq}$ is factorised in front of the 
integrated singularities, which means that the cancellation of $\ep$ poles 
between $\RVl$ and $I^{\one}$ occurs \emph{sector by sector} in the 
$(\npo)$-body phase space.

%%%%%%%%%%%%%%%%

\subsection{Double-unresolved counterterm}
\label{sec:2unrescntint}

The double-unresolved counterterm with $n$-body kinematics consists of two 
parts: the pure double-unresolved counterterm $\overline K^{\two}$, which must 
be integrated in the double-radiation phase space, and the mixed double-unresolved 
counterterm $\overline K^{\otwo}$ which must be integrated in a single-radiation 
phase space. From \secn{sec:NNLOgen} we see that, while their integration 
has to be performed independently, the non-integrated counterterms 
$\overline K^{\two}$ and $\overline K^{\otwo}$ appear only combined in the 
last line of \eq{eq:NNLO_structure}. Owing to the simplifications discussed at 
the end of \secn{sec:NNLOcntdef}, the sum $K^{\two} + K^{\otwo}$ is much simpler
than the two  terms taken separately, and it reads
\beq
  K^{\two} + K^{\otwo} & = & \sum_{i, \, j \neq i}
  \left( 1 - \bS{i} \right) \left( 1 - \bC{ij} \right) \sum_{k \neq i, j}
  \bigg\{ \Big[ \bS{ij} + \bC{ijk} \!\left( 1 - \bS{ij} \right)\!
  \Big] \, \W{ijjk}  \nnb\\
  && + \, \Big[ \bS{ik} + \bC{ijk}\! \left( 1 - \bS{ik} \right)\!
  \Big] \, \W{ijkj} \, + \!\!\! \sum_{l\neq i,j,k} \!\!
  \Big[ \bS{ik} + \bC{ijkl} \!\left( 1 - \bS{ik} \right) \!\Big] \, 
  \W{ijkl} \bigg\} \, \RR \, . \qquad \, \,
  \label{eq:sumK12K2}
\eeq
Exploiting Eqs.~(\ref{eq:propNNLOfact}), together with
\beq
  \bS{i} \, \bC{ij} \, \bS{ik} \, \RR & = & \bC{ij} \, \bS{ik} \, \RR  \, ,
\label{eq:SS.CC=SS.CC.S}
\nnb\\
  \bC{ij} \, \bC{ijkl} \, \RR & = & \bC{ijkl} \, \RR  \, ,
\label{eq:C.CC=CC}
\eeq
one is able to recast the above expression in a form that explicitly
features only sums of sector functions that add up to 1, according to
the sum rules of Eqs.~(\ref{eq:sec_prop4}),
(\ref{srule1})-(\ref{srule3}), and to
\beq
\label{eq:CC S Wab}
  \bS{i} \, \bC{ijk} \, \left( \Wab{ij} + \Wab{ik} \right) \, = \, 1 \, ,
\eeq
as well as
\beq
\label{eq: SS CC WW}
  \SSj\,\CCj \!\!\!\! \sum_{ab \, \in \, \pi(ij)} \!\!\!\!
  \left( \mc W_{abbk} + \mc W_{akbk} \right) \, = \, 1 \, ,
  \qquad \qquad
  \SSk \, \CCl \, \left(\mc W_{ijkl} + \mc W_{klij}\right) \, = \, 1 \, .
\eeq
Introducing remapped kinematics for the double-real matrix element and
for the sector functions $\W{ab}$, analogously to what done for the
single-unresolved counterterm, the double-unresolved counterterm finally
reads
\beq
\label{eq:K2+K12}
  \overline K^{\two} \! + \overline K^{\otwo} 
  & = & 
  \sum_{i, \, k > i}
  \bigg[ 
  \bS{ik} \, \Big( \sum_{j \neq i} \sum_{l \neq i,k} \W{ijkl} 
  +
  \sum_{j \neq k} \sum_{l \neq i,k} \W{kjil} \Big) 
  \bigg] \, 
  \bbS{ik} \, \RR
  \nnb \\
  & & 
  + \, 
  \sum_{i, \, j > i} \sum_{k > j}
  \bigg[ 
  \bC{ijk}
  \!\!\!\!\!
  \sum_{abc \, \in \,\pi(ijk)} \!\!\!\!\! 
  \big( \W{abbc} + \W{abcb} \big)
  \bigg]\, 
  \bbC{ijk}\,\RR
  \nnb \\
  & & 
  - \, 
  \sum_{i, \, j > i} \sum_{k \neq i,j}
  \bigg[
  \bS{ij} \, \bC{ijk}
  \!\!\!\!\!
  \sum_{ab \, \in \, \pi(ij)} \!\!\!\!\!
  \big( \W{abbk} + \W{akbk} \big)
  \bigg]
  \, \bbS{ij} \, \bbC{ijk} \, \RR
  \nnb \\
  & &
  + \,
  \sum_{i, \, j > i} \sum_{\substack{k > i \\ k \neq j}}
  \sum_{\substack{l > k \\ l \neq j}} \bigg[ \bC{ijkl}  \!\!\!\! 
  \sum_{\substack{ab \, \in \, \pi(ij) \\ cd \, \in \, \pi(kl)}} \!\!\!\!\!
  \big( \W{abcd} + \W{cdab} \big) \bigg] \,
  \bbC{ijkl} \, \RR
  \nnb \\
  & &
  - \,
  \sum_{i,\,j>i} \sum_{\substack{k > i \\ k \neq j}} 
  \sum_{\substack{ l > k \\ l \neq j}}
  \sum_{\substack{ab \, \in \, \pi(ij) \\ cd \, \in \, \pi(kl)}} \!
  \bigg[ \bS{ac} \, \bC{abcd} \big( \W{abcd} + \W{cdab} \big) \bigg] \,
  \bbS{ac} \, \bbC{ijkl} \, \RR \nnb
  \nnb\\
  & &
  -
  \sum_{i, \, j > i}
  \sum_{k\neq i,j}
  \bigg[ \bC{ij} \left( \Wab{ij} + \Wab{ji} \right) \bigg] \,
  \bigg\{ \Big[ \bC{jk} \left( \bW{jk} + \bW{kj} \right) \Big] \,
  \bbC{ij} \, \bbC{ijk}
  \nnb \\
  & &
  \hspace{4mm}
  + \!\!\!
  \sum_{l \neq i,j,k} \! 
  \left( \bC{kl} \, \bW{kl} \right)
  \bbC{ijkl}
  +
  \left( \bS{j} \bW{jk} \right)
  \bbC{ij} \, \bbS{ij}
  -
  \left( \bS{j} \, \bC{jk} \, \bW{jk} \right) \bbC{ij} \,
  \bbS{ij} \, \bbC{ijk}
  \bigg\} \,
  \RR
  \nnb\\
  & &
  +
  \sum_{i, \, j \neq i} \sum_{k \neq i,j} \!
  \Big[ \bS{i} \, \bC{ij} \, \Wab{ij} \Big]
  \bigg\{ \!
  \Big[ \bC{jk} \left( \bW{jk} + \bW{kj} \right) \! \Big] \,
  \bbS{i} \, \bbC{ij} \, \bbC{ijk}
  +
  \left( \bS{j} \, \bW{jk} \right) 
  \bbS{i} \, \bbC{ij} \, \bbS{ij}
  \nnb \\
  & & 
  \hspace{4mm}
  + \!
  \sum_{l \neq i,j,k} \!
  \left( \bS{k} \, \bC{kl} \, \bW{kl} \right) \,
  \bbS{i} \, \bbS{ik} \, \bbC{ijkl}
  -
  \left( \bS{j} \, \bC{jk} \, \bW{jk} \right) \,
  \bbS{i} \, \bbC{ij} \, \bbS{ij} \, \bbC{ijk}
  \bigg\} \,
  \RR
  \nnb \\
  & &
  -
  \sum_{i, \, j \neq i} \sum_{\substack{k \neq i \\ k > j}}
  \bigg[ \bS{i} \, \bC{ijk} \left( \Wab{ij} + \Wab{ik} \right) \bigg]
  \bigg\{
  \Big[ \bC{jk} \left( \bW{jk} + \bW{kj} \right) \Big] \,
  \bbS{i} \, \bbC{ijk} 
  \nnb \\ [-3pt]
  & &
  \hspace{8mm}
  -
  \left( \bS{j} \, \bC{jk} \, \bW{jk} \right) \, 
  \bbS{i} \, \bbS{ij} \, \bbC{ijk}
  - \,
  \left( \bS{k} \, \bC{jk} \, \bW{kj}\right) \,
  \bbS{i} \, \bbS{ik} \, \bbC{ijk}
  \bigg\} \,
  \RR
  \nnb \\
  & &
  -
  \sum_i \sum_{\substack{k \neq i \\ l \neq i,k}}
  \bigg[ \bS{i} \sum_{j \neq i} \Wab{ij} \bigg] \!
  \left( \bS{k} \, \bW{kl} \right) \, 
  \bbS{i} \, \bbS{ik} \, \RR \, .
\eeq
We stress that in each contribution the kinematics of the double-real matrix 
element undergoes a different mapping onto the Born one, so as to maximally 
adapt the parametrisation of the integrands to the kinematic invariants that 
naturally appear in the respective kernels. The explicit definition of the barred 
limits appearing in the first three lines of \eq{eq:K2+K12} is
\beq
\label{eq: SS mapped}
\bbS{ij} \, \RR 
&=&
\frac{\Norm^{\,2}}{2} \!
\sum_{\substack{c \neq i,j \\ d \neq i,j,c}} \!
\Bigg[
\sum_{\substack{e \neq i,j,c,d \\ f \neq i,j,c,d}} \! 
\mc I_{cd}^{(i)} \,
\mc I_{ef}^{(j)} 
{\Bn}_{cdef} \Big( \! \kkl{icd,jef} \! \Big) 
+
4
\sum_{\substack{e \neq i,j,c,d}} \! 
\mc I_{cd}^{(i)} \,
\mc I_{ed}^{(j)} 
{\Bn}_{cded} \Big( \! \kkl{icd,jed} \! \Big) 
\nnb\\
&&\quad
+ 
2\,
\mc I_{cd}^{(i)} \, 
\mc I_{cd}^{(j)}
{\Bn}_{cdcd} \Big( \! \kkl{ijcd} \! \Big) 
+
\left( 
\mc I_{cd}^{(ij)} - \frac12\,\mc I_{cc}^{(ij)} - \frac12\,\mc I_{dd}^{(ij)} 
\right) \!
{\Bn}_{cd} \Big( \! \kkl{ijcd} \! \Big) 
\Bigg]
\, ,  \\
\label{eq: CC mapped}
  \bbC{ijk} \, \RR & = & \frac{\Norm^{\, 2}}{s_{ijk}^2} \, P_{ijk}^{\mu\nu} \,
  {\Bn}_{\mu\nu} \Big( \! \kkl{ijkr} \! \Big) \, , \\
\label{eq: SSCC mapped}
  \bbS{ij} \, \bbC{ijk} \, \RR & = & \frac{\Norm^{\, 2}}2 
  C_{f_k} \bigg[ 8 \, \mc I_{rk}^{(i)} \, \mc I_{rk}^{(j)} \, C_{f_k} 
  + \mc I_{rr}^{(ij)} - 2 \, \mc I_{rk}^{(ij)} \, + \mc I_{kk}^{(ij)} \bigg] 
  {\Bn} \Big( \! \kkl{ijkr} \! \Big) \, ,
\eeq
where the same $r\neq i,j,k$ should be chosen for all permutations 
of $ijk$, and we have introduced the mapping
\beq
  \kkl{abcd} \, = \,  \left\{ \kk{h}{abcd} \right\}_{h \neq a,b} \, , & & \qquad
  \kk{n}{abcd} \, = \, k_n \, , \quad \, n \neq a,b,c,d \, , \nnb \\ [2pt]
  \kk{c}{abcd} \, = \, k_{a} + k_{b} + k_{c} - 
  \frac{s_{abc}}{s_{ad} + s_{bd} + s_{cd}} \, k_{d} \, , & & \qquad
  \kk{d}{abcd} \, = \, \frac{s_{abcd}}{s_{ad} + s_{bd} + s_{cd}} \, k_{d} \, ,
\label{eq: repar2 abcd}
\eeq
Notice that the second line in \eq{eq: SSCC mapped} would vanish by color 
conservation in the absence of phase-space mappings: its role is to ensure 
that the double-unresolved counterterm fulfil the proper limits also in the 
presence of the mappings. The definition of the barred limits in the fourth 
and fifth lines of \eq{eq:K2+K12} is
\beq
\label{eq: CCijkl mapped}
  \bbC{ijkl} \, \RR & = & \Norm^{\,2} \, 
  \frac{P^{\mu\nu}_{ij} \left(s_{il}, s_{jl} \right)}{s_{ij}} \,
  \frac{P^{\rho\sigma}_{kl} \left( \sk{kr}{ijl}, \sk{lr}{ijl} \right)}{\sk{kl}{ijl}} \,
  {\Bn}_{\mu\nu\rho\sigma} \Big( \! \kkl{ijl,klr} \! \Big) \, , \\
\label{eq: SSCCijkl mapped}
  \bbS{ac}\,\bbC{ijkl} \, \RR & = & 4 \, \Norm^{\,2} \, C_{f_b} \mc I_{bl}^{(a)} \,
  \delta_{f_c g} \, C_{f_d} \frac{\sk{dr}{ijl}}{\sk{cd}{ijl} \sk{cr}{ijl}} \,
  {\Bn} \Big( \! \kkl{ijl,klr} \! \Big) \, , \quad
  \begin{array}{l}
  ab \, \in \, \pi(ij) \, ,\\ 
  cd \, \in \, \pi(kl) \, ,
  \end{array}
\eeq
where the same $r \neq i, k, l$ should be chosen for all permutations
in $\pi(\pi(ij)\,\pi(kl))$. We have introduced the remapping
\beq
  \kkl{acd,bef} \, = \, \left\{ \kk{h}{acd,bef} \right\}_{h \neq a,b} \, ,
  \qquad
  \kk{n}{acd,bef} \, = \, \kk{n}{acd} \, ,
  \quad
  n \neq a,b,e,f \, ,
  \nnb
\eeq
\beq
\label{eq: repar2 acd,bef}
\hspace{-5mm}
  \kk{e}{acd,bef} \, = \, \kk{b}{acd} + \kk{e}{acd} - 
  \frac{\sk{be}{acd}}{\sk{bf}{acd} + \sk{ef}{acd}} \, \kk{f}{acd} \, ,  
  \qquad
  \kk{f}{acd,bef} \, = \, \frac{\sk{bef}{acd}}{\sk{bf}{acd} + \sk{ef}{acd}} \, 
  \kk{f}{acd} \, ,
\eeq
and it can be easily shown that the two remappings in \eq{eq: repar2 abcd}
and \eq{eq: repar2 acd,bef} satisfy
\beq
  \kkl{acd,bcd} \, = \, \kkl{abcd} \, , \qquad
  \kkl{abc,bcd} \, = \, \kkl{abcd} \, .
\eeq
The remaining composite limits of $\RR$ appearing in \eq{eq:K2+K12} are listed 
in \appn{app: composit limits RR}. 

%%%%%%%%%

\subsubsection{Integration of the mixed double-unresolved counterterm}
\label{sec:12unrescntint}

The mixed double-unresolved counterterm features $n$-body kinematics but, 
peculiarly, it needs to be integrated analytically only in the phase space of a 
single radiation. This operation is necessary to show that such an integral 
features the same explicit $\ep$ singularities as the $\overline K^{\RV}$ 
counterterm, and, at the same time, it features the same phase-space 
singularities is $I^{\one}$.

We start by considering the hard-collinear contribution to $K^{\otwo}$. Following Eqs.~(\ref{eq:NNLOcnt_final}) we have
\beq
\label{eq:K12hcdefinition}
  K^{\otwohc} & = & - \sum_{i, \, j \neq i} \sum_{k \neq i,j} 
  \bigg\{ \bC{ij}  \left( 1 - \bS{i} \right) \Big[ \bS{ij} + \bC{ijk} \left( 1 - \bS{ij} \right)
  \Big] \, \RR \, \W{ijjk} \\
  & &
  + \,\, \bigg[ \bC{ij}  \left( 1 - \bS{i} \right) \Big[ \bS{ik} + \bC{ijk} \, 
  \left( 1- \bS{ik} \right) \Big] + \bCS{ijk}  \left( 1 - \bS{ik} \right) 
  \left( 1 - \bC{ijk} \right) \bigg] \, \RR \, \W{ijkj} \nnb \\
  & &
  + \!\! \sum_{l\neq i,j,k}
  \bigg[ \bC{ij} \left( 1 - \bS{i} \right) \Big[ \bS{ik} + \bC{ijkl} \left( 1 - \bS{ik} \right)\!
  \Big] + \bCS{ijk} \left( 1 -  \bS{ik} \right) \!\left( 1 - \bC{ijkl} \right) \!
  \bigg] \, \RR \, \W{ijkl} \bigg\} \, . \nnb
\eeq
We stress that in the last expression we have kept the $\bCS{ijk}$ terms: these
cancel out in the sum $K^{\two} + K^{\otwo}$, but do contribute to the integrals 
$I^{\two}$ and $I^{\otwo}$, which have to be evaluated separately.

The explicit computations reported in \appn{app:intK12formulae} show that the 
phase-space integral $I^{\otwohc}$ of the hard-collinear contribution can be 
recast in the simple form
\beq
\label{eq:I12hcfinal}
  I^{\otwohc} & = & - \, \Norm \,\, \frac{\varsi_\npt}{\varsi_\npo}
  \sum_{i, \, j > i} \sum_{\substack{k \neq i \\ l \neq i,k}} \,
  J^{\hc}_{ij} \Big( \sk{jr}{ijr}, \eps \Big) \, \Big[ \bbS{k} + \bbC{kl}
  \left( 1 - \bbS{k} \right) \Big] \, R \Big( \! \kkl{ijr} \! \Big) \,\, \bW{kl}^{(ijr)} \, ,
\eeq
where the integral $J_{ij}^{\hc}$ is defined in \eq{eq:Jijhc}, and the barred
limits on $\Rl$ are given by
\beq
\label{eq:explicit singleunres}
  \bbS{k} \, R \Big( \! \kkl{ijr} \! \Big) & = & - \, \Norm \, 
  \sum_{\substack{c \neq k \\ d \neq k}} \, \delta_{f_kg}
  \frac{\sk{cd}{ijr}}{\sk{kc}{ijr} \sk{kd}{ijr}} \, 
  {\Bn}_{cd} \Big( \! \kkl{ijr,kcd} \! \Big) \, , \\
\label{eq:explicit singleunres C}
  \bbC{kl} \, R \Big( \! \kkl{ijr} \! \Big) & = & \frac{\Norm}{\sk{kl}{ijr}} \,\, 
  P^{\mu\nu}_{kl} \left( \sk{kr'}{ijr}, \sk{lr'}{ijr} \right)
  {\Bn}_{\mu\nu} \Big( \! \{\bar k\}^{(ijr,klr')} \! \Big) \, , \\
\label{eq:explicit singleunres SC}
  \bbS{k}\bbC{kl} \, R \Big( \! \kkl{ijr} \! \Big) & = & 2 \Norm \, \delta_{f_kg}
  \frac{\sk{lr'}{ijr}}{\sk{kl}{ijr} \sk{kr'}{ijr}} \, C_{f_l}
  \Bn \Big( \! \{\bar k\}^{(ijr,klr')} \! \Big) \, .
\eeq
In Eqs.~(\ref{eq:explicit singleunres})-(\ref{eq:explicit singleunres SC}) $r \neq i,j$ 
should be the same as in Eqs.~(\ref{eq:K1Cmapdef}) and (\ref{eq:K1SCmapdef}); 
if $k = j$, the same $r'$ should be chosen for all permutations of $ijl$, and 
analogously for the case $l = j$; if both $k \neq j$ and $l \neq j$,  the same $r'$
should be chosen for all permutations in $\pi(\pi(ij) \, \pi(kl))$. By comparing 
\eq{eq:I12hcfinal} with the second line of \eq{eq:IHC1totfin}, it is clear that, 
as desired, the $I^{\otwohc}$ integral contains all non-integrable phase-space 
singularities of $I^{\onehc}$. The leftover integrable logarithmic singularities, 
contained in the integral kernel $J^{\hc}_{ij} (\sk{jr}{ijr},\eps)$, do not hamper 
numerical integrability.

We now consider the $K^{\otwos}$ counterterm, which is obtained combining 
the soft contributions of the last three equations of (\ref{eq:NNLOcnt_final}). 
The result is
\beq
\label{eq:K12so1}
  K^{\otwos} & = & - \sum_{i, \, j \neq i} \sum_{k \neq i,j} 
  \bigg\{ \, \bigg[ \, \bS{i} \, \Big( \bS{ij} + \bC{ijk} \, \left( 1 - \bS{ij} \right) \Big) + 
  \bSC{ijk} \, \left( 1 - \bS{ij} \right) \left( 1 - \bC{ijk} \right) \bigg] \, \RR \, \W{ijjk}
  \nnb \\
  & & \hspace{-6mm} + \, \bigg[ \, \bS{i} \, \Big( \bS{ik} + \bC{ijk} \, 
  \left( 1 - \bS{ik} \right) \Big) + \bSC{ijk} \, \left( 1 - \bS{ik} \right) 
  \left( 1 - \bC{ijk} \right) \bigg] \, \RR \, \W{ijkj}
  \nnb \\
  & & \hspace{-6mm} + \! \sum_{l \neq i,j,k} \bigg[ \, \bS{i} \, 
  \Big( \bS{ik} + \bC{ijkl} \, \left( 1 - \bS{ik} \right) \Big) + \bSC{ikl} \, 
  \left( 1 - \bS{ik} \right) \left( 1 - \bC{ijkl} \right) \bigg] \, \RR \, \W{ijkl} \bigg\} \, .
\eeq
The explicit computations reported in \appn{app:intK12formulae} show that the
phase-space integral $I^{\otwos}$ of the soft contribution can be recast as
\beq
\label{eq:I12sfinal}
  I^{\otwos} & = & \Norm \,\, \frac{\varsi_\npt}{\varsi_\npo} \sum_i \, \delta_{f_i g}
  \sum_{\substack{k \neq i \\ l \neq i,k}} \sum_{\substack{a \neq i \\ b \neq i}}
  J^\so \Big( \bar s_{ab}^{(iab)}, \eps \Big) 
  \Big[ \, \bbS{k} + \bbC{kl} \, \left( 1 - \bbS{k} \right) \Big] 
  {\Rl}_{ab} \Big( \! \{\bar k\}^{(iab)} \! \Big) \, \bW{kl}^{(iab)} \, , \qquad
\eeq
where the integral $J^{\so}$ is defined in \eq{eq:Jsoft}, and the limits in this
case are defined by 
\beq
\label{explicitSforS}
  \bbS{k} \, R_{ab} \Big( \! \kkl{iab} \! \Big) & = & - \, \Norm \,
  \sum_{\substack{c\neq k\\d\neq k}} \, \delta_{f_kg} \, 
  \frac{\sk{cd}{iab}}{\sk{kc}{iab} \sk{kd}{iab}} \,
  {\Bn}_{abcd} \Big( \! \kkl{iab,kcd} \! \Big) \, , \\
\label{explicitCforS}
  \bbC{kl} \, R_{ab} \Big( \! \kkl{iab} \! \Big) & = & 
  \frac{\Norm}{\sk{kl}{iab}} \, P^{\mu\nu}_{kl} \Big( \sk{kr'}{iab}, \sk{lr'}{iab} \Big) \, 
  {\Bn}^{ab}_{\mu\nu} \Big( \! \{\bar k\}^{(iab,klr')} \! \Big) \, , \\
\label{explicitSCforS}
  \bbS{k} \, \bbC{kl} \, R_{ab} \Big( \! \kkl{iab} \! \Big) & = & 2 \, \Norm \,
  \delta_{f_kg} \, \frac{\sk{lr'}{iab}}{\sk{kl}{iab} \sk{kr'}{iab}} \, C_{f_l}
  \Bn_{ab} \Big( \! \{\bar k\}^{(iab,klr')} \! \Big) \, ,
\eeq
where $r' \neq k,l$, and the same $r'$ should be chosen for $kl$ and for $lk$.
The same considerations that were applied below \eq{eq:explicit singleunres} hold 
in this case as well, referring now to the comparison between \eq{eq:I12sfinal} and \eq{eq:Ionesoftintegrated}. Combining soft and hard-collinear contributions, the 
final expression for the integrated counterterm $I^{\otwo}$ is
\beq
\label{eq:intcnt12NNLOfin}
  I^{\otwo} \left( \{\bar k\} \right) & = & I^{\otwos} \left( \{\bar k\} \right) + 
  I^{\otwohc} \left( \{\bar k\} \right) \, = \, \sum_{h, \, q \neq h} I^{\otwo}_{hq}
  (\{\bar k\}) \nnb \\
  & = & - \, \frac{\as}{2\pi} \left( \frac{\mu^2}{s} \right)^{\eps} \,
  \sum_{h,\, q \neq h} \Big[ \bbS{h} + \bbC{hq} \left( 1 - \bbS{h} \right) \Big]
  \, \bW{hq} \\
  & & \times \, \Bigg\{ \Bigg[ \Rl \big( \{\bar k\} \big) \sum_{a}
  \Bigg( \frac{C_{f_a}}{\eps^2} + \frac{\gamma_a}{\eps} \Bigg) +
  \sum_{a, \, b \neq a} \, {\Rl}_{ab} \big( \{\bar k\} \big) \, \frac{1}{\eps}
  \ln \bar\eta_{ab} \Bigg] \nnb \\
  & & \hspace{3.4mm}
  + \Bigg[ \Rl \! \left( \{\bar k\} \right) \sum_{a} 
  \Bigg( \delta_{f_ag} \frac{C_A + 4 \, T_R \, N_f}6
  \left( \ln \bar\eta_{ar} - \frac{8}{3} \right) + 
  \delta_{f_ag} C_A \left( 6 - \frac72 \zeta_2 \right) \nnb \\
  & & \hspace{9mm}
  + \, \delta_{f_a \{q, \bar q\}} \, \frac {C_F}2
  \big(10 - 7 \zeta_2 + \ln \bar\eta_{ar} \big) \! \Bigg) + \!
  \sum_{a, \, b \neq a} \! {\Rl}_{ab} \! \left( \{\bar k\} \right)
  \ln \bar\eta_{ab} \left( 2 - \frac{1}{2} \ln \bar\eta_{ab} \right)
  \Bigg] \Bigg\} \, , \nnb
\eeq
where the soft and collinear limits are meant to be applied on matrix elements 
and on sector functions, but not on the logarithms $\ln \bar\eta_{ij}$, while barred 
momenta and invariants refer to NLO kinematics, and finally one must choose
$r \neq a$.

Since $I^{\otwo}$ collects the same phase-space singularities as $I^{\one}$, 
and $I^{\one}$ in turn features the same explicit $\ep$ poles as $\RVl$, it follows 
by construction that $I^{\otwo}$ also contains the same $\ep$ poles as 
$\overline K^{\RV}$, as necessary in order to compute the second line 
of \eq{eq:NNLO_structure} in $d = 4$. We stress that these considerations 
hold separately in each NLO sector $\bW{hq}$.

%%%%%%%%%

\subsubsection{Integration of the pure double-unresolved counterterm}
\label{sec:2unrescntintpure}

The candidate pure double-unresolved counterterm, summed over NNLO 
sectors, follows from \eq{eq:NNLOcnt_final} and reads
\beq
  {K}^{\two} & = & \sum_{i, \, j \neq i} \sum_{k \neq i,j} \bigg\{ \Big[ \,
  \bS{ij} + \bC{ijk} \left( 1 - \bS{ij} \right) + \bSC{ijk} \left( 1 - \bS{ij} \right)
  \left( 1 - \bC{ijk} \right) \Big] \, \RR \, \W{ijjk} \\
  & & + \, \Big[ \, \bS{ik} + \bC{ijk} \left(1 - \bS{ik} \right) + 
  \left( \bSC{ijk} + \bCS{ijk} \right) \left( 1 - \bS{ik} \right) 
  \left( 1 - \bC{ijk} \right) \Big] \, \RR \, \W{ijkj} \nnb \\
  & & + \!\!\! \sum_{l \neq i,j,k} \Big[ \, \bS{ik} + \bC{ijkl} \left( 1 - \bS{ik} \right) + 
  \left( \bSC{ikl} + \bCS{ijk} \right) \left( 1 - \bS{ik} \right) \left( 1 - \bC{ijkl} \right) 
  \Big] \, \RR \, \W{ijkl} \bigg\} \, . \nnb
\eeq
We work on this expression by symmetrising indices, and exploiting the sum 
rules in Eqs.~(\ref{srule1})-(\ref{srule3}), as well as \eq{eq: SS CC WW}, together 
with
\beq
  & & \hspace{-4.6mm} \bSC{ijk} \,\bS{ij} \, \sum_{b \neq i} \W{ibjk} \, = \, 1 \, ,
  \hspace{34.6mm}
  \bCS{ijk} \, \bS{ik} \, \sum_{d\neq i,k} \W{ijkd} \, = \, 1 \, , \nnb \\
  & & \hspace{-5mm} 
  \bCS{ijk} \, \bC{ijk} \, \left( \W{ijkj} + \W{jiki} \right) \, = \, 1 \, ,
  \hspace{22.8mm} \bCS{ijk} \, \bC{ijkl} \, \left( \W{ijkl} + \W{jikl} \right) \, = \, 1 \, ,
  \nnb \\ [8pt]
  & & \hspace{-5mm}
  \bCS{ijk} \, \bC{ijk} \, \bS{ik} \, \W{ijkj} \, = \, 1 \, ,
  \hspace{33.3mm}
  \bCS{ijk} \, \bC{ijkl} \, \bS{ik} \, \W{ijkl} \, = \, 1 \, , \nnb \\ [5pt]
  & & \hspace{-5mm}
  \bSC{ijk} \, \bC{ijk} \!\!\!\!\! \sum_{ab \, \in \, \pi(jk)} \!\!\!\!\!
  \left( \W{iaab} + \W{iaba} \right) \, = \, 1 \, ,
  \hspace{15.9mm}
  \bSC{ikl}\,\bC{ijkl} \, \left( \W{ijkl} + \W{ijlk} \right) \, = \, 1 \, , \nnb \\
  & & \hspace{-5mm}
  \bSC{ijk} \, \bC{ijk} \, \bS{ik} \, \left( \W{ijkj} + \W{ikkj} \right) \, = \, 1 \, ,
  \hspace{16.8mm} 
  \bSC{ijk} \, \bC{ijkl} \, \bS{ik} \, \W{ijkl} \, = \, 1 \, .
\eeq
Introducing remapped kinematics for the double-real matrix element, the pure 
double-unresolved counterterm can be finally cast in the form
\beq
\label{eq:K2noW}
  \overline K^{\two} & = & \sum_i \, \Bigg\{ \sum_{j > i} \, \bbS{ij} + \sum_{j > i} 
  \sum_{k > j} \, \bbC{ijk} \, \left( 1 - \bbS{ij} - \bbS{ik} - \bbS{jk} \right) \\
  & & \hspace{1cm}
  + \, \sum_{j > i} \, \sum_{\substack{k > i \\ k \neq j}} \, 
  \sum_{\substack{l > k \\ l \neq j}} \, \bbC{ijkl} 
  \left( 1 - \bbS{ik} - \bbS{jk} - \bbS{il} - \bbS{jl} \right)
  \nnb \\ [-2pt]
  & & \hspace{1cm}
  + \, \sum_{j \neq i} \sum_{\substack{k \neq i \\ k > j}} \, 
  \bbSC{ijk} \left( 1 - \bbS{ij} - \bbS{ik} \right) 
  \bigg( 1 - \bbC{ijk} - \!\!\! \sum_{l \neq i,j,k} \bbC{iljk} \bigg)
  \nnb \\ [-5pt]
  & & \hspace{1cm}
  + \, \sum_{j > i} \sum_{\substack{k \neq i,j}} \,
  \bbCS{ijk} \left( 1 - \bbS{ik} - \bbS{jk} \right)
  \bigg( 1 - \bbC{ijk} - \!\!\! \sum_{l \neq i,j,k} \bbC{ijkl} \bigg) \bigg\}
  \, \RR \, , \nnb
\eeq
which is manifestly free of NNLO sector functions. The counterterm in~\eq{eq:K2noW} 
is thus suitable for analytic integration over the double-unresolved phase space, 
upon definition of the barred limits. First, we note that the barred limits 
appearing in the first line of \eq{eq:K2noW} have already been defined in 
Eqs.~(\ref{eq: SS mapped})-(\ref{eq: SSCC mapped}). Next, we consider 
all terms in \eq{eq:K2noW} containing the four-particle double-collinear 
barred limits $\bbC{abcd}$. Their contribution can be rewritten as
\beq
\label{mysplit}
  \overline K^{\two}_{\rm cc4} & \equiv & \sum_i \, \left[
  \sum_{j > i} \, \sum_{\substack{k > i \\ k \neq j}} \, 
  \sum_{\substack{l > k \\ l \neq i,j}} \, 
  \Big( 1 - \bbS{ik} - \bbS{jk} - \bbS{il} - \bbS{jl} \Big) \right. \\
  & & \!\!\! \left. - \, \sum_{j \neq i}
  \sum_{\substack{k \neq i,j}} \sum_{\substack{l > k \\ l \neq i,j}}
  \bbSC{ikl} \Big( 1 - \bbS{ik} - \bbS{il} \Big) - \sum_{j > i}
  \sum_{\substack{k \neq i,j}} \sum_{\substack{l \neq i,j,k}} \!
  \bbCS{ijk} \Big( 1 - \bbS{ik} - \bbS{jk} \Big) \right] \bbC{ijkl} \, \RR \, . \nnb
\eeq
Defining the barred limits in terms of soft and collinear 
kernels, \eq{mysplit} becomes
\beq
  \overline K^{\two}_{\rm cc4} & = & \Norm^{\, 2} \, \sum_{i , \, j > i} \,
  \sum_{\substack{k > i \\ k \neq j}} \,
  \sum_{\substack{l > k \\ l \neq i,j}} \, \frac{P^{\hc \, \mu\nu}_{ij} 
  \left(s_{il},s_{jl} \right)}{s_{ij}} \, \frac{P^{\hc \, \rho\sigma}_{kl}
  \Big( \sk{kr}{ijl}, \sk{lr}{ijl} \Big)}{\sk{kl}{ijl}} \,
  {\Bn}_{\mu\nu\rho\sigma} \Big( \! \kkl{ijl,klr} \! \Big) \nnb \\
  & & - \, 2 \, \Norm^{\, 2} \sum_{i , \, j > i}
  \sum_{\substack{k < i \\ k \neq j}} \sum_{\substack{l > k \\ l \neq i,j}}
  \left[ \frac{P^{\hc \, \mu\nu}_{ij} \left(s_{il}, s_{jl} \right)}{s_{ij}}
  \left(\! C_{f_l} \, \delta_{f_kg} \, \frac{\sk{lr}{ijl}}{\sk{kl}{ijl} \sk{kr}{ijl}} +
  C_{f_k} \, \delta_{f_lg} \, \frac{\sk{kr}{ijl}}{\sk{kl}{ijl} \sk{lr}{ijl}} \right) \right.
  \nnb \\
  & & \hspace{20mm}
  + \, \left. \left(\! C_{f_j} \mc I^{(i)}_{jl} + C_{f_i} \mc I^{(j)}_{il} \right)
  \frac{P^{\hc \, \mu\nu}_{kl} \Big( \sk{kr}{ijl}, \sk{lr}{ijl} \Big)}{\sk{kl}{ijl}} \,
  \right] \! {\Bn}_{\mu\nu} \Big( \! \kkl{ijl,klr} \! \Big) \, .
\eeq
Finally, the remaining terms in \eq{eq:K2noW}, involving the limits $\bbSC{}$ 
and $\bbCS{}$, can be explicitly defined as 
\beq
\label{eq: bar SCijk (1-Sij-Sik) (1-Cijk) RR}
  \bbSC{ijk}  \left(1 - \bbS{ij} - \bbS{ik} \right) 
  \left(1 - \bbC{ijk} \right) \! \RR  & = & - \, \Norm^{\, 2} \, \!\!
  \sum_{\substack{c \neq i,j,k \\ d \neq i,j,k}} \!\!
  \mc I^{(i)}_{cd} \, \frac{P_{jk}^{\hc \, \mu\nu} 
  \Big( \sk{jr'}{icd}, \sk{kr'}{icd} \Big)}{\sk{jk}{icd}}
  B_{\mu\nu}^{cd} \Big( \! \kkl{icd,jkr'} \! \Big) \, ,
  \nnb \\ [-3mm] \\
\label{eq: bar CSijk (1-Sik-Sjk) (1-Cijk) RR}
  \bbCS{ijk}  \left( 1 - \bbS{ik} - \bbS{jk} \right) \left(1 - \bbC{ijk} \right) \! \RR
  & = & - \, \Norm^{\, 2} \, \!\! \sum_{\substack{c \neq i,j,k \\ d \neq i,j,k}} \!\!
  \frac{P_{ij}^{\hc \, \mu\nu} \left( s_{ir}, s_{jr} \right)}{s_{ij}} \,
  \frac{\delta_{f_kg} \, \sk{cd}{ijr}}{\sk{kc}{ijr} \sk{kd}{ijr}} \,
  B_{\mu\nu}^{cd} \Big( \! \kkl{ijr,kcd} \! \Big) \, .
  \nnb \\ [-3mm]
\eeq
Note that $\overline{K}^{\two}$ only involves simple combinations of soft and
collinear kernels, all remapped in an optimal manner so as to make their analytic
integration as straightforward as possible. The complete integration of the pure
double-unresolved counterterm, along with other details of the implementation,
will be presented in a forthcoming publication. Here we will limit ourselves,
for the sake of illustration, to the computation, in \secn{Proof}, of the subset
of the terms that enter the $T_RC_F$ contribution to $e^+e^-\to q \bar q$ at NNLO.

%%%%%%%%%%%%%%%%

\subsection{Real-virtual counterterm}
\label{sec:RVcntint}

The real-virtual NNLO contribution $\RVl$ features a structure of explicit $\eps$ 
poles dictated by its nature of virtual one-loop matrix element, namely
\beq
\label{eq:RVdefinition}
  \RVl & = & - \frac{\as}{2 \pi} \left( \frac{\mu^2}{s} \right)^{\eps}
  \Bigg[ \Rl \sum_{k} \Bigg( \frac{C_k}{\eps^2} + \frac{\gamma_k}{\eps} \Bigg) + 
  \sum_{k, \, l \neq k} \, {\Rl}_{kl} \, \frac{1}{\eps} \ln \eta_{kl} \, + \, G (\eps)
  \Bigg] \, ,
\eeq
where the indices $k$ and $l$ run over real-radiation multiplicities, and $G(\eps)$ 
denotes the collection of terms that are non-singular in the $\eps \to 0$ limit, 
encoding process-specific information.

The corresponding real-virtual counterterm $\overline K^{\RV}$ contains all 
phase-space singularities appearing in \eq{eq:RVdefinition}. Analogously to 
what done at NLO in \eq{eq:NLOKdef}, it is defined as
\beq
  \overline{K}^{\RV} \, = \, \sum_{i, \, j \neq i} \overline{K}_{ij}^{\RV} \, = \, 
  \sum_{i, \, j \neq i} \Big( \bSi + \bCj - \bSi \, \bCj \Big) \, \RVl \, \mc W_{ij} \, .
\label{eq:NNLOKRVdef}
\eeq
In this paper we do not aim at giving a final expression for the integrated 
real-virtual counterterm $I^{\RV}$, which will instead be detailed in a subsequent 
publication, together with the completion of the integrals contributing to $I^{\two}$; 
we limit ourselves to stressing that such an analytic integration is of a comparable 
or lower complexity with respect to that of the pure double-unresolved counterterms, 
hence it does not pose any new significant computational challenges. Indeed, as 
far as the $\eps$-singular contributions in \eq{eq:RVdefinition} are concerned, they 
are proportional to real or colour-connected real matrix elements, hence their IR limits 
in \eq{eq:NNLOKRVdef} involve single-soft and single-collinear kernels of NLO-level
complexity. The structure of the $\eps$-finite remainder $G(\eps)$ is slightly subtler:
it can be further split into the sum of a process-specific regular contribution, plus 
a universal phase-space-singular term. The IR limits of the latter, in particular, 
involve kernels which represent integrands of a higher complexity than the NLO 
ones, but still can be handled analytically in full generality. We leave the completion 
of these contributions to future work.

%%%%%%%%%%%%%%%%%%%%%%%%%%%%%%%%%

\section{Proof-of-concept calculation}
\label{Proof}

In order to demonstrate the validity of our local subtraction method, in this 
Section we apply it to di-jet production in electron-positron annihilation, as 
a test case. We consider radiative corrections up to NNLO, restricting our 
analysis to the contributions proportional to $T_RC_F$. The production 
channels available in this case are
\beq
  \Bn, \,\, \Vl, \,\, \VV : && \qquad e^+ \, e^- \,\, \to \,\, q \, \bar q \, , \nnb \\
  \Rl, \,\, \RVl : && \qquad e^+ \, e^-  \,\, \to \,\, q \, \bar q \, g \, , \nnb \\
  \RR : && \qquad e^+ \, e^-  \,\, \to \,\, q \, \bar q \, q' \bar q' \, .
\label{eq:eejjchannels}
\eeq

%%%%%%%%%%%%%%%%

\subsection{Matrix elements}
\label{maele}

The relevant $\mc O(\as^2)$ matrix elements are known analytically, and up to 
$\mc O(\eps^0)$ they yield \cite{GehrmannDeRidder:2004tv,Hamberg:1990np,
Ellis:1980wv}
\beq
  \VV & = & \Bn \left( \frac\as{2\pi} \right)^2 T_R \, C_F 
\label{eq:NNLOVV}\\
  & & \times \Bigg\{ \left( \frac{\mu^2}s \right)^{2 \eps} 
  \left[ \frac1{3 \eps^3} + \frac{14}{9 \eps^2} + 
  \frac1\eps \left( - \frac{11}{18} \pi^2 + \frac{353}{54} \right) + 
  \left( - \frac{26}9 \zeta_3 - \frac{77}{27} \pi^2 + \frac{7541}{324} \right) 
  \right] \nnb \\
  & & \hspace{2.8mm} + \, \left( \frac{\mu^2}s \right)^\eps 
  \left[ - \frac4{3 \eps^3} - \frac2{\eps^2} + \frac1\eps 
  \left( \frac79\pi^2 - \frac{16}3 \right) + \left( \frac{28}9 \zeta_3 + \frac76\pi^2 -
  \frac{32}3 \right) \right] \Bigg\} \, , \nnb \\ [1pt]
  \int d \Phi_{\rm{rad}} \, \RVl & = & \frac\as{2\pi} \, \frac{1}{\eps} \, \frac23 \, T_R 
  \int d \Phi_{\rm{rad}} \, \Rl 
\label{eq:NNLORV} \\ 
  & = & \Bn \left( \frac\as{2\pi} \right)^2 T_R \, C_F \nnb \\
  & & \times \left( \frac{\mu^2}s \right)^\eps \left[ \frac4{3\eps^3} + \frac2{\eps^2} + 
  \frac1\eps \left( - \frac79\pi^2 + \frac{19}3 \right) + \left( - \frac{100}9 \zeta_3 - 
  \frac76\pi^2 + \frac{109}6 \right) \right] \, , \nnb \\ [4pt]
  \int d \Phi_{\rm rad, 2} \, \RR & = &
  \Bn \left( \frac\as{2\pi} \right)^2 T_R \, C_F 
\label{eq:NNLORR} \\
  & & \times \left( \frac{\mu^2}s \right)^{2 \eps}
  \left[ - \frac1{3\eps^3} - \frac{14}{9 \eps^2} + \frac1\eps 
  \left( \frac{11}{18} \pi^2 - \frac{407}{54} \right) + \left( \frac{134}{9} \zeta_3 + 
  \frac{77}{27} \pi^2 - \frac{11753}{324} \right) \right] \, , \nnb
\eeq
where, in this case, $d \Phi_{\rm{rad}} = d \Phi_3/d \Phi_2$, $d \Phi_{\rm{rad, 1}} = 
d \Phi_4/d \Phi_3$, and $d \Phi_{\rm{rad, 2}} = d \Phi_4/d \Phi_2$. The $T_R C_F$ 
contribution to the $\mc O(\as^2)$ coefficient of the total cross section is thus
\beq
  \sig_{\NNLO} & = & \sig_{\LO} \left(\frac\as{2\pi} \right)^2 T_R \, C_F
  \left( - \frac{11}2 + 4 \, \zeta_3 - \ln\frac{\mu^2}s \right) \, .
\label{eq:sigtotNNLO}
\eeq
We now proceed to compute and integrate the local counterterms relevant 
for this particular process.

%%%%%%%%%%%%%%%%

\subsection{Local subtraction}
\label{subtrapoc}

The non-zero double-real singular limits for the process we are considering 
are $\bS{34}$, $\bC{134}$, $\bC{234}$ (double-unresolved), and $\bC{34}$ 
(single-unresolved), where labels 1 and 2 refer to $q$ and $\bar q$, while 
labels 3 and 4 refer to $q'$ and $\bar q'$, according to the process definitions 
in \eq{eq:eejjchannels}. The integrated pure double-unresolved counterterm, 
according to \secn{sec:2unrescntint}, is 
\beq
  I^{\two} & = & \int d \Phi_{\rm rad, 2} \, \Big[ \, \bbS{34} + \bbC{134} \,
  \left( 1 - \bbS{34} \right) + \bbC{234} \, \left( 1 - \bbS{34} \right) \Big]
  \RR \, .
\label{eq:I2poc}
\eeq
In the case we are considering, thanks to the simple 
singularity structure of the process, only the parametrisation (\ref{eq: repar2 abcd}),
involving four parton indices, is required. We introduce, therefore, the phase-space
measure 
\beq
  d \Phi_{\npt} & = & d \Phi_n^{\, (abcd)} \, d \Phi_{\rm rad, 2}^{\,(abcd)} \, ,
\label{factophsp2}
\eeq
where $a$ and $b$ are the unresolved partons, while $c$ and $d$ are two 
massless partons, other than $a$ and $b$ (which in the present case of course
exhaust the list of final-state particles). Using \eq{eq: repar2 abcd}, the
double-radiation 
phase space $d \Phi_{\rm rad,2}^{(abcd)}$ depends explicitly on the invariant 
$s_{abcd} = {\bar s}_{cd}^{(abcd)}$ and can be parametrised as
\beq
  \int d \Phi_{\rm rad, 2}^{\,(abcd)} & = & \int d \Phi_{\rm rad, 2} 
  \left( s_{abcd}; \, y, z, \phi, y', z', x' \right) \, \nnb \\ 
  & = & N^2 (\eps) \, {(s_{abcd})}^{2 - 2 \eps} \!\!
  \int_0^1 \!\! d x'  \int_0^1 \!\! d y' \int_0^1 \!\! d z'
  \int_0^\pi \!\! d \phi \, \left(\sin \phi \right)^{- 2 \eps}
  \int_0^1 \!\! d y \int_0^1 \!\! d z \, \nnb \\
  & & \times \, \Big[ 4 \, x' \left( 1 - x' \right) \, y' \left( 1 - y' \right)^2 \, 
  z' \left( 1 - z' \right) \, y^2 (1 - y)^2 \, z \left( 1 - z \right) \Big]^{- \eps} \nnb \\
  & & \times \, \big[ x' \left( 1 - x' \right) \big]^{-1/2} \left( 1 - y' \right) \, 
  y \left( 1 - y \right) \, ,
\label{phspproco}
\eeq
where $y'$ and $z'$ are the Catani-Seymour variables relative to the secondary-radiation
phase space, and $x'$ parametrises the azimuth between subsequent emissions.
In the chosen parametrisation, four out of the six involved binary invariants have 
simple expressions, while the remaining two involve square roots related to 
azimuthal dependence. The explicit expressions are
\beq
  s_{ab} & = & y' \, y \, s_{abcd} \, ,\nnb \\
  s_{ac} & = & z' \left( 1 - y' \right) \, y \, s_{abcd} \, ,\nnb \\
  s_{bc} & = & \left( 1 - y' \right) \left( 1  - z' \right) \, y \, s_{abcd} \, ,\nnb \\
  s_{cd} & = & \left( 1 - y' \right) \left( 1 - y \right) \left( 1 - z \right) \, s_{abcd} \, ,\nnb \\
  s_{ad} & = & \left( 1 - y \right) \left[ y' \left( 1 - z' \right) \left( 1 - z \right) 
  + z' z - 2 \left( 1 - 2 x' \right) \sqrt{y' z' \left( 1 - z' \right ) \, 
  z \left( 1 - z \right)} \, \right] s_{abcd} \, ,\nnb \\
  s_{bd} & = & \left( 1 - y \right) \left[ y' z' \left( 1 - z \right) + 
  \left( 1 - z' \right) z + 2 \left( 1 - 2 x' \right) \sqrt{ y' z' \left( 1 - z' \right) \,
  z \left( 1 - z \right)} \, \right] s_{abcd} \, ,
\label{invaproco}
\eeq
where, for the process at hand, the invariant $s_{abcd} = 
{\bar s}_{cd}^{(abcd)}$ coincides with the squared centre-of-mass energy 
$s$. In this parametrisation, all integrations for the process we are considering
are straightforward. For the case of double-soft radiation the relevant integral
is~\cite{Catani:1999ss}
\beq
  \hspace{-7mm} \int d \Phi_{\rm rad, 2} \, \bbS{ij} \, \RR
  & = & \Norm^{\,2} \,T_R \! \sum_{l, m = 1}^2 {\Bn}_{lm} \! 
  \Big( \{ \bar k \}^{(ijlm)} \Big) \int d \Phi_{\rm rad, 2}^{(ijlm)} \,\,
  \frac{s_{il} s_{jm} + s_{im} s_{jl} - s_{ij} s_{lm}}{s_{ij}^2 \left( s_{il} + s_{jl} \right)
  \left( s_{im} + s_{jm} \right)} \, ,
\label{dossoproco}
\eeq
where $\{ij\} = \{34\}$, according to \eq{eq:I2poc}. Different terms in the eikonal 
sum can be remapped to the same Born kinematics, and, performing the relevant 
colour algebra, the result is
\beq
  \int d \Phi_{\rm rad, 2} \, \bbS{ij} \, \RR & = & \Norm^{\,2} \, B \, T_R \, C_F \,
  \frac8{s^2} \int d \Phi_{\rm rad, 2} \left( s; y, z, \phi, y', z', x' \right) \, 
  \frac{ z' \left( 1 - z' \right)}{y^2 y'^2} 
  \frac{y' \left( 1 - z \right)}{y' \left( 1 - z \right) + z} \nnb \\
  & = & \Bn \left( \frac{\as}{2 \pi} \right)^2 T_R \, C_F
  \left( \frac{\mu^2}{s} \right)^{2 \eps} \, \Bigg[ - \frac{1}{3 \eps^3}
  - \frac{17}{9 \eps^2} + \frac1{\eps} \left( \frac{7}{18} \pi^2 - \frac{232}{27} \right)
\label{dossoprocofin} \\
  & & \hspace{41.5mm} + \left( \frac{38}{9} \zeta_3 + \frac{131}{54} \pi^2 - 
  \frac{2948}{81} \right) + \mc O(\eps) \, \Bigg] \, . \nnb
\eeq
The double-collinear contribution (before the subtraction of the soft-collinear 
region) can be similarly computed, and it yields
\beq
  \int d \Phi_{\rm rad, 2}^{(ijkr)} \, \bbC{ijk} \, \RR & = & \Norm^{\, 2}
  \, \Bn \, T_R \, C_F \int d \Phi_{\rm rad,2}^{(ijkr)} \,\,  \frac{1}{2 s_{ijk} s_{ik}}
\label{docolloproco} \\ 
  && \hspace{5mm} \times \, \bigg[ - \frac{t^{\, 2}_{ik, j}}{s_{ik} s_{ijk}} + 
  \frac{4 z_j + \left( z_i - z_k \right)^2}{z_i + z_k} + \left( 1 - 2 \eps \right)
  \left( z_i + z_k - \frac{s_{ik}}{s_{ijk}} \right) \bigg] \nnb \\
  & = & \Bn \left( \frac{\as}{2 \pi} \right)^2 T_R \, C_F 
  \left( \frac{\mu^2}{s} \right)^{2 \eps} \, \Bigg[ - \frac{1}{3 \eps^3}
  - \frac{31}{18 \eps^2} + \frac{1}{\eps} \left( \frac{1}{2} \pi^2 - 
  \frac{889}{108} \right) \nnb \\
  & & \hspace{41.5mm} + \, \left(\frac{80}{9} \zeta_3 + \frac{31}{12} \pi^2 - 
  \frac{23941}{648} \right) + \mc O(\eps) \, \Bigg] \, , \nnb
\eeq
where, following \cite{Catani:1998nv,Catani:1999ss}, we have set
\beq
  t_{ik,j} & = & 2 \, \frac{z_i s_{kj} - z_ks_{ij}}{z_i + z_k}
  + \frac{z_i - z_k}{z_i + z_k} \, s_{ik} \, .
\label{tcatgra}
\eeq
Note that the result in \eq{docolloproco} applies to the configurations 
$\{ijk\} = \{134\}$ and $\{ijk\} = \{234\}$, as seen from \eq{eq:I2poc}. 
The subtraction of the double-counted soft-collinear limit is very simple 
in this case, since one has
\beq
  \int d \Phi_{\rm rad, 2} \, \bbS{ij} \, \bbC{ijk} \, \RR & = & 
  \int d \Phi_{\rm rad, 2} \, \bbS{ij} \, \RR \, ,
\label{socoeso}
\eeq
as can be deduced from \eq{eq: SS mapped} and \eq{eq: SSCC mapped} 
in the case of two soft quarks, in a process featuring only two partons 
at Born level, identified here with $k$ and $r$. Adding up all contributions
to the pure double-unresolved integrated counterterm, we get
\beq
  I^{\two} & = & \Bn \left( \frac{\as}{2\pi} \right)^2 \, T_R \, C_F
  \left( \frac{\mu^2}{s} \right)^{2 \eps} \nnb \\
  & & \,\,\, \times \left[ - \frac{1}{3 \eps^3} - \frac{14}{9 \eps^2}
  + \frac{1}{\eps} \left( \frac{11}{18} \pi^2 - \frac{425}{54} \right)
  + \left( \frac{122}{9} \zeta_3 + \frac{74}{27} \pi^2 - \frac{12149}{324}
  \right) \right] +\mc O(\eps)\, .
\label{totdocoint}
\eeq
Next, we consider the integration of the single-unresolved counterterm, applying 
the general formula, \eq{eq:intcnt1NNLO}, and restricting our analysis to the 
case in which only the single-collinear limit is non-zero. We find
\beq
  I^{\one}_{hq} & = & - \, \frac{\as}{2 \pi} \left( \frac{\mu^2}{s} \right)^{\eps}
  \frac{2}3 \, T_R \left( \frac{1}{\eps} - \ln \bar\eta_{\, [34] r} + \frac{8}{3} \right) \,
  \Rl \, \bW{hq} \, + \, \mc O(\eps) \, ,
\label{ionehq}
\eeq
where the real-radiation matrix element $\Rl$ involves $\npo = 3$ particles,
the indices $h$ and $q$ take values in the set $\{1, 2, 3 \equiv [34] \}$, 
and we can choose $r=1$ or $r=2$ when $h=1$, $q=2$, while $r=3-h$ in the other
cases. The result in \eq{ionehq} must be combined with the 
$\RVl$ contribution, and we can explicitly check that their sum is finite in 
$d = 4$, sector by sector in the NLO phase space. Indeed
\beq
  \RVl\, \bW{hq} \, + \, I^{\one}_{hq} & = & \frac{\as}{2 \pi} \, \frac{2}3 \, T_R \, 
  \frac{1}{\eps} \, \Rl \, \bW{hq} \, - \, \frac{\as}{2 \pi} 
  \left( \frac{\mu^2}{s} \right)^{\eps} \frac{2}3 \,T_R \left( \frac{1}{\eps}
  - \ln \bar\eta_{\, [34] r} + \frac{8}{3} \right) \, \Rl \, \bW{hq} \, + \,
  \mc O(\eps) \, \nnb \\
  & = & - \, \frac{\as}{2 \pi} \, \frac{2}3 \, T_R \left( \ln \frac{\mu^2}{s_{34r}}
  + \frac{8}{3} \right) \, \Rl \, \bW{hq} \, + \, \mc O(\eps) \, .
\label{finonecomb}
\eeq
The next ingredient is the mixed double-unresolved contribution, which can 
be read off the general formula, \eq{eq:intcnt12NNLOfin}. In sector $hq$ it 
reads
\beq
  I^{\otwo}_{hq} & = & \frac{\as}{2 \pi} \left( \frac{\mu^2}{s} \right)^{\eps}
  \frac{2}3 \, T_R \left( \frac{1}{\eps} - \ln \bar\eta_{\, [34] r} + \frac{8}{3} \right)
  \Big[ \bar{\bf S}_h + \bbC{hq} \left( 1 - \bar{\bf S}_h \right) \Big] \Rl \,
  \bW{hq} \, + \, \mc O(\eps) \, .
\label{i12toint}
\eeq
The combination of \eq{i12toint} with the real-virtual local counterterm in the
same NLO sector must be finite in $d=4$. Indeed we find that 
\beq
  \overline{K}^{\RV}_{hq} \, - \, I^{\otwo}_{hq} & = & \frac{2}3 \, T_R \, \frac{1}{\eps} \,
  \Big[ \bar{\bf S}_h \, + \, \bbC{hq} \left( 1 - \bar{\bf S}_h \right) \Big]
  \Rl \, \bW{hq} \nnb \\
  & & - \, \frac{\as}{2 \pi} \left( \frac{\mu^2}{s} \right)^\eps \frac{2}3 \, T_R
  \left( \frac{1}{\eps} - \ln \bar\eta_{\, [34] r} + \frac{8}{3} \right)
  \Big[ \bar{\bf S}_h \, + \, \bbC{hq} \left( 1 - \bar{\bf S}_h \right) \Big] \Rl \,
  \bW{hq} \, + \, \mc O(\eps) \, \nnb \\
  & = & - \, \frac{\as}{2 \pi} \, \frac{2}3 \, T_R \left( 
  \ln \frac{\mu^2}{s_{34r}} + \frac{8}{3} \right) \Big[ \bar{\bf S}_h \, + \, 
  \bbC{hq} \left( 1 - \bar{\bf S}_h \right) \Big] \Rl \, \bW{hq} \, + \, 
  \mc O(\eps) \, .
\label{finrv12}
\eeq
The final ingredient for subtraction is the integral of the real-virtual counterterm.
In the present case, it is given by
\beq
  I^{\RV} & = & \frac\as{2 \pi} \, \frac23 \, \frac1\eps \, T_R \int d \Phi_{\rm rad} \,
  \Big[ \bbS{[34]} + \bbC{1[34]} \left( 1 - \bbS{[34]} \right) + 
  \bbC{2[34]} \left( 1 - \bbS{[34]} \right) \Big] \, \Rl \nnb \\ [2pt]
  & = & \frac\as{2 \pi} \, \frac23 \, \frac1\eps \, T_R \, \times \, I \big|_{C_F, \, n = 2} 
  \label{finintrv} \\ [3pt]
  & = & \Bn \left( \frac\as{2 \pi} \right)^2 T_R \, C_F \! \left( \frac{\mu^2}{s} \right)^\eps 
  \left[ \frac4{3 \eps^3} + \frac2{\eps^2} - \frac1\eps \left( \frac79\pi^2 - \frac{20}3 \right)
  - \left(\frac{100}9 \zeta_3 + \frac{7}{6} \pi^2 - 20 \right) \right] + \mc O(\eps) \, , \nnb
\eeq
where $I\big|_{C_F, \, n = 2}$ denotes the NLO counterterm given in \eq{eq:intcntNLO}, 
considered in the particular case of two non-gluon final-state partons at Born level. 
All required ingredients for NNLO subtraction for the process at hand are now 
assembled, and we can proceed to a numerical consistency check.

%%%%%%%%%%%%%%%%

\subsection{Collection of results}
\label{colleresu}

The heart of the subtraction procedure is the combination of analytic results 
with numerical integration of the finite remainder of the real-radiation squared 
matrix element, to get physical distributions and cross sections. For this proof 
of concept, we will simply reconstruct numerically the total cross section for 
the production of two quark pairs of different flavours. We emphasise however 
that the formalism we constructed is completely general and local: a detailed 
numerical implementation for all processes involving only final state massless 
partons is being developed and will be presented in forthcoming work.

\begin{figure}[htp!]
  \centering
  \includegraphics[width=0.95\columnwidth]{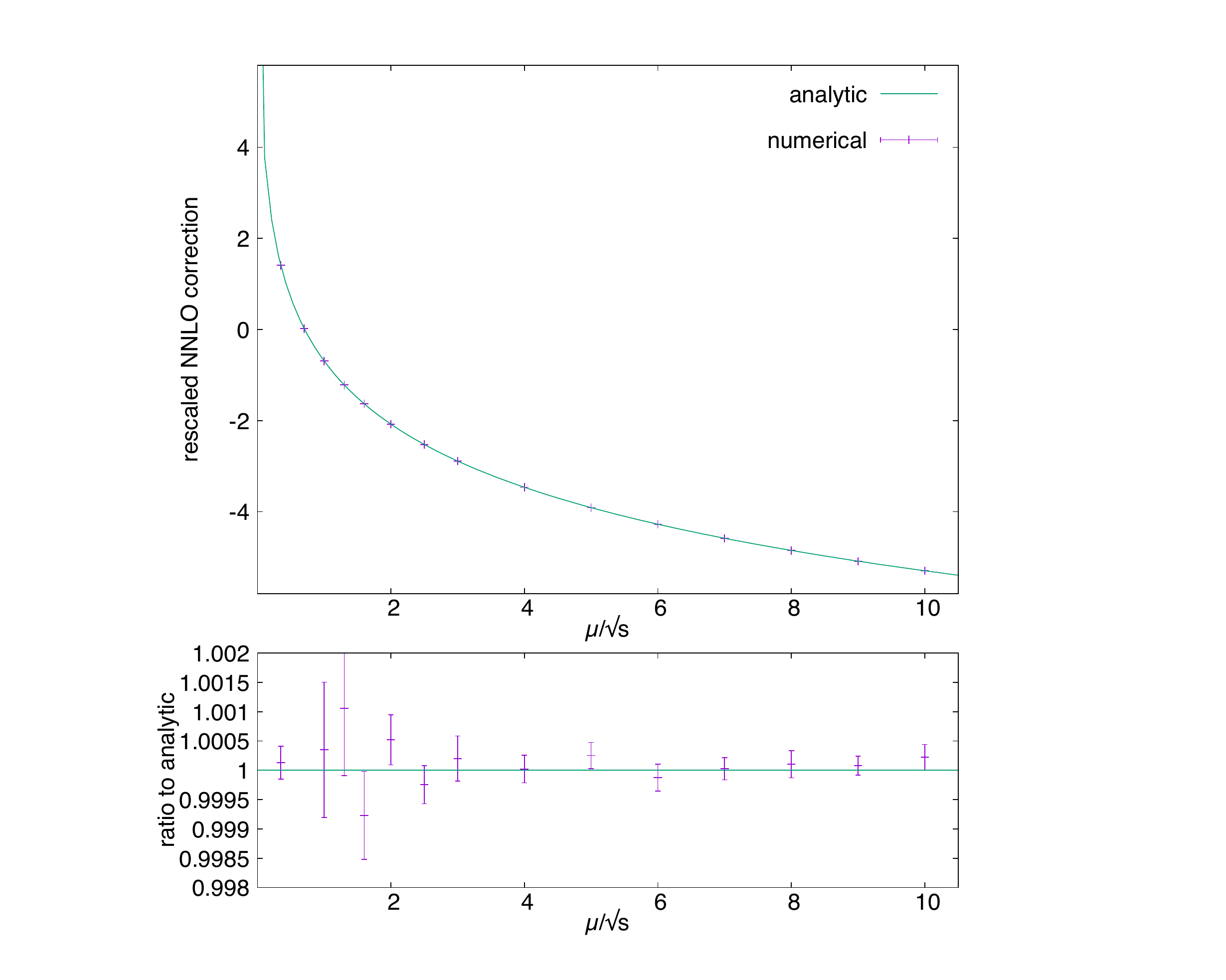} 
  \caption{Rescaled NNLO correction as a function of the renormalisation scale.}
  \label{fig:sigma}
\end{figure}

The cross section is constructed in general, as shown in \eq{eq:NNLO_structure},
as a sum of three finite and integrable contributions, given by
\beq
  \VV^{\sub} & = & \VV + I^{\two} + I^{\RV} \, , \nnb \\
  \RVl^{\sub} & = & \left( \RVl + I^{\one} \right) - 
  \left( \overline{K}^{\RV} - I^{\otwo} \right) \, , 
\label{threefin} \\
  \RR^{\sub} & = & \RR - \overline{K}^{\one} - \overline{K}^{\two} -
   \overline{K}^{\otwo} \, . \nnb 
\eeq
The subtracted double-virtual contribution is computed analytically, and is finite 
in $d=4$. In this case, it is given by
\beq
  \VV^{\, \sub} & = & B \left( \frac{\as}{2 \pi} \right)^2 \, T_R \, C_F 
  \left( \frac83 \zeta_3 - \frac19 \pi^2 - \frac{44}9 - \frac43 \ln \frac{\mu^2}s \right) \\
  & = & B \left( \frac{\as}{2\pi} \right)^2 \,T_R \, C_F \times 0.01949914 \, . \nnb
\label{numvirt}
\eeq
where, for definiteness, in the second line we have randomly chosen 
$\mu^2/s = 0.35$. For real radiation, we have written a Monte Carlo code 
to integrate numerically the remaining two terms in \eq{threefin}, obtaining 
\beq
  \int d \ppon \, \RVl^{\, \sub} & = & \Bn \left( \frac{\as}{2 \pi} \right)^2 T_R \, C_F
  \times \big( - \, 0.90635 \, \pm \, 0.00011 \big) \, , \nnb \\
  \int d \ppon \, \RR^{\, \sub} & = & \Bn \left( \frac{\as}{2 \pi} \right)^2 T_R \, C_F
  \times \big( + \, 2.29491 \, \pm \,  0.00038 \big) \, .
\label{numreal}
\eeq
The rescaled NNLO correction, evaluated numerically by means of the subtraction 
method, is then
\beq
  K^{\rm num.}_{\NNLO} \, \equiv \, 
  \frac{\sig_{\NNLO}}{\left( \frac{\as}{2 \pi} \right)^2 T_R \, C_F \, \sig_{\LO}} \, = \,
  1.40806 \, \pm \, 0.00040 \, ,
\label{numresc}
\eeq
to be compared with the analytical result
\beq
  K^{\rm an.}_{\NNLO} \, = \, \left( - \frac{11}2 + 4 \zeta_3 - \ln \frac{\mu^2}s \right)
  \, = \, 1.40787186 \, .
\label{anaresc}
\eeq
For completeness, we also show in Fig.~\ref{fig:sigma} that also the logarithmic 
renormalisation-scale dependence is correctly reproduced with the same accuracy.

%%%%%%%%%%%%%%%%

%%%%%%%%%%%%%%%%%%%%%%%%%%%%%%%%%

\section{Conclusions}
\label{sec:conclusions}

In this work we have presented a new scheme to perform local analytic 
subtraction of infrared divergences up to NNLO in QCD. The method has 
for now been developed and applied to processes featuring only massless 
partons, and not involving coloured partons in the initial state, as a first 
significant step towards a general formulation. Our subtraction procedure 
is conceived with the aim of minimising complexity in the definition of the 
local IR counterterms, aiming for their complete analytic integration in 
the unresolved phase space, and working towards an optimal organisation 
of the numerical integration of the observable cross section.

Our local IR counterterms are defined through a unitary partition of the 
phase space into sectors, in such a way as to isolate in each sector a 
minimal number of phase-space singularities, associated with soft and 
collinear configurations of an identified set of partons (up to two at NLO, 
and up to four at NNLO). In each sector, the counterterms are built out 
of a collection of universal kernels, written in terms of kinematic invariants, 
which can be defined in terms of gauge-invariant operator matrix elements,
as detailed in~\cite{Magnea:2018ebr}, or can be obtained as limits of radiative 
matrix elements in the dominant soft and collinear configurations. Overlapping 
singularities are fully taken into account by suitable compositions of such 
singular limits, with no need to resort to sector-decomposition techniques.

The sector functions that realise the phase-space partition are engineered 
in such a way as to satisfy fundamental relations that allow to achieve the 
main goals of the method. A number of sum rules, stemming from the definition 
of the sector functions, allow one to recombine various subsets of sectors, 
prior to performing counterterm integration, eventually yielding integrands 
that in all cases are solely made up by sums of elementary infrared and 
collinear kernels. Moreover, through factorisation relations, NNLO sector 
functions reproduce the complete structure of NLO sectors in all relevant 
single-unresolved limits, allowing to subtract, sector by sector in the NLO 
phase space, the singularities of the NNLO contributions featuring NLO 
kinematics.

The kinematic mappings necessary for phase-space factorisation, as well 
as the parametrisations of the radiation phase space over which the 
counterterms are integrated, are devised by maximally exploiting the freedom 
one has in their definition. They are not only chosen differently for different 
sectors, but also, importantly, for different counterterm contributions in the 
same sector. This allows us to employ parametrisations that are naturally 
adapted to the kinematic invariants that appear in each singular contribution, 
yielding simple integrands to be evaluated analytically.

In this article we have integrated all needed counterterms over the exact
phase-space measures, without exploring the possibility of approximating 
the latter in the relevant soft and collinear limits. While this possibility would 
not have resulted in any analytic simplification in the cases considered here, 
this might instead be the case for general hadronic reactions (for example when
including initial-state partons, or for a generalisation to the massive case). 
This possibility will be investigated in dedicated future studies, which are 
beyond the scope of the present paper. 

At NLO, we have shown that the proposed subtraction method works in the 
general case of massless QCD final states, with the integrated counterterms 
reproducing analytically the full structure of virtual one-loop singularities. 
Moreover, as a test of the power of the method, we have shown that the 
NLO counterterm integration can be performed exactly to all orders in the 
dimensional regulator $\eps$, which bears witness to the extreme simplicity 
of the integrands involved.

At NNLO, we have deduced the structure of the subtraction scheme in full
generality for massless QCD final states. All single-unresolved and mixed
double-unresolved counterterms of double-real origin have been integrated
analytically to all orders in $\eps$,  as simply as in the NLO case, and
the properties of sector functions have  allowed us to show that these 
integrals correctly reproduce, sector by sector,  the explicit $\eps$ 
poles and phase-space singularities of real-virtual contributions. 
We stress that this is a highly non-trivial test of the consistency of the scheme, 
and of the delicate organisation of different contributions to the cross section. 
As for double-unresolved counterterms, we have deduced their structure in 
general, and performed the relevant integrations in a proof-of-concept case, 
the $T_R C_F$ contribution to $e^+e^-\to q \bar q$ at NNLO, which has been 
detailed explicitly. 

While in this paper we have concentrated on the general structure of our method, 
in particular concerning sector functions and phase-space mappings, and we have 
given only a simple example of implementation, we emphasize that we do not 
expect significant further technical difficulties for the extension of our algorithm
to a general massless final states at NNLO: indeed, an important advantage of
our method is that the required local counterterms are essentially combinations 
of the (re-mapped) NNLO splitting kernels. The corresponding integrals are 
therefore closely related to integrals known in the literature (see, for 
example~\cite{Gehrmann-DeRidder:Thesis,Gehrmann-DeRidder:2003pne}), 
and they are not expected to pose an obstacle for a general application of the 
method. The inclusion of initial state radiation is expected to require more 
work, in order to design and test appropriate sector functions and dedicated
phase-space mappings, as well as implementing collinear factorization, but
no new conceptual problems are expected to arise.

To summarize, this article represents a first step towards the formulation of a 
general, local, analytic, and minimal subtraction scheme, relevant for generic 
multi-particle hadronic processes at NNLO in QCD. To reach this goal, a 
number of important steps still need to be taken, including the analytic integration 
of the remaining double-unresolved counterterms for final-state processes, 
the generalisation to include initial-state massless partons, and the extension 
to the massive case, as well as the completion of an efficient computer code 
implementing the subtraction method in a fully differential framework. We believe 
however that the present work lays a solid foundation for these future developments.

%%%%%%%%%%%%%%%%%%%%%%%%%%%%%%%%%

\section*{Acknowledgements}

\noindent We would like to thank S.~Frixione, F.~Herzog, V.~Hirschi and
N. Deutschmann for very useful discussions.  The work of PT was funded 
by the European Union Seventh Framework programme for research and 
innovation under the Marie Curie grant agreement N.~609402-2020 researchers: 
Train to Move (T2M).

%%%%%%%%%%%%%%%%%%%%%%%%%%%%%%%%%

\appendix

%%%%%%%%%%%%%%%%%%%%%%%%%%%%%%%%%

\section{Commutation of soft and collinear limits at NLO}
\label{app:commutationNLO}

In this Appendix, as an example, we explicitly show the commutation of the soft 
and collinear limits $\Si$  and $\Cj$, and, in the process, deduce the form of the 
soft-collinear kernel $\Si\,P_{ij}$ appearing in \eq{eq:SiCijR1}. The action of 
operators $\Si$ and $\Cj$ on ratios of elementary massless invariants $s_{ij}$ 
is given by
\beq
  \Si \, \frac{s_{ia}}{s_{ib}} & \ne & 0 \, , \qquad \quad \,\,\, 
  \Si \, \frac{s_{ia}}{s_{bc}} \, = \, 0 \, ,  
  \qquad \forall \, a, b, c \ne i \, , 
\label{eq:soft_prop} \\
  \Cj \, \frac{s_{ij}}{s_{ab}} & = & 0 \, , \qquad \quad \Cj \, \frac{s_{ia}}{s_{ja}} \, = \,
  \mbox{independent of $a$} \, , \qquad \forall \, a b \notin \pi(ij) \, .
\label{eq:coll_prop}
\eeq
We start by verifying that the sequential action of the singular projectors on sector 
functions does not depend on their ordering. To this end note that
\beq
  \Si \, \mc W_{ij} \, = \, \frac{1/w_{ij}}{\sum \limits_{l \neq i}1/w_{il}} \qquad 
  & \Longrightarrow & \qquad \Cj \, \Si \,\mc W_{ij} \, = \, 1 \, ,
\label{eq:commutSfun1} \\
  \Cj \, \mc W_{ij} \, = \, \frac{e_j}{e_i + e_j}  \qquad & \Longrightarrow & \qquad 
  \Si \, \Cj \,\mc W_{ij} \, = \, 1 \, ,
\label{eq:commutSfun2}
\eeq
where in \eq{eq:commutSfun1} we used the fact that only $l = j$ gives rise to a 
singular contribution $1/w_{il}$ in the collinear limit, while in \eq{eq:commutSfun2} 
we have noted that $e_i \to 0$ in the soft limit.

Next, we consider the action of the composite projector $\Si \, \Cj$ on the 
physical real-radiation amplitude squared, where, without loss of generality, 
we drop all kinematic dependences in the real and Born-like matrix elements. 
Starting from \eq{eq:CijR1} we find
\beq
  \Si \, \Cj \, \Rl \, = \, \frac{\Norm}{s_{ij}} \, \Big[ \Si \, P_{ij} \, \Bn + \Si \, Q_{ij}^{\mu\nu}
  {\Bn}_{\mu\nu} \Big] \, .
\label{SC1}
\eeq
We now note that $Q_{ij}^{\mu\nu}$, defined in \eq{eq:APkernels2}, is not singular in 
the soft limit for parton $i$, hence $\Si \, Q_{ij}^{\mu\nu} = 0$. The same happens for 
all terms in $P_{ij}$ which do not contain a denominator $1/x_i$. We now rewrite the 
remaining contributions in terms of Mandelstam invariants, using the definition of $x_i$ 
and $x_j$ in \eq{translong}, with the result
\beq
  P_{ij} & = & \delta_{f_ig} \, \delta_{f_jg} \, 2 C_A \, \frac{x_j}{x_i} +
  \delta_{f_ig} \, \delta_{f_j \{q, \bar q\}} \, C_F \, \frac{1 + x_j^2}{x_i} + \, \ldots \, , \nnb \\
  & = & \delta_{f_ig} \, \delta_{f_jg} \, 2 C_A \, \frac{s_{jr}}{s_{ir}} + \delta_{f_ig} \, 
  \delta_{f_j \{q, \bar q\}} \, C_F \, \frac{1 + \big[ s_{jr}/\left( s_{ir} + s_{jr} \right) 
  \big]^2}{s_{ir}/\left( s_{ir} + s_{jr} \right)} + \, \ldots \, ,
\label{takecoll}
\eeq
where the ellipses denote terms that remain regular as parton $i$ becomes soft. 
Taking now the $\Si$ limit according to \eq{eq:soft_prop}, we get
\beq
  \Si \, P_{ij} & = & \delta_{f_ig} \, \delta_{f_jg} \, 2 C_A \, \frac{s_{jr}}{s_{ir}} + 
  \delta_{f_ig} \, \delta_{f_j \{q, \bar q\}} \, C_F \, \frac{2 s_{jr}}{s_{ir}} \nnb \\
  & = & \delta_{f_ig} \, \delta_{f_jg} \, 2 C_A \, \frac{x_j}{x_i} + 
  \delta_{f_i g} \, \delta_{f_j \{q, \bar q\}} \, C_F \, \frac{2 x_j}{x_i} \, ,
\label{takeS}
\eeq
which is \eq{softcollker}. In particular, we note that the soft limit $\Si$ does 
\emph{not} correspond to taking $x_i \to 0$, rather to taking $s_{ir}\to 0$ (the 
two definitions differ by subleading soft terms).  The soft-collinear limit is thus
\beq
  \Si \, \Cj \, \Rl \, = \, \Bn \,\, \frac{\Norm}{s_{ij}} \left(
  \delta_{f_ig} \, \delta_{f_jg} \, 2 C_A \, \frac{s_{jr}}{s_{ir}} + 
  \delta_{f_ig} \, \delta_{f_j \{q, \bar q\}} \, C_F \, \frac{2 s_{jr}}{s_{ir}} \right) \, .
\label{eq:SCR}
\eeq
We can now verify commutation by considering the two singular limits in reversed 
order. We find
\beq
  \Cj \, \Si \, R \, = \,  - \, \Norm \, \Cj \sum_{k \neq i \, , l \neq i} \!
  \mc I_{kl}^{(i)} \, {\Bn}_{kl} \, .
\label{softfirst}
\eeq
Among all the terms in the double sum, only those with $k = j$ or $l = j$ are singular 
in the collinear limit, hence
\beq
  \Cj  \, \Si \, \Rl & = & - \, \Norm \, \frac 2{s_{ij}} \, \Cj \, \sum_{l \neq i}
  \frac{s_{jl}}{s_{il}} \, \Bn_{jl} \, .
\eeq
According to property (\ref{eq:coll_prop}), in the collinear limit $\Cj$ the ratio 
$s_{jl}/s_{il}$ is independent of $l$: we can therefore set $l = r$ and get
\beq
  \Cj \, \Si \, \Rl & = & - \, \Norm \, \delta_{f_ig} \, \frac 2{s_{ij}} \, \frac{s_{jr}}{s_{ir}}
  \sum_{l \neq i}{\Bn}_{jl} \, = \, \Norm \, \frac 2{s_{ij}} \frac{s_{jr}}{s_{ir}} \, 
  C_{f_j} \Bn \nnb \\
  & = & \Bn \,\, \frac{\Norm}{s_{ij}} \left( \delta_{f_ig} \, \delta_{f_jg} \, 2 C_A \, 
  \frac{s_{jr}}{s_{ir}} + \delta_{f_ig} \, \delta_{f_j \{q, \bar q\}} \, C_F \, 
  \frac{2 s_{jr}}{s_{ir}} \right) \, ,
\label{eq:CSR}
\eeq
where in the last two steps we have used colour algebra, and the definition of the 
Casimir operator $C_{f_j} = C_A \delta_{f_j g} + C_F \delta_{f_j \{q \bar q\}}$. The 
equality of \eq{eq:CSR} and \eq{eq:SCR}, together with relations (\ref{eq:commutSfun1}) 
and (\ref{eq:commutSfun2}), shows the desired commutation of limits in each 
sector $ij$.

%%%%%%%%%%%%%%%%%%%%%%%%%%%%%%%%%

\section{Soft and collinear limits of sector functions}
\label{app:SfunNNLOprop}

In this Appendix we explore the properties of the NNLO sector functions defined in 
Eqs.~(\ref{eq:sec_fun_NNLO}) and (\ref{eq:sigmadef}), including their relation to
the NLO-like functions defined in \eq{eq:Wab_def}. We begin by establishing
which limits, among $\bS{a}$, $\bC{ab}$, $\bS{ab}$, $\bC{abc}$, $\bC{abcd}$, 
$\bSC{abc}$, and $\bCS{abc}$, are non-vanishing in the three sector topologies 
$\W{ijjk}$, $\W{ijkj}$ and $\W{ijkl}$. To this end, we start by analysing the 
behaviour of the sector-function denominator $\sigma$ (see \eq{eq:sec_fun_NNLO}), 
in these limits. We find
\beq
  {\bf S}_i \, \sigma & = & \sum_{b \neq i} \sum_{c \neq i} \sum_{d \neq i,c} 
  {\bf S}_i \, \sigma_{ibcd} \, = \,  
  \sum_{b \neq i} \sigma_{ib}^{(\alpha\beta)} 
  \sum_{c \neq i} \sum_{d \neq i,c} \sigma_{cd} \, , \nnb \\
  \bC{ij} \, \sigma & = & \sum_{c \neq i} \sum_{d \neq i,c} \sigma_{ijcd} +
  \sum_{c \neq j} \sum_{d \neq j,c} \sigma_{jicd} \nnb \\
  & = & \left[ \sigma_{ij}^{(\alpha\beta)} + \sigma_{ji}^{(\alpha\beta)} \right]
  \bigg[ \sum_{c\neq i,j} \sigma_{c[ij]} + \sum_{d\neq i,j} \sigma_{[ij]d} +
  \sum_{c\neq i,j}\sum_{d\neq i,j,c} \sigma_{cd} \bigg] \, , \nnb \\
  \bS{ij}\,\sigma & = & \sum_{b \neq i}\sum_{d \neq i,j} \sigma_{ibjd}
  + \sum_{b\neq j}\sum_{d\neq j,i} \sigma_{jbid} \, , \nnb \\
  \bC{ijk}\,\sigma & = & \sigma_{ijjk} + \sigma_{ijkj} + \sigma_{ikkj} +
  \sigma_{ikjk} + \sigma_{jiik} + \sigma_{jiki} \nnb \\
  && + \, \sigma_{jkki} + \sigma_{jkik} + \sigma_{kiij} + \sigma_{kiji} +
  \sigma_{kjji} + \sigma_{kjij} \, , \nnb \\
  \bC{ijkl}\,\sigma & = & \sigma_{ijkl} + \sigma_{ijlk} + \sigma_{jikl} +
  \sigma_{jilk} + \sigma_{klij} + \sigma_{klji} + \sigma_{lkij} + \sigma_{lkji} \, , \nnb \\
  \bSC{ijk} \, \sigma & = & \sum_{b \neq i} \, \bS{i} 
  \left( \sigma_{ibjk} + \sigma_{ibkj} \right) \, = \, 
  \sum_{b \neq i} \sigma_{ib}^{(\alpha\beta)} 
  \left( \sigma_{jk} + \sigma_{kj} \right) \, , \nnb \\
  \bCS{ijk}\,\sigma & = & \sum_{d \neq i,k} \sigma_{ijkd} + 
  \sum_{d \neq j,k} \sigma_{jikd} \, = \,  
  \sigma_{ij}^{(\alpha\beta)}\sum_{d\neq i,k} \sigma_{kd} \, + \,
  \sigma_{ji}^{(\alpha\beta)}\sum_{d\neq j,k} \sigma_{kd} \, ,
\label{eq: limitsonsigma}
\eeq
where $[ij]$ denotes the parent parton of $i$ and $j$ and we have used the
definition of the NLO-like sector functions in \eq{eq:Wab_def}. Now we note that 
a singular limit $\bf L$ gives a non-zero result, when applied to the sector functions 
$\W{abcd}$, only if the numerator of the latter, $\sigma_{abcd}$, appears as 
one of the addends of $\bf{L} \, \sigma$. Inspection of \eq{eq: limitsonsigma} 
then proves that the limits reported in \eq{eq:NNLOprop2} exhaust the surviving 
ones in each sector.

Next, we show that all of the limits in \eq{eq:NNLOprop2} commute when acting 
on $\sigma$. This is a crucial step for our method, since commutation of limits
drastically reduces the number of independent configurations one needs to 
explore. Furthermore, one must note that, while commutation can be understood 
from physical considerations when limits are taken on squared matrix elements,
sector functions are a crucial but artificial ingredient of our method, and 
commutation of limits is non-trivial in this case. We list below all relevant
ordered limits, acting on the denominator function $\sigma$, beginning with 
those involving the single-soft limit $\bS{i}$.
\beq
  \bS{i} \, \bC{ij} \, \sigma \, = \, \bC{ij} \, \bS{i} \, \sigma & = &  
  \sum_{c \neq i} \sum_{d \neq i,c} {\bf S}_i \, \sigma_{ijcd} \, = \,
  \sigma_{ij}^{(\alpha\beta)} \sum_{c \neq i} \sum_{d \neq i,c} \sigma_{cd} \, ,
  \nnb \\
  \bS{i}\,\bS{ij}\,\sigma \, = \, \bS{ij} \, \bS{i} \, \sigma & = &
  \sum_{b \neq i} \sum_{d \neq i,j} \bS{i} \, \sigma_{ibjd} \, = \,
  \sum_{b \neq i} \sigma_{ib}^{(\alpha\beta)} \sum_{d \neq i,j} \sigma_{jd} \, ,
  \nnb \\
  \bS{i}\,\bC{ijk}\,\sigma \, = \, \bC{ijk} \, \bS{i} \, \sigma & = & 
  \bS{i} \left( \sigma_{ijjk} + \sigma_{ijkj} + \sigma_{ikkj} + \sigma_{ikjk} \right)
  \, = \, \left[ \sigma_{ij}^{(\alpha\beta)} + \sigma_{ik}^{(\alpha\beta)} \right]
  \big( \sigma_{jk} + \sigma_{kj} \big) \, ,
  \nnb \\
  \bS{i} \, \bC{ijkl} \, \sigma \, = \, \bC{ijkl} \, \bS{i} \, \sigma & = & 
  \sigma_{ijkl} + \sigma_{ijlk} \, = \, \sigma_{ij}^{(\alpha\beta)}
  \left( \sigma_{kl} + \sigma_{lk} \right) \, ,
  \nnb \\
  \bS{i} \, \bSC{ijk} \, \sigma \, = \, \bSC{ijk} \, \bS{i} \, \sigma & = &
  \bSC{ijk} \, \sigma \, = \, \sum_{b \neq i} \bS{i} 
  \left( \sigma_{ibjk} + \sigma_{ibkj} \right) \, = \,
  \sum_{b \neq i} \sigma_{ib}^{(\alpha\beta)} 
  \left( \sigma_{jk} + \sigma_{kj} \right) \, ,
  \nnb \\
  \bS{i} \, \bCS{ijk} \, \sigma \, = \, \bCS{ijk} \, \bS{i} \, \sigma & = &
  \sum_{d \neq i,k} \sigma_{ijkd} \, = \, \sigma_{ij}^{(\alpha\beta)}
  \sum_{d\neq i,k} \sigma_{kd} \, .
\label{eq: commuting relations 1}
\eeq
Next, we list ordered limits involving the single-collinear limit $\bC{ij}$, and not 
considered above.
\beq
  \label{eq: commuting relations 2} 
  \bC{ij} \, \bS{ij} \, \sigma \, = \, \bS{ij} \, \bC{ij} \, \sigma 
  & = & 
  \sum_{d \neq i,j} \left( \sigma_{ijjd} + \sigma_{jiid} \right)
  \, = \,
  \left[ \sigma_{ij}^{(\alpha\beta)} + \sigma_{ji}^{(\alpha\beta)} \right]
  \sum_{d \neq i,j} \sigma_{[ij]d} \, ,
  \nnb \\
  \bC{ij} \, \bS{ik} \, \sigma \, = \, \bS{ik} \, \bC{ij} \, \sigma 
  & = & 
  \sum_{d \neq i,k} \sigma_{ijkd} \, = \, \sigma_{ij}^{(\alpha\beta)} 
  \bigg[ \sigma_{k[ij]} + \sum_{d\neq i,j,k} \sigma_{kd} \bigg] \, ,
  \nnb \\
  \bC{ij} \, \bC{ijk} \, \sigma \, = \, \bC{ijk} \, \bC{ij} \, \sigma
  & = & 
  \sigma_{ijjk} + \sigma_{ijkj} + \sigma_{jiik} + \sigma_{jiki}
  \, = \,
  \left[ \sigma_{ij}^{(\alpha\beta)} + \sigma_{ji}^{(\alpha\beta)} \right]
  \left( \sigma_{[ij]k} + \sigma_{k[ij]} \right) \, ,
  \nnb \\
  \bC{ij} \, \bC{ijkl} \, \sigma
  \, = \,
  \bC{ijkl} \, \bC{ij} \, \sigma
  & = &
  \sigma_{ijkl} + \sigma_{ijlk} + \sigma_{jikl} + \sigma_{jilk}
  \, = \,
  \left[ \sigma_{ij}^{(\alpha\beta)} + \sigma_{ji}^{(\alpha\beta)} \right]
  \big( \sigma_{kl} + \sigma_{lk} \big) \, ,
  \nnb \\
  \bC{ij} \, \bSC{ijk} \, \sigma
  \, = \,
  \bSC{ijk} \, \bC{ij} \, \sigma
  & = &
  \bS{i} \left( \sigma_{ijjk} + \sigma_{ijkj} \right)
  \, = \,
  \sigma_{ij}^{(\alpha\beta)} \big( \sigma_{jk} + \sigma_{kj} \big) \, ,
  \\
  \bC{ij} \, \bSC{ikl} \, \sigma \, = \, \bSC{ikl} \, \bC{ij} \, \sigma
  & = &
  \sigma_{ijkl} + \sigma_{ijlk} \, = \, \sigma_{ij}^{(\alpha\beta)} 
  \big( \sigma_{kl} + \sigma_{lk} \big) \, ,
  \nnb \\
  \bC{ij} \, \bCS{ijk} \, \sigma \, = \, \bCS{ijk} \, \bC{ij} \, \sigma
  & = &
  \bCS{ijk} \, \sigma
  \, = \,
  \! \sum_{d \neq i,k} \! \sigma_{ijkd}
  +
  \! \sum_{d \neq j,k} \! \sigma_{jikd}
  \, = \,
  \! \sigma_{ij}^{(\alpha\beta)} \!\! \sum_{d \neq i,k} \! \sigma_{kd}
  + 
  \sigma_{ji}^{(\alpha\beta)} \!\! \sum_{d \neq j,k} \! \sigma_{kd} \, .
  \nnb
\eeq
Moving on to ordered limits involving the double-soft limit $\bS{ab}$, and not 
considered above, we find
\beq
  \bS{ij} \, \bC{ijk} \, \sigma \, = \, \bC{ijk} \, \bS{ij} \, \sigma & = & 
  \sigma_{ijjk} + \sigma_{jiik} + \sigma_{ikjk} + \sigma_{jkik} \, ,
  \nnb \\
  \bS{ik} \, \bC{ijkl} \, \sigma \, = \, \bC{ijkl} \, \bS{ik} \, \sigma & = &  
  \sigma_{ijkl} + \sigma_{klij} \, = \, \sigma_{ij}^{(\alpha\beta)} \sigma_{kl} +
  \sigma_{kl}^{(\alpha\beta)} \sigma_{ij} \, ,
  \nnb \\
  \bS{ij} \, \bSC{ijk} \, \sigma \, = \, \bSC{ijk} \, \bS{ij} \, \sigma & = &
  \sum_{b \neq i} \, \bS{i} \, \sigma_{ibjk} \, = \,
  \sum_{b \neq i} \sigma_{ib}^{(\alpha\beta)} \, \sigma_{jk} \, ,
\label{eq: commuting relations 3} \\  
  \bS{ik} \, \bCS{ijk} \, \sigma \, = \, \bCS{ijk} \, \bS{ik} \, \sigma & = & 
  \bS{i} \, \bCS{ijk} \, \sigma \, = \, \bCS{ijk} \, \bS{i} \, \sigma \, = \,
  \sum_{d \neq i,k} \! \sigma_{ijkd} \, = \, \sigma_{ij}^{(\alpha\beta)}
  \sum_{d \neq i,k} \! \sigma_{kd} \, .
\nnb
\eeq
Coming to double-collinear limits of type $\bC{ijk}$ and $\bC{ijkl}$, we get
\beq
  \bC{ijk} \, \bSC{ijk} \, \sigma \, = \, \bSC{ijk} \, \bC{ijk} \, \sigma & = &
  \bS{i} \, \bC{ijk} \, \sigma \, = \, \bC{ijk} \, \bS{i} \, \sigma
  \nnb \\
  & = & \bS{i} \big( \sigma_{ijjk} + \sigma_{ijkj} + \sigma_{ikjk} + \sigma_{ikkj}
  \big) \, = \, \left[ \sigma_{ij}^{(\alpha\beta)} + \sigma_{ik}^{(\alpha\beta)} \right]
  \big( \sigma_{jk} + \sigma_{kj} \big) ,
  \nnb \\
  \bC{ijkl} \, \bSC{ikl} \, \sigma \, = \, \bSC{ikl} \, \bC{ijkl} \, \sigma & = & 
  \bS{i} \, \bC{ijkl} \, \sigma \, = \, \bC{ijkl} \, \bS{i} \, \sigma \, = \,
  \sigma_{ijkl} + \sigma_{ijlk} \, = \, \sigma_{ij}^{(\alpha\beta)} 
  \big( \sigma_{kl} + \sigma_{lk} \big) \, ,
  \nnb \\ 
  \bC{ijk} \, \bCS{ijk} \, \sigma \, = \, \bCS{ijk} \, \bC{ijk} \, \sigma & = &
  \sigma_{ijkj} + \sigma_{jiki} \, ,
  \nnb \\
  \bC{ijkl} \, \bCS{ijk} \, \sigma \, = \, \bCS{ijk} \, \bC{ijkl} \, \sigma & = &
  \sigma_{ijkl} + \sigma_{jikl} \, . 
\label{eq: commuting relations 4}
\eeq
Finally, the mixed soft-collinear limits $\bSC{ijk}$ and $\bCS{ijk}$ satisfy
\beq
  \bSC{ijk} \, \bCS{ijk} \, \sigma \, = \, \bCS{ijk} \, \bSC{ijk} \, \sigma & = &
  \sigma_{ijkj} \, ,
  \nnb\\
  \bSC{ikl} \, \bCS{ijk} \, \sigma \, = \, \bCS{ijk} \, \bSC{ikl} \, \sigma & = &
  \sigma_{ijkl} \, .
\label{eq: commuting relations 5}
\eeq
The relations in Eqs.~(\ref{eq: commuting relations 1})-(\ref{eq: commuting relations 5}), 
where the limits are applied to the sector-function denominator $\sigma$, are 
sufficient to prove that all non-vanishing limits in the different topologies commute 
when acting on the sector functions. The same commutation relations hold when 
applied to the physical double-real matrix elements, as can be proved analogously 
to what was done in \appn{app:commutationNLO}.

The next step in our analysis is to prove that the compositions of the limits given 
in \eq{eq:NNLOprop2} exhaust all single- and double-unresolved configurations 
in each sector. In other words, there are no leftover singular phase-space regions 
after all combinations of limits in \eq{eq:NNLOprop2} have been applied.
We start by denoting with ${\bf L}_i$ a generic set of soft and collinear limits, 
corresponding to configurations where some physical quantities $\lam_i$, 
which could be collections of energies, or angles, or similar, approach zero.
Compositions of two (or more) such limits can be either `uniform' or `ordered', 
with the two cases being defined as
\beq
  [{\bf L}_j {\bf L}_i] \, = \, [{\bf L}_i {\bf L}_j] & : & \left\{
  \begin{array}{l}
  \lambda_i \, , \, \lambda_j  \, \to  \, 0 \\
  \lambda_i/\lambda_j \, \to  \, \mbox{const.}
  \end{array}
  \right.
  \quad \Longleftrightarrow \quad
  \mbox{uniform composition of ${\bf L}_i$ and ${\bf L}_j$ ;}
  \nnb \\
  {\bf L}_j {\bf L}_i & : &
  \left\{
  \begin{array}{l}
  \lambda_i \, , \, \lambda_j  \, \to  \, 0 \\
  \lambda_i/\lambda_j \, \to  \, 0
  \end{array}
  \right.
  \quad\Longleftrightarrow\quad
  \mbox{ordered composition of ${\bf L}_i$ (first) and ${\bf L}_j$ .}
\label{orderL}
\eeq
All single- and double-unresolved configurations in each sector can then be
systematically generated by combining in all possible ways the single-soft and 
single-collinear limits selected by the sector functions, namely $\bS{a}$, 
$\bS{c}$, $\bC{ab}$, and $\bC{cd}$\footnote{\label{footnote Cik} Note that 
compositions of limits involving both $\bC{ij}$ and $\bC{jk}$ automatically 
also involve the limit $\bC{ik}$. Indeed
\beq
  [\bC{jk}\,\bC{ij}] \, = \, [\bC{ik} \, \bC{jk} \, \bC{ij}] \, , \qquad
  \bC{jk}\,\bC{ij} \, = \, [\bC{ik} \, \bC{jk}] \, \bC{ij} \, , \qquad
  \bC{ij}\,\bC{jk}  \, = \,  [\bC{ik} \, \bC{ij}] \, \bC{jk}\, . \nnb
\eeq
} in sector $\W{abcd}$.

Owing to the prescription $\alpha > \beta > 1$ in \eq{eq:sigmadef}, the action on 
$\sigma$ of a uniform composition involving soft and collinear limits is equivalent 
to the corresponding ordered composition where the soft limits act first:
\beq
  {\bf L}' \, [\bL{}{\rm c} \, \bL{}{\so}] \, {\bf L} \, \sigma \, = \,    
  {\bf L}' \, [\bL{}{\rm c}] [\bL{}{\so}] \, {\bf L} \, \sigma \, ,
\label{LLLL}
\eeq
where $\bL{}{\so}$ ($\bL{}{\rm c}$) are collections of soft (collinear) limits, while 
${\bf L}$, and ${\bf L}'$ are generic combinations of limits. The remaining uniform 
compositions involve either a pair of collinear or a pair of soft limits\footnote{
\label{footnote repetition} Repeated limits can in all cases be readily simplified. 
Given a generic limit $\bf L$, one has for example
\beq
  [{\bf L}_i \, {\bf L} \, {\bf L}_i] \, = \, [{\bf L} \, {\bf L}_i] \, , \qquad
  {\bf L}_i \, {\bf L} \, {\bf L}_i \, = \, {\bf L} \, {\bf L}_i \, . \nnb
\label{secfooteq}
\eeq
}, which can be directly identified with the limits given in \eq{eq:NNLOprop2}: 
\beq
  [\bS{i} \, \bS{j}] \, = \, \bS{ij} \, , \qquad [\bC{ij} \, \bC{jk}] \, = \, \bC{ijk} \, , \qquad
  [\bC{ij} \, \bC{kl}] \, = \, \bC{ijkl} \, .
\label{complim}
\eeq
We conclude that all possible single- and double-unresolved singular 
configurations can be  obtained as ordered compositions without 
repetition\textsuperscript{\ref{footnote repetition}} of the limits 
\\[-4mm]
\begin{itemize}
\item $\bS{i}$,  \, $\bS{j}$,  \, $\bC{ij}$,   \, $\bC{jk}$,  \, $\bS{ij}$, \, 
and \, $\bC{ijk}$ \, for topology   $\W{ijjk}$  ;
\\[-5mm]
\item $\bS{i}$,  \, $\bS{k}$,  \, $\bC{ij}$,  \, $\bC{jk}$,  \, $\bS{ik}$, \,
and \, $\bC{ijk}$ \, for topology   $\W{ijkj}$  ; 
\\[-5mm]
\item $\bS{i}$,  \, $\bS{k}$,  \, $\bC{ij}$,  \, $\bC{kl}$,  \,  $\bS{ik}$, \, 
and \, $\bC{ijkl}$ \, for topology   $\W{ijkl}$  . 
\end{itemize}
To conclude, we reduce this list of limits, topology by topology, to that
given in \eq{eq:NNLOprop2}.

\begin{itemize}

\item \underline{Topology $\W{ijjk}$}
\\[6pt]
According to Eqs.~(\ref{eq: commuting relations 1})-(\ref{eq: commuting relations 5}), 
the $\bS{j}$ limit commutes with all other limits in the list except $\bS{i}$. Therefore, 
when appearing in a generic composition of limits, it can be moved to the right until 
it encounters $\bS{i}$. At this point one can use 
\beq
  {\bf L}' \, \bS{j} \, \bS{i} \, {\bf L} \, \W{ijjk} \, = \, 
  {\bf L}' \, \bS{ij} \, \bS{i} \, {\bf L} \, \W{ijjk} \, ,
\label{traslSi}
\eeq
valid for generic limits ${\bf L}$ and ${\bf L}'$, to remove $\bS{j}$. If $\bS{i}$ is not 
present at the right of $\bS{j}$, the latter can be moved to the rightmost position, where 
it vanishes:
\beq
  {\bf L} \, \bS{j} \, \W{ijjk} \, = \, 0 \, .
\label{Sjvan}
\eeq
Since the action of $\bS{j}$ either gives zero or can be replaced by that of $\bS{ij}$, 
$\bS{j}$ can be simply removed from the list.

Considering now $\bC{jk}$, we note that it commutes with $\bC{ijk}$ and with 
$\bS{ij}$, and it satisfies
\beq
  {\bf L}' \, \bC{jk} \, \bS{i} \, {\bf L} \, \W{ijjk}  & = & 
  {\bf L}' \, \bSC{ijk} \, {\bf L} \, \W{ijjk} \, , \nnb \\
  {\bf L}' \, \bC{jk} \, \bC{ij} \, {\bf L} \, \W{ijjk} & = & 
  {\bf L}' \, \bC{ijk} \, \bC{ij} \, {\bf L} \, \W{ijjk} \, , \nnb\\
  {\bf L}' \, \bC{jk} \, \W{ijjk} & = & 0 \, ,
\label{LCij}
\eeq
so that $\bC{jk}$ can either be moved to the rightmost position, where it gives zero, 
or replaced by $\bC{ijk}$ or $\bSC{ijk}$. Consequently, one can remove $\bC{jk}$ 
from the list of limits, and add $\bSC{ijk}$ in its stead. The list of singular limits is 
thus reduced to the first line of \eq{eq:NNLOprop2},
\beq
  \W{ijjk} &:& \, \, \bS{i} \, , \, \bC{ij} \, , \, \bS{ij} \, , \,  \bC{ijk} \, , \, \bSC{ijk} \, .
\label{listlim1}  
\eeq

\item \underline{Topology $\W{ijkj}$}
\\[6pt]
Besides commuting with $\bC{jk}$, $\bS{ik}$, and $\bC{ijk}$, the single-soft limit
$\bS{k}$ satisfies
\beq
  {\bf L}' \, \bS{k} \, \bS{i} \, {\bf L} \, \W{ijkj}  & = & 
  {\bf L}' \, \bS{ik} \, \bS{i} \, {\bf L} \, \W{ijkj} \, , \nnb \\
  {\bf L}' \, \bS{k} \, \bC{ij} \, {\bf L} \, \W{ijkj}  & = & 
  {\bf L}' \, \bCS{ijk} \, {\bf L} \, \W{ijkj} \, , \nnb \\
  {\bf L}'\,\bS{k}\,\W{ijkj} & = & 0 \, .
\label{eq:prop Sk Wijkj}
\eeq
Since $\bS{k}$ can be either moved to the rightmost position, where it gives zero, 
or replaced by $\bS{ik}$ or $\bCS{ijk}$, one can remove it from the list of contributing 
limits. A similar statement holds for $\bC{jk}$, which commutes with $\bS{ik}$,
and $\bC{ijk}$, and satisfies 
\beq
  {\bf L}' \, \bC{jk} \, \bS{i} \, {\bf L} \, \W{ijkj}  & = & 
  {\bf L}' \, \bSC{ijk} \, {\bf L} \, \W{ijkj} \, , \nnb \\
  {\bf L}' \, \bC{jk} \, \bC{ij} \, {\bf L} \, \W{ijkj} & = & 
  {\bf L}' \, \bC{ijk} \, \bC{ij} \, {\bf L} \, \W{ijkj} \, , \nnb \\
  {\bf L}' \, \bC{jk} \, \bCS{ijk} \, {\bf L} \, \W{ijkj} & = & 
  {\bf L}' \, \bC{ijk} \, \bCS{ijk} \, {\bf L} \, \W{ijkj} \, , \nnb \\
  {\bf L}' \, \bC{jk} \, \W{ijkj} & = & 0 \, .
\label{eq:prop Cjk Wijkj}
\eeq
As a consequence, $\bC{jk}$ can either be moved to the rightmost position, where it 
gives zero, or replaced by $\bC{ijk}$ or $\bSC{ijk}$. The list of singular limits in sector 
$\W{ijkj}$ can thus be reduced to the second line of \eq{eq:NNLOprop2},
\beq
  \W{ijkj} & : & \,\, \bS{i}\, , \, \bC{ij}\, , \, \bS{ik}\, , \, \bC{ijk}\, , \, \bSC{ijk}\, , \, 
  \bCS{ijk} \, .
\label{listWijkj}
\eeq

\item \underline{Topology $\W{ijkl}$}
\\[6pt]
The discussion of the $\bS{k}$ and $\bC{kl}$ limits holds unchanged with respect 
to the one relevant for $\bS{k}$ and $\bC{kj}$ in topology $\W{ijkj}$. These limits 
can either be moved to the rightmost position, where they yield zero, or be 
replaced by limits that are already present in the list, ($\bS{ik}$ or $\bCS{ijk}$ in 
the case of $\bS{k}$, $\bC{ijkl}$ or $\bSC{ikl}$ in the case of $\bC{kl}$). The final
list of contributing limits thus coincides with the third line of \eq{eq:NNLOprop2},
\beq
  \W{ijkl} & : & \,\, \bS{i} \, , \, \bC{ij} \, , \, \bS{ik} \, , \, \bC{ijkl} \, , \, 
  \bSC{ikl} \, , \, \bCS{ijk} \, .
\label{listWijkl}
\eeq

\end{itemize}

%%%%%%%%%%%%%%%%%%%%%%%%%%%%%%%%%

\section{Composite IR limits of the double-real matrix element}
\label{app: composit limits RR}

In this Appendix we list the composite soft and collinear limits of the double
real-radiation squared matrix element needed for the evaluation of the 
double-unresolved counterterm $\overline K^{\two} + \overline K^{\otwo}$ in
\eq{eq:K2+K12}, including the detailed dependence on the remapped phase-space 
variables described in \secn{sec:2unrescntint}. The remappings described in the
following apply also to the corresponding sector functions $\bW{ab}$ in \eq{eq:K2+K12}.
First, we consider composing a double-collinear limit and a collinear limit. We find
\beq
  \bbC{ij} \, \bbC{ijk} \RR & = & \Norm \,
  \frac{P_{ij}^{\mu\nu} \left( s_{ir}, s_{jr} \right)}{s_{ij}} \,\, \bbC{jk} \,
  \Rl_{\mu\nu} \! \left( \kkl{ijr} \right) 
\label{compCijCijk} \\  
  & = & \frac{\Norm^2}{s_{ij}\sk{jk}{ijr}} \, \Bigg\{ \bigg[
  \left( P_{ij} + Q_{ij} \right)
  \left( \overline P_{jk}^{(ijr)}  + \overline Q_{jk}^{(ijr)}  \right)
  - \frac{d - 2}{2} \, C_F \, Q_{ij} \, \delta_{f_k \{q, \bar q\}} \, x_j'
  \nnb \\ && \qquad \qquad
  - \, \frac{d - 2}{2} \, Q_{ij} \, C_{f_k} \,  \frac{x_k'}{x_j'} \, 
  \frac{(2 \kt \cdot \kt')^2}{\kt^2 \, \kt'^2}  \bigg] \Bn \! \left( \kkl{ijr,jkr} \right)
  \nnb \\ && \qquad \qquad
  + \, (d - 2) \bigg[ \Big( \left( P_{ij} + Q_{ij} \right) \, \overline Q_{jk}^{(ijr)}  
  + Q_{ij} \, C_A \, \delta_{f_k g} \left( 2 x_j' x_k' \right) \Big)
  \frac{\kt'^{\mu}\kt'^{\nu}}{\kt'^2}
  \nnb \\ && \qquad \qquad \qquad \qquad \, \, + \, 
  Q_{ij} \, C_A \, \delta_{f_k g} \,\frac{2 x_j'}{x_k'} \, \frac{\kt^{\mu}\kt^{\nu}}{\kt^2}
  \bigg] \Bn_{\mu\nu} \! \left( \kkl{ijr,jkr} \right) \Bigg\} \, , \nnb
\eeq
where $r \neq i, j, k$, and we introduced the shorthand notations
\beq
\label{pbar}
  \overline P_{jk}^{(ijr)} \, = \, P_{jk} \left( \sk{jr}{ijr}, \sk{kr}{ijr} \right) \, ,
  \qquad \overline Q_{jk}^{(ijr)}  \, = \, Q_{jk} \left( \sk{jr}{ijr},\sk{kr}{ijr} \right) \, .
\eeq
The primed variables in \eq{compCijCijk} are defined in analogy to \eq{translong}, as
\beq
\label{primvar}
  \kt' & = & x_k'\,\kk{j}{ijr} - x_j' \,\kk{k}{ijr} - \left( 1 - 2 x_j' \right) 
  \frac{\sk{jk}{ijr}}{\sk{jr}{ijr} + \sk{kr}{ijr}} \, \kk{r}{ijr} \, , \\
  && x_j' \, = \, \frac{\sk{jr}{ijr}}{\sk{jr}{ijr} + \sk{kr}{ijr}} \, ,
  \qquad
  x_k' \, = \, \frac{\sk{kr}{ijr}}{\sk{jr}{ijr} + \sk{kr}{ijr}} \, .
\eeq
Acting with further soft limits leads to
\beq
  \bbC{ij} \, \bbS{ij} \, \bbC{ijk} \RR & = & 
  \Norm \, \frac{P_{ij}^{\mu\nu} \left( s_{ir}, s_{jr} \right)}{s_{ij}} \,\, 
  \bbS{j} \, \bbC{jk} \, \Rl_{\mu\nu} \! \left( \kkl{ijr} \right) \\
  & = & \frac{\Norm^{\,2}}{s_{ij}} \, 2 \, C_{f_k} \, 
  \frac{\sk{kr}{ijr}}{\sk{jk}{ijr} \sk{jr}{ijr}} (\delta_{f_i g} \delta_{f_j g}
+ \delta_{ \{f_i f_j\} \{q \bar q\} }) \nnb\\
  && \, \times \, 
  \bigg[ P_{ij} + Q_{ij} \left(1 - \frac{d - 2}{4} \, 
  \frac{(2 \kt \cdot \kt')^2}{\kt^2 \, \kt'^2} \right) \bigg] \!
  \Bn \! \left( \kkl{ijr,jkr} \right) \, ,
\label{compCijSijCijk} 
  \nnb \\
  \bbS{i} \, \bbC{ij} \, \bbC{ijk} \RR & = & 2 \, \Norm \, C_{f_j} \, \mc I_{jr}^{(i)} \,
  \bbC{jk} \, \Rl \! \left( \! \kkl{ijr} \! \right) \\ 
  & = & 2 \, \Norm^{\,2} \, C_{f_j} \, \mc I_{jr}^{(i)} \,
  \frac{P_{jk}^{\mu\nu} \left( \sk{jr}{ijr}, \sk{kr}{ijr} \right)}{\sk{jk}{ijr}} \,
  \Bn_{\mu\nu} \! \left( \! \kkl{ijr,jkr} \! \right) \, ,
\label{compSiCijCijk}
  \nnb \\
  \bbS{i} \, \bbC{ij} \, \bbS{ij} \, \bbC{ijk} \RR & = & 2 \, \Norm \, C_{f_j} \,
  \mc I_{jr}^{(i)} \, \bbS{j} \, \bbC{jk} \, \Rl \! \left( \! \kkl{ijr} \! \right) \\
  & = & 4 \, \Norm^{\,2} \, \delta_{f_jg} \, C_{f_j} \, C_{f_k} \, \mc I_{jr}^{(i)} \,
  \frac{\sk{kr}{ijr}}{\sk{jk}{ijr} \sk{jr}{ijr}} \, \Bn \! \left( \! \kkl{ijr,jkr} \! \right) \, ,
  \nnb
\label{compSiCijSijCijk}
\eeq
where the same $r \neq i, j, k$ should be chosen for all permutations of $ijk$. 
Composition of a double-soft limit and a collinear limit (on the same pair of 
particles) yields
\beq
\label{compCijSij} 
  \bbC{ij} \, \bbS{ij} \RR & = & \Norm \, 
  \frac{P_{ij}^{\mu\nu} \left(s_{ir}, s_{jr} \right)}{s_{ij}} \, 
  \bbS{j} \, \Rl_{\mu\nu} \! \left( \! \kkl{ijr} \!\right)
   =  \frac{\Norm^{\, 2}}2 \,
  \sum_{\substack{c \neq i,j \\ d \neq i,j,c}} \!
  \mc J_{cd}^{(ij)} {\Bn}_{cd} \Big( \! \kkl{ijr,jcd} \! \Big) 
 \, ,
\eeq
where the same $r \neq i, j$ should be chosen for all permutations of $ijk$ (recall 
that the index $k$ appears in the sector function associated with these contributions 
in \eq{eq:K2+K12}), and we have defined the quantities
\beq
\label{slicum2}
  \mc J_{cd}^{(ij)} & = &
- 
\left[
\delta_{f_i g} \delta_{f_j g} \, 2 \, C_A 
\left( \frac{x_i}{x_j} + \frac{x_j}{x_i} \right) 
+ 
\delta_{ \{f_i f_j\} \{q \bar q\} } \, T_R
\right]
\frac{2\,\sk{cd}{ijr}}{s_{ij}\,\sk{jc}{ijr} \sk{jd}{ijr}} \,
\nnb\\
&&\qquad\qquad\quad
- \,
(d\!-\!2)\,
\frac{Q_{ij}\left( s_{ir}, s_{jr} \right)}{2\,\kt^2\,s_{ij}} \,
\Bigg[
\frac{2\kt\cdot \kk{c}{ijr}}{\sk{jc}{ijr}}
-
\frac{2\kt\cdot \kk{d}{ijr}}{\sk{jd}{ijr}}
\Bigg]^2 \, .
\eeq
Further applying the soft 
limit $\bbS{i}$ leads to
\beq
  \bbS{i} \, \bbC{ij} \, \bbS{ij} \RR & = & 2 \, \Norm \, C_{f_j} \, \mc I_{jr}^{(i)} \,
  \bbS{j} \, \Rl \! \left( \! \kkl{ijr} \! \right) \\
  & = & - \, 2 \, \Norm^{\,2} \, \delta_{f_jg} \, C_{f_j} \, \mc I_{jr}^{(i)} \,
  \sum_{\substack{c \neq i,j \\ d \neq i,j}} \frac{\sk{cd}{ijr}}{\sk{jc}{ijr} \sk{jd}{ijr}} \,
  \Bn_{cd} \! \left( \! \kkl{ijr,jcd} \! \right) \, .
\label{compSiCijSij}
\nnb
\eeq
As in \eq{compCijSij}, the same $r\neq i,j$ should be chosen for all permutations 
of $ijk$. Next, we consider the composition of the double-collinear limit $\bbC{ijkl}$
with a soft limit. We get
\beq
\label{compSiCijkl}
  \bbS{i} \, \bbC{ijkl} \RR & = & 2 \, \Norm \, C_{f_j} \, \mc I_{jl}^{(i)} \,
  \bbC{kl} \, \Rl \Big( \! \kkl{ijl} \! \Big) \\
  & = & 
  2 \, \Norm^{\,2} \, C_{f_j} \mc I_{jl}^{(i)} \,
  \frac{P^{\mu\nu}_{kl} \left( \sk{kr}{ijl}, \sk{lr}{ijl} \right)}{\sk{kl}{ijl}} \,
  {\Bn}_{\mu\nu} \Big( \! \kkl{ijl,klr} \! \Big) \, , \nnb
\eeq
where the same $r \neq i, k, l$ should be chose for all permutations
in $\pi \left(\pi(ij) \pi(kl) \right)$. Acting with a further double soft limit $\bbS{ik}$ leads to
\beq 
  \bbS{i} \, \bbS{ik} \, \bbC{ijkl} \RR  & = &
\label{compSiSikCijkl}
4 \,  \Norm^{\, 2} \, \delta_{{f_k} g} \, C_{f_j} \, C_{f_l} \, \mc I_{jl}^{(i)} \, 
  \frac{\sk{lr}{ijl}}{\sk{kl}{ijl} \sk{kr}{ijl}} \, 
  {\Bn} \Big( \! \kkl{ijl,klr} \! \Big) \\
  &&
  + \, 2 \,  \Norm^{\, 2} \, \delta_{{f_k} g}  \, C_{f_l} \, \mc I_{jl}^{(i)} \, 
  \left[ \frac{\sk{lr}{ijl}}{\sk{kl}{ijl} \sk{kr}{ijl}} \, 
  {\Bn}_{jl} \Big( \! \kkl{ijl,klr} \! \Big) \,
  - \, 
  \frac{\sk{lr}{ilj}}{\sk{kl}{ilj} \sk{kr}{ilj}} \, 
  {\Bn}_{lj} \Big( \! \kkl{ilj,klr} \! \Big) \right]
  \, , \nnb
\eeq
where the same $r\neq i,k,l$ should be chosen for all permutations
in $\pi(\pi(ij)\,\pi(kl))$.

Taking successively a double-soft limit and a single-soft limit, we get
\beq
\label{compSiSij} 
  \bbS{i} \, \bbS{ij} \RR & = & - \, \Norm \sum_{c \neq i \, d \neq i}
  \mc I_{cd}^{(i)} \,\, \bbS{j} \, \Rl_{cd} \Big( \! \kkl{icd} \! \Big) \\
&=&
\frac{\Norm^2}{2} \!
\sum_{\substack{c \neq i,j \\ d \neq i,j,c}} \!
\Bigg[
\sum_{\substack{e \neq i,j,c,d \\ f \neq i,j,c,d}} \! 
\mc I_{cd}^{(i)} \,
\delta_{f_jg}\,\frac{\sk{ef}{icd}}{\sk{ej}{icd} \sk{fj}{icd}}
{\Bn}_{cdef} \Big( \! \kkl{icd,jef} \! \Big) \nnb \\
&& \qquad
+\,
2
\sum_{\substack{e \neq i,j,c,d}} \! 
\mc I_{cd}^{(i)} \,
\delta_{f_jg}\,\frac{\sk{ed}{icd}}{\sk{ej}{icd} \sk{dj}{icd}}
{\Bn}_{cded} \Big( \! \kkl{icd,jed} \! \Big) \nnb \\
&& \qquad
+\,
2
\sum_{\substack{e \neq i,j,c,d}} \! 
\mc I_{cd}^{(i)} \,
\delta_{f_jg}\,\frac{\sk{ed}{idc}}{\sk{ej}{idc} \sk{dj}{idc}}
{\Bn}_{cded} \Big( \! \kkl{idc,jed} \! \Big) 
\nnb\\
&&\qquad
+\, 
2\, \, 
\mc I_{cd}^{(i)} \, 
\delta_{f_jg}\,\frac{\sk{cd}{icd}}{\sk{cj}{icd} \sk{dj}{icd}}
{\Bn}_{cdcd} \Big( \! \kkl{ijcd} \! \Big) 
+
\overline{\mc I}_{cd}^{(ij) \, {\rm s.o.}}
{\Bn}_{cd} \Big( \! \kkl{ijcd} \! \Big) 
\Bigg]
\, , \nnb
\eeq
where $\overline{\mc I}_{cd}^{(ij) \, {\rm s.o.}}$ is the strongly-ordered limit, $(k_i \ll k_j) \to 0$, 
of the kernel in Eq.~(111) of Ref.~\cite{Catani:1999ss}, after an appropriate remapping, defined by
\beq
\overline{\mc I}_{cd}^{(ij) \, {\rm s.o.}} \, \equiv \,
- \, 2 \, C_A \delta_{f_j g} \left[
\mc I_{cj}^{(i)}  \;
\frac{\sk{cd}{icj}}{\sk{jc}{icj} \sk{jd}{icj}}+
\mc I_{jd}^{(i)}  \;
\frac{\sk{cd}{ijd}}{\sk{jc}{ijd} \sk{jd}{ijd}}-
\mc I_{cd}^{(i)}  \;
\frac{\sk{cd}{icd}}{\sk{jc}{icd} \sk{jd}{icd}}
\right] 
\, .
\eeq
Composing a double-collinear limit and a single-soft limit, we get
\beq
\label{compSiCijk}
  \bbS{i}\,\bbC{ijk} \, RR & = &
 \Norm^{\,2} \sum_{\substack{a,b = j,k,r}} \, C_{a b}^{\, r} \, 
  \mc I^{(i)}_{ab} \, \frac{P^{\mu\nu}_{jk} 
  \big( \bar s_{jr}^{(iab)}, \bar s_{kr}^{(iab)} \big)}{\bar s_{jk}^{(iab)}} \, 
  B_{\mu\nu}\!\left(\{\bar k\}^{(iab,jkr)} \right) \, ,
\eeq
where the color structure is given by the combinations
\beq
\label{newCas}
C^r_{ab}\,&\equiv&\,\frac{1}{2}\,\Big[
  (C_{f_{[jk]}}+C_{f_j}-C_{f_k})(\delta_{aj}\delta_{br}+\delta_{ar}\delta_{bj})\nnb\\
  &&\hspace{2mm}+\,(C_{f_{[jk]}}+C_{f_k}-C_{f_j})(\delta_{ak}\delta_{br}+\delta_{ar}\delta_{bk})\nnb\\
  &&\hspace{2mm}-\,(C_{f_{[jk]}}-C_{f_j}-C_{f_k})(\delta_{aj}\delta_{bk}+\delta_{ak}\delta_{bj})
\Big] \, .
\eeq
Finally, inserting a further double soft limit on partons $i$ and $j$ yields
\beq
  \bbS{i}\,\bbS{ij}\,\bbC{ijk} \, \RR  & = & \Norm^{\,2} 
  \sum_{\substack{a , b = j,k,r}} \, C_{a b}^{\, r} \, \mc I^{(i)}_{ab} \, 2 \, C_{f_k} \,\delta_{f_jg} \,  
   \frac{\bar s_{kr}^{(iab)}}{\bar s_{jr}^{(iab)}\bar s_{jk}^{(iab)}}\,  
  B \! \left(\{\bar k\}^{(iab,jkr)} \right) \, ,
\label{compSiSijCijk}
\eeq
which completes our list of relevant nested limits.

%%%%%%%%%%%%%%%%%%%%%%%%%%%%%%%%%

\section{Results for the mixed double-unresolved counterterm}
\label{app:intK12formulae}

In this Appendix we show explicitly how the terms in the integrated mixed 
double-unresolved counterterm organise themselves in the form of 
single-unresolved limits in the NLO phase space. Starting from 
\eq{eq:K12hcdefinition}, using Eqs.~(\ref{eq:propNNLOfact}), 
and (\ref{eq:SS.CC=SS.CC.S}), and introducing remapped kinematics
for the double-real matrix element and for the sector functions $\W{ab}$, the
hard-collinear contribution to the mixed double-unresolved counterterm can
be cast in the form
\beq
\label{eq:K12hcWabini}
  \overline K^{\otwohc} & = & - \sum_{i, \, j > i} \sum_{k \neq i,j}
  \bigg[ \bC{ij} \left( \Wab{ij} + \Wab{ji} \right) \bigg]
  \bigg\{ \Big[ \bC{jk} \left( \bW{jk} + \bW{kj} \right) \Big] \,
  \bbC{ij} \, \bbC{ijk} \nnb \\
  & & \hspace{-1.3cm}
  + \, \sum_{l \neq i,j,k} \left( \bC{kl} \, \bW{kl} \right) \, \bbC{ijkl} +
  \left( \bS{j} \, \bW{jk} \right) \, \bbC{ij} \, \bbS{ij} - 
  \left( \bS{j} \, \bC{jk} \, \bW{jk} \right) \, \bbC{ij} \, \bbS{ij} \, \bbC{ijk} \nnb \\
  & & \hspace{-1.3cm}
  + \, \sum_{l \neq i,k} \left( \bS{k} \, \bW{kl} \right) \, \bbCS{ijk} - 
  \left( \bS{k} \, \bC{jk} \, \bW{kj} \right) \, \bbCS{ijk} \, \bbC{ijk} - 
  \sum_{l \neq i,j,k} \left( \bS{k} \, \bC{kl} \, \bW{kl} \right) \, \bbCS{ijk} \, 
  \bbC{ijkl} \bigg\} \RR \nnb \\
  & & + \,  \sum_{i, \, j \neq i} \sum_{k \neq i,j}
  \Big[ \bS{i} \, \bC{ij} \, \Wab{ij} \Big] \bigg\{ \Big[ \bC{jk} \, 
  \left( \bW{jk} + \bW{kj} \right) \Big] \, \bbS{i} \, \bbC{ij} \, \bbC{ijk}
  + \left( \bS{j} \, \bW{jk} \right) \, \bbS{i} \, \bbC{ij} \, \bbS{ij} \nnb \\
  & & \hspace{-1.3cm} + \sum_{l\neq i,j,k} \left( \bC{kl} \, \bW{kl} \right) \,
  \bbC{ijkl} \, \bbS{i} - \left( \bS{j} \, \bC{jk} \, \bW{jk} \right) \,
  \bbS{ij} \, \bbC{ijk} \, \bbS{i} \, \bbC{ij} + \sum_{l \neq i,k}
  \left( \bS{k} \, \bW{kl} \right) \, \bbCS{ijk} \, \bbS{ik} \nnb \\
  & & \hspace{-1.3cm} - \, \left( \bS{k} \, \bC{jk} \, \bW{kj} \right) \, \bbCS{ijk} \, 
  \bbS{ik} \, \bbC{ijk} - \sum_{l \neq i,j,k} \left( \bS{k} \, \bC{kl} \, \bW{kl} \right) \, 
  \bbCS{ijk} \, \bbS{ik} \, \bbC{ijkl} \bigg\} \RR \, .
\eeq
Using the NLO sector-function sum rules, and appropriate symmetrisations, 
the latter becomes
\beq
\label{eq:K12hcsymm}
  \overline K^{\otwohc} & = & - \sum_{i, \, j > i} \sum_{k \neq i,j} \Bigg\{
  \bigg[ \Big( \bC{jk} \, \left( \bW{jk} + \bW{kj} \right) \Big) \, 
  \bbC{ij} \, \bbC{ijk} + \! \sum_{l \neq i,j,k} \left( \bC{kl} \, \bW{kl} \right) \,
  \bbC{ijkl} \nnb\\
  & & \hspace{22mm} 
  + \, \left( \bS{j} \, \bW{jk} \right) \, \bbC{ij} \, \bbS{ij} - 
  \left( \bS{j} \, \bC{jk} \, \bW{jk} \right) \,
  \bbC{ij} \, \bbS{ij} \, \bbC{ijk} \bigg] \Big(1 - \bbS{i} - \bbS{j} \Big)
  \nnb \\ [5pt]
  & & \hspace{18mm} 
  + \, \bigg[ \sum_{l \neq i,k} \left( \bS{k} \, \bW{kl} \right) \, \bbCS{ijk} -
  \left( \bS{k} \, \bC{jk} \, \bW{kj} \right) \, \bbCS{ijk} \, \bbC{ijk}
  \nnb \\ [-3pt]
  & & \hspace{22mm}
  - \sum_{l \neq i,j,k} \left( \bS{k} \, \bC{kl} \, \bW{kl} \right) \, 
  \bbCS{ijk} \, \bbC{ijkl} \bigg] \Big( 1 - \bbS{ik} - \bbS{jk} \Big) \,
  \Bigg\} \, \RR \,.
\eeq
The singular limits in \eq{eq:K12hcsymm}, as well as their phase-space integrals, 
are explicitly computed in the following. For brevity, in all contributions to the 
hard-collinear counterterm we do not display kinematic dependences, writing 
$P^{\hc}_{ij}$ for $P^{\hc}_{ij} (s_{ir}, s_{jr})$, and similarly for $Q^{\mu\nu}_{ij}$, 
while the real matrix element is written as $\Rl \equiv \Rl \left( \kkl{ijr }\right)$, 
and similarly for $R_{\mu\nu}$. We note that all limits are accompanied by single-
and double-soft subtractions, guaranteeing the hard-collinear character of the
counterterm. Terms containing $\bbC{ij} \, \bbS{ij}$ give, upon integration
\beq
  \int d \Phi_{\rm rad,1} \, \bbC{ij} \, \bbS{ij} \, \Big( 1 - \bbS{i} - \bbS{j} \Big) \, \RR 
  & = & \Norm \, \frac{\varsi_\npt}{\varsi_\npo} \,\, \bbS{j} \, \Rl
  \int d \Phi_{\rm rad,1}^{(ijr)} \, \frac{P_{ij}^{\hc} (z,1 - z)}{y \, \sk{jr}{ijr}}
  \nnb \\
  & = & \Norm \, \frac{\varsi_\npt}{\varsi_\npo} \,\, 
  J^{\hc}_{ij} \! \left( \sk{jr}{ijr}, \eps \right) \, \bbS{j}\, \Rl \, ,
\label{eq: int bar Cij(1-Si-Sj) Sij RR}
\eeq
where the hard-collinear integral $J^{\hc}_{ij}$ is defined in \eq{eq:Jijhc}, and we
exploited the fact that azimuthal terms integrate to zero in the radiation 
phase space. The soft-collinear limit $\bbCS{ijk}$ contributes to the integrated counterterm
\beq
  \int d \Phi_{\rm rad,1} \, \bbCS{ijk} \, \Big( 1 - \bbS{ik} - \bbS{jk} \Big) \, \RR 
  & = & \Norm \, \frac{\varsi_\npt}{\varsi_\npo} \,\, \bbS{k} \, \Rl
  \int \! d \Phi_{\rm rad,1}^{(ijr)} \, \frac{P_{ij}^{\hc} (z,1 - z)}{y \, \sk{jr}{ijr}}
  \nnb \\
  & = & \Norm \, \frac{\varsi_\npt}{\varsi_\npo} \,\, 
  J^{\hc}_{ij} \! \left( \sk{jr}{ijr}, \eps \right) \, \bbS{k}\, \Rl \, .
\label{eq: int bar SCkij(1-Sik-Sjk) RR}
\eeq
The nested collinear limit $\bbC{ij} \, \bbC{ijk}$ contributes
\beq
  \int \! d \Phi_{\rm rad,1} \, \bbC{ij} \, \bbC{ijk} \, 
  \Big( 1 - \bbS{i} - \bbS{j} \Big) \, \RR  & = &
  \Norm \, \frac{\varsi_\npt}{\varsi_\npo} \,\, \bbC{jk} \, \Rl
  \int \! d \Phi_{\rm rad,1}^{(ijr)} \, \frac{P_{ij}^{\hc} (z,1 - z)}{y \, \sk{jr}{ijr}}
  \nnb \\
  & = & \Norm \, \frac{\varsi_\npt}{\varsi_\npo} \,\, 
  J^{\hc}_{ij} \! \left( \sk{jr}{ijr}, \eps \right) \, \bbC{jk} \, \Rl \, .
\label{eq: int bar Cij(1-Si-Sj) Cijk RR}
\eeq
The nested collinear limit $\bbC{ij}  \, \bbC{ijkl}$ contributes
\beq
  \int d \Phi_{\rm rad,1} \,\bbC{ij}\,\bbC{ijkl} \, 
  \Big( 1 - \bbS{i} - \bbS{j} \Big) \, \RR  & = & \Norm \, 
  \frac{\varsi_\npt}{\varsi_\npo} \, \, \bbC{kl} \, \Rl
  \int d \Phi_{\rm rad,1}^{(ijr)} \, \frac{P_{ij}^{\hc} (z, 1 - z)}{y \, \sk{jr}{ijr}}
  \nnb \\
  & = & \Norm \, \frac{\varsi_\npt}{\varsi_\npo} \,\,
  J^{\hc}_{ij} \! \left( \sk{jr}{ijr}, \eps \right) \, \bbC{kl} \, \Rl \, .
\label{eq: int bar Cij(1-Si-Sj) Cijkl RR}
\eeq
Combining \eq{eq: int bar Cij(1-Si-Sj) Cijk RR} with a double-soft limit we get
\beq
  \int d \Phi_{\rm rad,1} \, \bbC{ij} \, \bbC{ijk} \, \bbS{ij} \,  
  \Big( 1 - \bbS{i} - \bbS{j} \Big) \, \RR & = & \Norm \,  
  \frac{\varsi_\npt}{\varsi_\npo} \,\, \bbS{j} \, \bbC{jk} \, \Rl
  \int d \Phi_{\rm rad,1}^{(ijr)} \, \frac{P_{ij}^{\hc} (z, 1 - z)}{y \, \sk{jr}{ijr}}
  \nnb \\
  & = & \Norm \, \frac{\varsi_\npt}{\varsi_\npo} \,\, 
  J^{\hc}_{ij} \! \left( \sk{jr}{ijr}, \eps \right) \, \bbS{j} \, \bbC{jk} \, \Rl \, .
\label{eq: int bar Cij(1-Si-Sj) Sij Cijk RR}
\eeq
Acting on \eq{eq: int bar SCkij(1-Sik-Sjk) RR} with a three-particle double-collinear
limit one finds
\beq
  \int d \Phi_{\rm rad,1} \, \bbCS{ijk} \, \bbC{ijk} \,
  \Big( 1 - \bbS{ik} - \bbS{jk} \Big) \, \RR & = & \Norm \,
  \frac{\varsi_\npt}{\varsi_\npo} \,\, \bbS{k} \, \bbC{jk} \, \Rl
  \int d \Phi_{\rm rad,1}^{(ijr)} \, \frac{P_{ij}^{\hc} (z, 1 - z)}{y \, \sk{jr}{ijr}}
  \nnb \\
  & = & \Norm \, \frac{\varsi_\npt}{\varsi_\npo} \,\,
  J^{\hc}_{ij} \! \left( \sk{jr}{ijr}, \eps \right) \, \bbS{k} \, \bbC{jk} \, \Rl \, .
\label{eq: int bar SCkij Cijk (1-Sik-Sjk) RR}
\eeq
Finally, replacing the three-particle double-collinear limit in 
\eq{eq: int bar SCkij Cijk (1-Sik-Sjk) RR} with the four-particle one we get
the result
\beq
  \int d \Phi_{\rm rad,1} \, \bbCS{ijk} \, \bbC{ijkl} \, 
  \Big( 1 - \bbS{ik} - \bbS{jk} \Big) \, \RR  & = & \Norm \, 
  \frac{\varsi_\npt}{\varsi_\npo} \,\, \bbS{k}\,\bbC{kl} \, \Rl
  \int d \Phi_{\rm rad,1}^{(ijr)} \, \frac{P_{ij}^{\hc} (z, 1 - z)}{y \, \sk{jr}{ijr}}
  \nnb \\
  & = & \Norm \, \frac{\varsi_\npt}{\varsi_\npo} \,\,
  J^{\hc}_{ij} \! \left( \sk{jr}{ijr}, \eps \right) \, \bbS{k}\,\bbC{kl} \, \Rl \, .
\label{eq: int bar SCkij Cijkl (1-Sik-Sjk) RR}
\eeq
Collecting all the above integrated terms, the resulting integral $I^{\otwohc}$ is
\beq
  I^{\otwohc} & = & \frac{\varsi_\npt}{\varsi_\npo} \, \int d \Phi_{\rm rad,1}^{(ijr)} \,
  \overline K^{\otwohc}
  \nnb \\
  & = & - \, \Norm \, \frac{\varsi_\npt}{\varsi_\npo} \sum_{i, \, j > i} \, 
  \sum_{k \neq i,j} \, J^{\hc}_{ij} \! \left( \sk{jr}{ijr}, \eps \right) \, \bigg\{ \,
  \Big[ \bC{jk} \, \left( \bW{jk}^{(ijr)} + \bW{kj}^{(ijr)} \right) \Big] \, \bbC{jk}
  \nnb\\ [2pt]
  & & \hspace{10mm}
  + \sum_{l \neq i,j,k} \left( \bC{kl} \, \bW{kl}^{(ijr)} \right) \, \bbC{kl}   + \, \left( \bS{j} \, \bW{jk}^{(ijr)} \right) \, \bbS{j} + \sum_{l \neq i,k}
  \left( \bS{k} \, \bW{kl}^{(ijr)} \right) \, \bbS{k}
  \nnb \\ [2pt]
  & & \hspace{10mm}
 -  \left( \bS{j} \, \bC{jk} \, \bW{jk}^{(ijr)} \right) \,
  \bbS{j} \, \bbC{jk}  - \, \left( \bS{k} \, \bC{jk} \, \bW{kj}^{(ijr)} \right) \, 
  \bbS{k} \, \bbC{jk}
  \nnb \\ [2pt]
  & & \hspace{10mm} - \sum_{l \neq i,k} \left( \bS{k} \, \bC{kl} \, \bW{kl}^{(ijr)} \right) \,
  \bbS{k} \, \bbC{kl} \bigg\} \, \Rl \left( \! \kkl{ijr} \! \right) \, ,
\label{eq:I12hcsemifinal}
\eeq
which can be straightforwardly rewritten as \eq{eq:I12hcfinal}.

We next turn to the soft term in \eq{eq:K12so1}. Using \eq{eq:propNNLOfact}, together 
with
\beq
  \bSC{ikl} \, \bC{ijkl} \, \RR \, = \, \bS{i} \, \bC{ijkl} \, \RR \, ,
\label{somecomm}
\eeq
and introducing, as usual, remapped kinematics for the sector functions and for the
limits of the matrix element, we obtain the expression
\beq
\label{eq:K12so}
  \overline K^{\otwos} & = &  - \sum_{i, \, k \neq i} \sum_{l \neq i,k} \,
  \bigg[ \bS{i} \sum_{j \neq i} \Wab{ij} \bigg] \nnb \\ [2pt]
 & & \hspace{7mm}
  \times \, \bigg\{  \left( \bS{k} \, \bW{kl} \right) \, \bbS{i} \, \bbS{ik} + 
  \left(\bC{kl} \, \bW{kl} \right) \, \bbSC{ikl} - \left( \bS{k} \, \bC{kl} \, \bW{kl} \right)\, 
  \bbSC{ikl} \, \bbS{ik} \bigg\} \RR \nnb \\ [2pt]
  & & - \sum_{i, \, j \neq i} \sum_{\substack{k \neq i \\ k > j}}
  \bigg[ \bS{i} \, \bC{ijk} \left( \Wab{ij} + \Wab{ik} \right) \bigg] \,
  \bigg\{ \Big[ \bC{jk} \left( \bW{jk} + \bW{kj} \right) \Big]
  \nnb \\ [-5pt]
  && \hspace{7mm}
  - \, \left( \bS{j} \, \bC{jk} \, \bW{jk} \right) \bbS{ij} - 
  \left(\bS{k} \, \bC{jk} \, \bW{kj} \right) \bbS{ik} \bigg\}
  \big( \bbS{i} - \bbSC{ijk} \big) \, \bbC{ijk}\, \RR \, .
\eeq
By means of \eq{eq:CC S Wab}, and renaming indices, we finally get
\beq
\label{eq:K12ssymm}
  \overline K^{\otwos} & = & - \sum_{i,\, j \neq i} \sum_{k \neq i,j}
  \bigg\{ \left( \bS{j} \, \bW{jk} \right) \, \bbS{i} \, \bbS{ij} +
  \left (\bC{jk} \, \bW{jk} \right) \, \Big[ \, \bbS{i} \, \bbC{ijk} + 
  \bbSC{ijk} \left( 1 - \bbC{ijk} \right) \Big]
  \nnb\\
  && \hspace{21mm}
  - \, \left( \bS{j} \, \bC{jk} \, \bW{jk} \right) \, \bbS{ij} \,
  \Big[ \, \bbS{i} \, \bbC{ijk} + \bbSC{ijk} \left( 1 - \bbC{ijk} \right) \Big]
  \bigg\} \, \RR \, .
\eeq
The singular limits in \eq{eq:K12ssymm}, as well as their phase-space integrals, 
are explicitly computed below. For brevity, in the following we set $\Rl_{ab} \equiv 
\Rl_{ab} \left(\kkl{iab}\right)$ unless explicitly stated otherwise. Let us begin by 
considering the iteration of a soft limit and a double-soft limit. We find 
\beq
  \int d \Phi_{\rm rad,1} \, \bbS{i} \, \bbS{ik} \, \RR & = & - \, \Norm \, 
  \frac{\varsi_\npt}{\varsi_\npo} \, \sum_{c \neq i, d \neq i} \!
  \bbS{k} \, \Rl_{cd} \, \frac{1}{\sk{cd}{icd}} \int d \Phi_{\rm rad,1}^{(icd)} \,
  \frac{1 - z}{y z}
  \nnb \\
  & = & - \, \Norm \, \frac{\varsi_\npt}{\varsi_\npo} \,\, 
  \delta_{f_ig} \, \sum_{c \neq i, d \neq i} \!\! J^{\so} \Big( \sk{cd}{icd}, \eps \Big) \,
  \bbS{k} \, \Rl_{cd} \, ,
\label{eq: int bar Si Sik RR}
\eeq
where the soft integral $J^{\so}$ is defined in \eq{eq:Jsoft}. Next, we note that collinear
contributions to \eq{eq:K12ssymm} are proportional to the combination $\big [ \, \bbS{i} \, \bbC{ijk} + 
  \bbSC{ijk} \big( 1 - \bbC{ijk} \big) \big] \RR $, which integrates to
\beq
\label{eq: int bar (C+CC-CCC)S RR}
  & & \int d \Phi_{\rm rad,1} \Big[ \, \bbS{i} \, \bbC{ijk} + 
  \bbSC{ijk} \big( 1 - \bbC{ijk} \big) \Big] \RR  \, = \, - \, \Norm \, 
  \frac{\varsi_\npt}{\varsi_\npo} \,\, \sum_{\substack{c \neq i \\ d \neq i}} \,
  \bbC{jk}\, \Rl_{cd} \, \frac{1}{\sk{cd}{icd}} \int d \Phi_{\rm rad,1}^{(icd)} \,
  \frac{1-z}{yz}
  \nnb \\ & & \hspace{4cm} = \,
  - \, \Norm \, \frac{\varsi_\npt}{\varsi_\npo} \,\, \delta_{f_ig}
  \sum_{\substack{c \neq i \\ d \neq i}} \, J^{\so} \Big( \sk{cd}{icd}, \eps \Big) \,
  \bbC{jk} \, \Rl_{cd} \, .
\eeq
Further acting with a double-soft limit 
$\bbS{ij}$ we get
\beq
\label{eq: int bar (C+CC-CCC)S.SS RR}
  & & \int d \Phi_{\rm rad,1} \, \bbS{ij} \, \Big[ \, \bbS{i} \, \bbC{ijk} + 
  \bbSC{ijk} \left( 1 - \bbC{ijk} \right) \Big] \, \RR \nnb \\ 
  & & \hspace{3cm} = \, - \, \Norm \, 
  \frac{\varsi_\npt}{\varsi_\npo} \sum_{c \neq i, d \neq i}
  \bbS{j} \, \bbC{jk} \, \Rl_{cd} \, \frac{1}{\sk{cd}{icd}}
  \int d \Phi_{\rm rad,1}^{(icd)} \, \frac{1 - z}{y z}
  \nnb \\ 
  & & \hspace{3cm} = \, - \, \Norm \, \frac{\varsi_\npt}{\varsi_\npo} \,\, \delta_{f_ig} \!
  \sum_{c \neq i, d \neq i} \! J^{\so} \Big( \sk{cd}{icd}, \eps \Big) \,
  \bbS{j}\,\bbC{jk} \, \Rl_{cd} \, .
\eeq
Collecting all the integrated contributions, the resulting integrated soft counterterm is
\beq
  I^{\otwos} & = & \frac{\varsi_\npt}{\varsi_\npo} \int d \Phi_{\rm rad,1} \,
  \overline K^{\otwos}
  \nnb\\
  & = & - \, \Norm \, \frac{\varsi_\npt}{\varsi_\npo} \, \sum_i \, \delta_{f_ig} \,
  \sum_{\substack{k \neq i \\ l \neq i,k}} \sum_{\substack{a \neq i \\ b \neq i}} \,
  J^{\so} \Big( \sk{ab}{iab}, \eps \Big) 
\label{fin} \\
  & & \hspace{0mm} \times \Big[ \left( \bS{k} \, \bW{kl}^{(iab)} \right) \, \bbS{k} + 
  \left( \bC{kl} \, \bW{kl}^{(iab)} \right) \, \bbC{kl} - \left( \bS{k} \, \bC{kl} \, \bW{kl}^{(iab)} \right) \,
  \bbS{k} \, \bbC{kl} \Big] \, R_{ab} \Big( \kkl{iab} \Big) \, , \nnb
\eeq
which can be straightforwardly rewritten as \eq{eq:I12sfinal}.

%%%%%%%%%%%%%%%%%%%%%%%%%%%%%%%%%

\bibliographystyle{JHEP}
\bibliography{subt}

%%%%%%%%%%%%%%%%%%%%%%%%%%%%%%%%%

\end{document}